%% file: main.tex
\documentclass{article}

\usepackage{microtype}
\usepackage{graphicx}
\usepackage{subfigure}
\usepackage{booktabs} 
\usepackage{adjustbox}
\usepackage{makecell}
\usepackage{hyperref}

\usepackage[accepted]{icml2024}

\usepackage{amsmath}
\DeclareMathOperator{\Tr}{Tr}

\usepackage{amssymb}
\usepackage{mathtools}
\usepackage{amsthm}
\usepackage{braket}
\usepackage{multirow}
\usepackage{scalerel}
\usepackage[capitalize,noabbrev]{cleveref}

\usepackage{algorithm}
\usepackage{algorithmic}
\usepackage{listings}

\usepackage{xcolor}
\usepackage{longtable}
\usepackage{arydshln}

\usepackage{tikz}
\usetikzlibrary{quantikz2}

\usepackage{subcaption}

\lstdefinelanguage{json}{
    basicstyle=\normalfont\ttfamily,
    numberstyle=\scriptsize,
    stepnumber=1,
    numbersep=8pt,
    showstringspaces=false,
    breaklines=true,
    frame=lines
}

\definecolor{Tommaso}{rgb}{0.8,0.2,0.2}
\newcommand{\tom}[1]{\textcolor{Tommaso}{[TF] #1}}

\newcommand{\gab}[1]{\todo[color=green]{#1}}

\newcommand{\andres}[1]{\textcolor{blue}{[AR] #1}}

\usepackage[normalem]{ulem}
\usepackage{url}
\theoremstyle{plain}

\theoremstyle{definition}

\theoremstyle{remark}

\usepackage[textsize=tiny]{todonotes}

\icmltitlerunning{Quantum feature encoding optimization}

\begin{document}



\twocolumn[
\icmltitle{Quantum feature encoding optimization}



\icmlsetsymbol{equal}{*}

\begin{icmlauthorlist}
\icmlauthor{Tommaso Fioravanti}{zzz}
\icmlauthor{Brian Quanz}{yyy}
\icmlauthor{Gabriele Agliardi}{yyy}
\icmlauthor{Edgar Andres Ruiz Guzman}{yyy}
\icmlauthor{Ginés Carrascal}{yyy}
\icmlauthor{Jae-Eun Park}{yyy}
\end{icmlauthorlist}

\icmlaffiliation{yyy}{IBM Quantum}
\icmlaffiliation{zzz}{IBM Italy}

\icmlcorrespondingauthor{Tommaso Fioravanti}{tommaso.fioravanti@ibm.com}

\icmlkeywords{Machine Learning, ICML}

\vskip 0.3in
]



\printAffiliationsAndNotice{}  

\input{sections/abstract}

\input{sections/introduction}

\input{sections/related_work}

\input{sections/methodology}

\input{sections/experiments}

\input{sections/conclusion}

\clearpage
\nocite{langley00}
\bibliography{ref}
\bibliographystyle{icml2024}

\newpage
\input{sections/appendix}


\end{document}

%% file: sections/abstract.tex
\begin{abstract}
Quantum Machine Learning (QML) holds the promise of enhancing machine learning modeling in terms of both complexity and accuracy. A key challenge in this domain is the encoding of input data, which plays a pivotal role in determining the performance of QML models. In this work, we tackle a largely unaddressed aspect of encoding that is unique to QML modeling -- rather than adjusting the ansatz used for encoding, we consider adjusting how data is conveyed to the ansatz.  We specifically implement QML pipelines that leverage classical data manipulation (i.e., ordering, selecting, and weighting features) as a preprocessing step, and evaluate if these aspects of encoding can have a significant impact on QML model performance, and if they can be effectively optimized to improve performance. Our experimental results, applied across a wide variety of data sets,  ansatz, and circuit sizes, with a representative QML approach, demonstrate that by optimizing how features are encoded in an ansatz we can substantially and consistently improve the performance of QML models, making a compelling case for integrating these techniques in future QML applications.  Finally we demonstrate the practical feasibility of this approach by running it using real quantum hardware with 100 qubit circuits and successfully achieving improved QML modeling performance in this case as well.  
\end{abstract}

%% file: sections/introduction.tex
\section{Introduction}
\label{sec: intro}
In recent years, quantum computing has gained increasing interest, driven by  advancements in quantum error correction~\cite{Bravyi_2024,acharya2024quantumerrorcorrectionsurface} and mitigation~\cite{cai2023quantum,temme2017error,nation2021scalable,berg:2022,van2023probabilistic}, and improvements in hardware capabilities~\cite{HW_improvement_IBM,HW_improvement_IBM_2,HW_improvement_IBM_3,HW_improvement_IBM_4}. Researchers are exploring various domains to identify where this paradigm could fulfill its promise of performing certain tasks more efficiently or effectively than classical computers. One such domain is machine learning, and quantum computing applied to machine learning, or Quantum Machine Learning (QML), holds the potential to enhance or out-perform classical machine learning approaches in terms of both computational efficiency and accuracy~\cite{Schuld_2014,Cerezo2022}. 

Among the various paradigms currently investigated in QML, two main classes of near-term implementable approaches have emerged as particularly prominent.
These are Quantum Kernel methods~\cite{Huang_2021, miyabe2023quantummultiplekernellearning, Jerbi_2023, schuld2021supervisedquantummachinelearning}, and the broader class of Variational Quantum Classifiers (VQC)~\cite{Havl_ek_2019} which includes Quantum Neural Networks~\cite{Abbas_2021,farhi2018classification}.
Both of these nearer-term quantum algorithms use quantum circuits as feature maps to encode classical input data into a quantum state, with the latter also including data-independent variational gates that depend on some parameters to be optimized through a classical optimization procedure, and state or operator measurement. 

The first step of data encoding is a crucial phase of QML algorithms that affects their computational capabilities~\cite{bowles2023contextualityinductivebiasquantum, kübler2021inductivebiasquantumkernels, hw_effects_on_vqc, alami2025comparativeperformanceanalysisquantum, Park_2023}; numerous encoding strategies can be explored, ranging from straightforward methods like basis encoding, amplitude encoding, and angle encoding~\cite{LaRose_2020}. Angle encoding has become the most dominant approach in the recent years for near-term applicability, with commonly used parametrized quantum feature maps such as EfficientSU2 and Pauli feature map~\cite{Havl_ek_2019}. In addition, more complicated encoding schemes have recently been explored. LCU-based feature maps induce a non-Fourier spectrum through measurement by preparing quantum states as linear combinations of unitaries (LCUs)~\cite{LCUfm_1, LCUfm_2} while ground state-based feature maps embed data into highly specific quantum states, acting as quasi-Fourier models with exponentially scaling spectrum~\cite{groundstatefm}.

Due to the importance of the data encoding, one key area of research has been on how to find the right feature map for a given dataset.  Most research has focused on searching or optimizing over the particular parameterized circuit used, for example, with kernel alignment~\cite{Gentinetta_2023},  evolutionary algorithms~\cite{zhang2023evolutionary}, or reinforcement learning~\cite{dai2024quantum}.

However, one complementary aspect that has received less attention is how the method of feeding data into a particular feature map affects overall QML model performance.  
For example, this could correspond to which features are included in the encoding (feature selection), what order the features are fed into the encoding circuit, or how they are scaled / weighted.  The latter two of these examples are more uniquely applicable to QML modeling, as most unstructured / general-purpose classical models are insensitive to the ordering of the features, and are either insensitive to the scaling or include feature normalization as part of their ML pipelines.  

In this work, we explore this aspect in particular and propose a general framework for 
optimizing how input data is manipulated before encoding into a given feature map to improve QML model performance.




\comment{Other works \gab{"other works" gives the impression you have to fill out a list. say why the paper is important to your research, instead} present new data features selection techniques exploiting QSVC~\cite{9915517} or treating this task as a QUBO problem~\cite{hellstern2023quantumcomputerbasedfeature},~\cite{M_cke_2023}.}

\comment{In this work, we integrate various data manipulation techniques into the QML processes, aiming to explore and analyze how preprocessing input data before encoding them into the ansatz affects model performance; we call this procedure Quantum Feature Encoding Optimization (QFEO).}

Our contributions are the following:
\begin{itemize}
    \comment{\item We conduct an in-depth investigation into how classical data should be injected into a given quantum feature map, focusing on the impact of data manipulation prior to encoding.
    we enable effective and efficient optimizing of realistic estimates of test scores unlike past related approaches most other work – with our approach of efficient black box (e.g., Bayesian) optimization combined with both tuning and cross-validation evaluation of the given pipeline to accurately capture inference time performance of the complete model
    }
    \item We introduce an innovative and general framework -- referred to as Quantum Feature Encoding Optimization (QFEO) -- that tunes input features before encoding to maximize model performance within a fixed feature map structure. In our framework, unlike past related approaches, cross-validation is used both for hyperparameter tuning and for estimating the QML test performance --  i.e., importantly, we estimate true model performance under the applied data manipulation.
    \item We suggest a subset of simple and interpretable data manipulation techniques in our experimental study that are interesting as being mostly unique to QML models and not yet explored in the literature. However, the framework remains general and can support any data manipulation method and any QML model.
    \item We demonstrate the consistent and significant impact of input feature manipulation and our QFEO optimization strategy across a wide range of experiments.
    \item We validate the practical feasibility of our approach by executing it on actual quantum hardware at a significant scale, showing that the QFEO approach is feasible and works in practice on real quantum hardware.
\end{itemize}

The rest of the paper is organized as follows: in Section~\ref{sec:related_work}, we report relevant literature reviews; in Section~\ref{sec:method}, we describe our framework QFEO starting from a general overview and then describing in more detail some of the possible optimization methodologies; in Section~\ref{sec:experiments}, we provide a detailed explanation of all the choices made regarding the feature map and classifier for conducting the experiments. Additionally, we report both numerical and visual results for the most significant cases across three different feature maps implemented, accompanied by relevant analyses for  noiseless simulator execution. In Section~\ref{sec: real_hw_experiments}, we present experiments on real hardware to showcase the effectiveness of our procedures even on the current noisy Quantum Processing Units (QPU).
In Section~\ref{sec:conclusion}, we summarize the conclusions by highlighting the key findings derived from the results.

%% file: sections/related_work.tex
\section{Related Work}
\label{sec:related_work}
\comment{In the field of Quantum Machine Learning, the Quantum Kernel approach is a popular method, which leverages quantum computers to more estimate the kernel matrix in a quantum feature space that is hard to simulate classically. The Projected Quantum Kernel approach~\cite{Huang_2021} simplifies the quantum kernel computation by approximating the quantum state density matrix using One-Particle Reduced Density Matrix (1-RDM) \gab{why? how is this connected to your topic?}. However, this method is still limited to SVC/kernel method \gab{SVC is not introduced -- define the acronym. how does it relate to SVC? avoid column and semicolumn, only use full stop}: with our Projected Quantum Feature Map (PQFM) extension \gab{is this a new contrib? if so, it must be listed in the intro. I would not describe further your contrib, in the section of related work. move what follows somewhere else!}, we unlock quantum feature maps to be used with any kind of Machine Learning model \gab{any kind sounds very broad -- are you sure?}. PQFM is part of the Quantum Machine Learning model that we want to optimize by exploiting and combining classical data feature engineering techniques, such as Feature Selection or Feature Weighting, before feeding the data into the ansatz.}
One of the foundational components of QML is the encoding of classical data into quantum states. While various types of quantum feature maps have been explored in recent literature -- including both problem specific constructions and general purposes encoding~\cite{LCUfm_1, LCUfm_2, groundstatefm, Havl_ek_2019} -- the mainstream approach remains angle encoding~\cite{10493306}, due to its hardware efficiency and compatibility with parametrized quantum circuits~\cite{Havl_ek_2019}.

Although many studies explore encoding optimization ~\cite{review_modifying_encoding_circuit}, there is currently a lack of research on optimizing QML models by manipulating input data with classical techniques such as feature selection, ordering, weighting, and feature combining as is the focus here.  Furthermore the possibility of such pre-processing manipulations affecting QML model performance has been largely unexplored and unrealized, aside from mainly some works considering feature selection in particular. 
Driven by the promise of quantum algorithms for tackling computationally expensive combinatorial optimization tasks, prior works have been applied to develop new feature selection-based techniques, and some approaches have incorporated feature selection with QML modeling for specific models. 
In~\cite{9915517}, the feature importance is analyzed for use with a QSVM model using classical dimensionality reduction techniques ranging from Principal Component Analysis (PCA) to XGBoost integrated feature importance. The authors also proposed using a sequential, greedy forward feature selection approach \cite{Aha1996} with a given QSVM model to evaluate feature sets to iteratively build up a good set of features to use with it. In~\cite{wang2023novelfeatureselectionmethod}, Quantum Support Vector Machine feature selection (QSVMF) integrates QSVM with multi-objective genetic algorithms to select the best features reducing feature covariance and minimizing quantum circuit cost. They compare quantum feature selection against classical approaches showing the QSVMF superior performances. Moreover,~\cite{M_cke_2023} represents a line of other, unrelated work on feature selection where quantum optimization is used to select features for classical / independent of modeling -- in this case with a Quadratic Unconstrained Binary Optimization (QUBO) formulation.

Other research has devoted attention on enhancing particular parametrized circuits by utilizing a Variational Quantum Classifier approach (i.e., optimizing data-independent rotation parameters of a quantum circuit)  to try to find the most similar quantum kernel with respect to the target one (defined by the prediction targets) -- also referred to as Quantum Kernel Alignment ~\cite{Gentinetta_2023}. Since this approach is computationally expensive,~\cite{miyabe2023quantummultiplekernellearning} propose the quantum Multiple Kernel Learning approach, which builds upon the classical multiple kernel learning approach~\cite{cortes2024algorithmslearningkernelsbased} used in classical kernel-based machine learning methods, to enhance model performance when the data is difficult to model. However, these do not consider how changing the way the data is fed into the circuit impacts QML performance, or optimizing this aspect of the QML model, and hence is complementary to our work.

Moreover, with the rapid progress of Generative Artificial Intelligence (GenAI), large language models (LLMs)~\cite{zhao2025surveylargelanguagemodels} have recently been investigated as tools to automate the design of high-performance quantum circuits. In~\cite{QAS_llms}, Quantum GPT-Guided Architecture Search (QGAS) framework is presented, where GPT-4~\cite{openai2024gpt4technicalreport}, guided through iterative prompting, is employed to generate ansatz structures for variational quantum algorithms. Circuit blocks are selected from predefined design spaces and translated into Quantum Assembly Language (QASM)~\cite{Cross_2022}, with candidate architectures benchmarked on quantum chemistry and finance task. A related line of research focuses more broadly on Quantum Architecture Search (QAS) / Quantum Circuit Search (QCS)~\cite{qcs_1, qcs_2, qcs_3} designed to automatically discover circuits with high performance and robustness to noise. A recent work~\cite{QCS_aws} introduces a resource- and noise-aware QCS framework that systematically optimizes the search space, search strategy, and candidate evaluation. This framework rapidly prunes low-fidelity candidates reducing circuit evaluation costs and identifies circuits that achieve high classification accuracy across multiple QML benchmark. To the best of our knowledge, no QAS / QCS work focuses on finding the best ansatz by optimizing the way data is fed through classical manipulation techniques. Even QCS works employing neural networks, which could potentially be trained to learn data manipulation function such as feature selection / ordering / weighting, do not take this aspect into account, which remains unique to our work. 

All previous works share the common goal of enhancing the performance of QML models by proposing different type of feature maps to encode data or by incorporating feature selection with specific QML models. In our work, we contribute to the same objective but introducing a general framework to support optimizing how features are fed into the feature map by integrating classical feature engineering techniques into the QML process. We propose a comprehensive study which relies not only on feature selection but encompasses a broader set of approaches including other previously undiscovered and unexplored data manipulation approaches we identify and evaluate -- such as additional simple, interpretable manipulation schemes, like feature weighting and ordering.
\comment{\tom{TBD
The idea behind Projected Quantum Feature Map comes from Quantum Kernel method (\tom{references}) and its limitations. Quantum Kernel is not currently scalable to practical data with large number of data points $n$ and features $p$ since each kernel function must be evaluated for every pair of data points and typical feature encoding used requires one qubit per feature. Projected Quantum Kernel (\tom{references}) promises to reduce the complexity by approximating quantum kernel computation using an approximation of the quantum state density matrix.
Given a classical input data point $x_i$ and its corresponding quantum encoding $\ket{x_i}=\mathbf{U}(x_i)\ket{0}^{\otimes n}$, the corresponding quantum state density matrix is defined as $\rho(x_i)=\mathbf{U}(x_i)\ket{0}^{\otimes n}\bra{0}^{\otimes n}\mathbf{U}(x_i)^{\dag}$. The one-particle reduced density matrix (1-RDM) via partial trace could be used to approximate the density matrix as $\rho_k(x_i)=\Tr_{j\neq k}[\rho(x_i)]$ and define a new kernel function $k^{PQ}(x_i,x_j)=\exp{(-\gamma\sum_k ||\rho_k(x_i)-\rho_k(x_j)||^2_F)}$.
The problem is that this new kernel definition is still limited to SVM model and kernel methods and still does not address large number of features. From here our extension: to compute the 1-RDM for projected kernel calculation, we measure expectations for Pauli X,Y, and Z observables per qubit which contain the same information as the 1-RDMs representation. We use these set of measurements as new set of classical, real values features (unlike 1-RDMs with imaginary numbers) and we call this procedure Projected Quantum Feature Map (PQFM). In Figure~\ref{fig: PQFM}, we highlight the main components of PQFM: supposing $\ket{0}^{\otimes n}$ as the initial quantum state, a unitary $U$ acts as a dense quantum feature map encoding input data $x\in\mathbb{R}^{d\times p}$ into quantum registers. We measure expectations for Pauli X, Y, and Z observables per qubit obtaining three different real values per qubit, as $m_1, m_2,$ and $m_3$ for the first qubit. Finally, we stack real values $m_1, m_2, \dots, m_k$ as new features set for the input data $x$ which means projecting $x: \mathbb{R}^{d\times p}\rightarrow \mathbb{R}^{d\times 3n}$. If $x$ has large dimension $p$ and the number of qubits used to encode is significantly lower (so $n\ll p$), PQFM allows for a reduction to a low-dimensional classical feature space which could generalize better since it is obtained by projecting from the high-dimensional Hilbert space carrying information and properties related to the evolution of the quantum state.}}

%% file: sections/methodology.tex
\section{Methodology}
\label{sec:method}
\subsection{Preliminaries and Notations}
\label{subsec: background_methodology}
In QML, there are two main classes of near-term implementable approaches: Quantum Kernel (QK) methods and Variational Quantum Classifiers (VQC) (also commonly referred to as quantum neural networks, QNNs) \cite{jerbi2023quantum,jager2023universal}. In both approaches, classical data must be encoded into quantum states before further processing. This is typically done via a parameterized quantum circuit $U_{enc}$, known as an ansatz or feature map~\cite{Cerezo_2021, Havl_ek_2019}, which governs how information propagates through the model. 
\begin{figure}[h]
\centering
\includegraphics[width=0.48\textwidth]{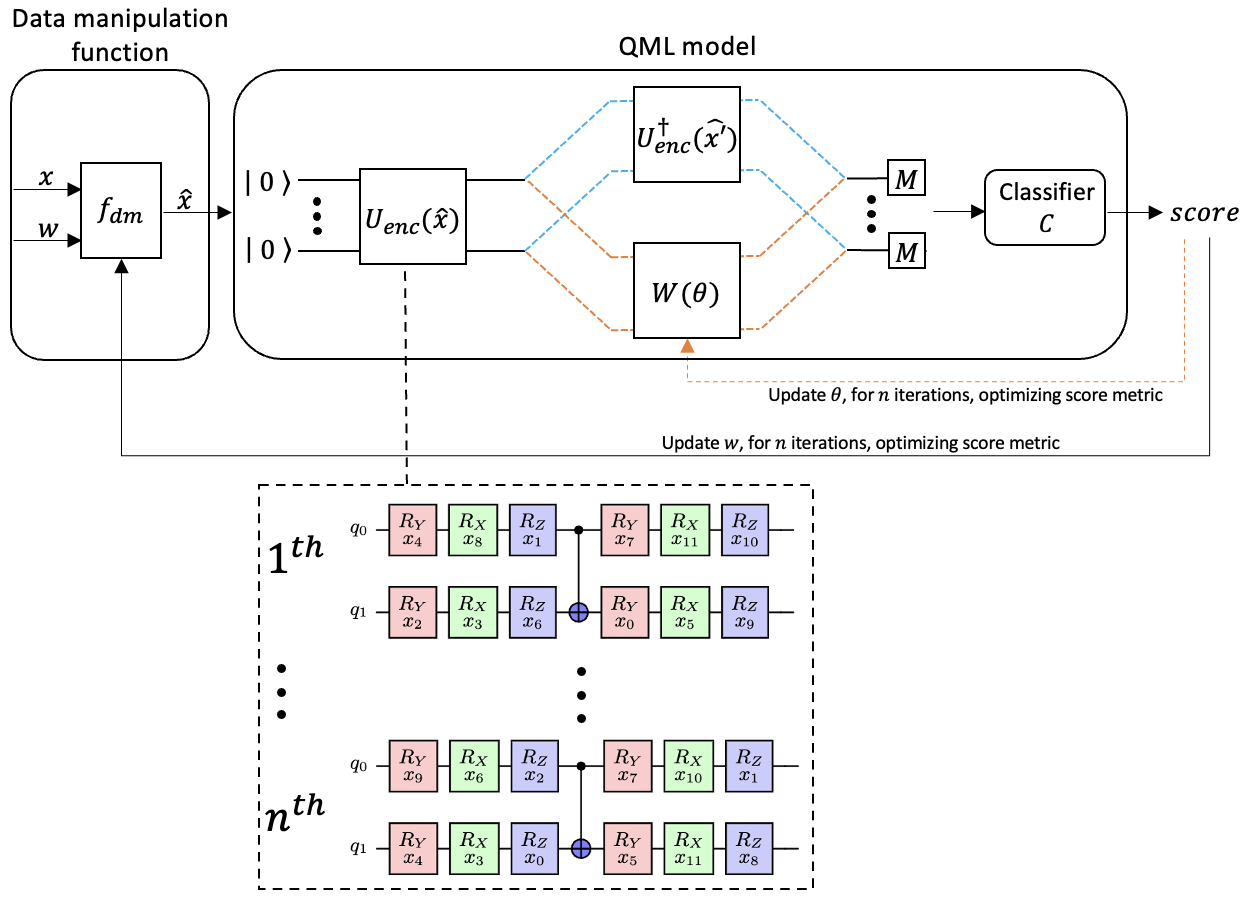}
\caption{Quantum Kernels (blue path) and Variational Quantum Classifiers (orange path) expect static initial encodings $U_{enc}(x)$. Our contribution is to optimize the way we encode the features in the initial feature map (e.g., by changing the order of the features over $n$ iterations)}
\label{fig: core_idea}
\end{figure}
\comment{In Section~\ref{subsec: qefo_framework}, we detail our approach which can be regarded as a variation of standard Variational Quantum Classifier (VQC) framework. Let us define a classical dataset $\mathcal{D} = \{(\vec{x_i}, y_i)\}_{i=1}^N$, where each input $\vec{x_i}\in\mathbb{R}^p$ is a real-valued feature vector and $y_i$ is a two-class target label}Standard VQCs schemes, differently from QKs, also expect a variational parametrized circuit $W(\vec{\theta})$ whose parameters $\theta$ need to be optimized through a classical optimization procedure. Indeed, given $\vec{x}\in\mathbb{R}^p$ as a real-valued input feature vector, the data encoding circuit block $U_{enc}(\vec{x})$ prepares the reference state $\ket{0}^{\otimes n} \xrightarrow{U_{enc}(\vec{x})} U_{enc}(\vec{x})\ket{0}^{\otimes n} = \ket{\psi}$. This encoded state is then passed through the parametrized circuit $W(\vec{\theta})$, producing the final state $\ket{\psi}\xrightarrow{W(\vec{\theta})} W(\vec{\theta})\ket{\psi}=\ket{\Psi(\vec{x}, \vec{\theta})}$. Measurements are then performed on this final state to compute the output and evaluate a score used to drive the optimization. The optimization focuses solely on the trainable parameters $\vec{\theta}$, while the encoding remains static as in QKs.

In contrast, as represented in Figure~\ref{fig: core_idea}, our method directly optimizes the model performance score without changing $U_{enc}$ or $W$ by manipulating weights $w$ that are used by a black-box function $f_{dm}$ to produce the new input $\hat{x}$ to be encoded with  $U_{enc}$.  Thus our method results in a modified $\ket{\psi}$ and may drive different $\vec{\theta}$ to be selected, with the goal of achieving a better overall QML model without changing the particular encoding or QML model procedure used.   In fact, our approach is generic with respect to the QML model used, acting directly on the encoding (as pre-processing), a necessary part in any QML model. As mentioned, encoding can be performed in several ways, ranging from basis and amplitude encoding to angle encoding, but since angle encoding has become the predominant choice in part due to its suitability for near-term implementations~\cite{LaRose_2020}, for simplicity we focus on this encoding for the remainder of this paper, i.e., to describe our framework and for use in experiments. Specifically, given a feature $x_j$, we encode it with a predefined set of quantum rotations $R_X = e^{-i\frac{x_j}{2}X}, R_Y=e^{-i\frac{x_j}{2}Y},$ and $R_Z = e^{-i\frac{x_j}{2}Z}$. These basic operations are combined in various way and integrated with different entanglement strategies to define distinct feature map encoding schemes. The specific architectures we adopt are detailed in Section~\ref{subsub: ansatz}.

\subsection{Framework Overview}
\label{subsec: qefo_framework}
Algorithm~\ref{alg:feat_opt} describes the proposed Quantum Feature Encoding Optimization (QFEO) framework, which optimizes the way data is fed into a given feature map improving the performance of an underlying QML model. The QFEO framework provides a unique-to-quantum aspect of manipulating input features to improve QML modeling; we expand on this in the next section.

Starting from a classical dataset $\mathcal{D}\in\mathbb{R}^{d\times p}$ with features set $\mathcal{X}=\{x_0, x_1, \dots, x_p\}$, the core idea of the framework is to iteratively optimize a set of input weights through a specific data manipulation process -- of which we provide several examples in Section~\ref{subsec: data_manipulation}. The weights drive the manipulation and, thus, the modified input data features to be encoded in the QML model.

\begin{algorithm}
    \caption{Quantum Feature Encoding Optimization}
    \label{alg:feat_opt}
    \begin{algorithmic}[1]
    \STATE \textbf{Input}: training set $\mathcal{D}\in\mathbb{R}^{d_D\times p}$, test set $\mathcal{T}\in\mathbb{R}^{d_T\times p}$, feature map $\mathcal{F}$, number of iterations $i_{\mathcal{O}}$ of a classical optimization procedure $\mathcal{O}$, data manipulation $f_{dm}$, performance metric $\mathcal{M}$
    \STATE \textbf{Output}: Finalized QML model pipeline (including weights $\mathcal{W}$ for the data manipulation function $f_{dm}$) and test performance
    \STATE Generate (e.g., randomly) initial weights $\mathcal{W}=\{w_0, w_1, \dots\}\in \mathbb{R}^m$ -- e.g., a weight for each feature ($m=p$) in the case of basic common manipulation such as feature selection, ordering, or weighting. These weights are optimized during the process and drive the manipulation to obtain a new modified set of input features 
    \REPEAT 
    \STATE Based on the initialized weights, perform data manipulation $f_{dm}$ to obtain a remapping / modification of the features and encode them into the feature map $\mathcal{F}$
    \STATE Evaluate the performance of the QML model on the training set $\mathcal{D}$ for any particular scoring function -- e.g., in practice, to estimate the generalization performance of the QML model, we use cross-validation based on the specified metric $\mathcal{M}$
    \STATE Update weights $\mathcal{W}$ based on the QML model performance, according to  $\mathcal{O}$
    \UNTIL reached $i_{\mathcal{O}}$ iterations 
    \STATE Select the best QML model found after $i_{\mathcal{O}}$ iterations
    \STATE Use the best weights found after $i_{\mathcal{O}}$ iterations to perform data manipulation $f_{dm}$ to obtain a remapping of the features on the test set $\mathcal{T}$ and encode them into the feature map $\mathcal{F}$  
    \STATE Evaluate the performance of the QML model on the test set based on the specified metric $\mathcal{M}$ 
    \end{algorithmic}
\end{algorithm}
\subsection{Data Manipulation Step}
\label{subsec: data_manipulation}
In classical Machine Learning, data preprocessing steps such as feature reordering (changing the order of the features) or feature scaling typically have no effect on model performance for most standard algorithms.
For example, learning for tree-based, linear models, and traditional neural nets is invariant to the order of input features (as weights or parameters can just be permuted without changing outputs), and many of them are also insensitive to feature scaling either by design (tree-based models) or by standard preprocessing like normalization necessary for the algorithms to learn effectively (linear and neural net models). 

In contrast, in QML, data must be encoded into quantum states using an initial feature map. The manner in which classical data is mapped into quantum states can significantly impact the expressivity of the quantum model, the nature of the quantum states created, and the ability of the quantum ML model to generalize. As such, we put forth that operations that are innocuous in classical settings -- like permuting the order of features or applying simple multiplicative weights -- can have nontrivial consequences in QML due to the nonlinear, often non-symmetric, and entangled structure of the quantum circuits used in feature maps and QML models. While our focus is on simple and interpretable discrete and continuous data manipulation techniques -- illustrated in Figure~\ref{fig: QFO} and detailed in the following subsections -- our framework is general and can accommodate a wider spectrum of strategies, paving the way for future extensions.
In Figure~\ref{fig: circuit expressibility} in Appendix ~\ref{sec: expressibility}, we also report an analysis on the expressibility of the circuit when QFEO is engaged. We demonstrate that different data manipulation techniques influence the expressiveness of a quantum circuit -- that is, its ability to generate a rich variety of quantum states, thus exploring a broader portion of the Hilbert space. 

\begin{figure*}[h]
    \centering
    \begin{tabular}{cc}
        \subfigure[No Feature Optimizations / Manipulations]{
        \scalebox{0.8}{
        \begin{quantikz}[column sep=2mm]
            \lstick{$q_0$}& \gate[style={fill=red!20}]{\shortstack{$R_Y$ \\ $ \alpha x_0$}} & \gate[style={fill=green!20}]{\shortstack{$R_X$ \\ $\alpha x_2$}} & \gate[style={fill=blue!20}]{\shortstack{$R_Z$ \\ $ \alpha x_4$}} & \ctrl{1} & \gate[style={fill=red!20}]{\shortstack{$R_Y$ \\ $ \alpha x_6$}} & \gate[style={fill=green!20}]{\shortstack{$R_X$ \\ $ \alpha x_8$}} & \gate[style={fill=blue!20}]{\shortstack{$R_Z$ \\ $ \alpha x_{10}$}}&&    \\
            \lstick{$q_1$}& \gate[style={fill=red!20}]{\shortstack{$R_Y$ \\ $ \alpha x_1$}} & \gate[style={fill=green!20}]{\shortstack{$R_X$ \\ $ \alpha x_3$}} & \gate[style={fill=blue!20}]{\shortstack{$R_Z$ \\ $ \alpha x_5$}} & \targ[style={fill=blue!50,draw=black}]{} & \gate[style={fill=red!20}]{\shortstack{$R_Y$ \\ $ \alpha x_7$}} & \gate[style={fill=green!20}]{\shortstack{$R_X$ \\ $ \alpha x_9$}} & \gate[style={fill=blue!20}]{\shortstack{$R_Z$ \\ $ \alpha x_{11}$}} &&    
        \end{quantikz}
        }
        \label{subfig: nfo}
        }
        &
        \subfigure[Feature Selection]{
        \scalebox{0.8}{
        \begin{quantikz}[column sep=2mm]
            \lstick{$q_0$}& \gate[style={fill=red!20}]{\shortstack{$R_Y$ \\ $\alpha x_0$}} & \gate[style={fill=green!20}]{\shortstack{$R_X$ \\ $\alpha x_3$}} & \gate[style={fill=blue!20}]{\shortstack{$R_Z$ \\ $\alpha x_5$}} & \ctrl{1} & \gate[style={fill=red!20}]{\shortstack{$R_Y$ \\ $\alpha x_8$}} &&    \\
            \lstick{$q_1$}& \gate[style={fill=red!20}]{\shortstack{$R_Y$ \\ $\alpha x_2$}} & \gate[style={fill=green!20}]{\shortstack{$R_X$ \\ $\alpha x_4$}} & \gate[style={fill=blue!20}]{\shortstack{$R_Z$ \\ $\alpha x_7$}} & \targ[style={fill=blue!50,draw=black}]{} & \gate[style={fill=red!20}]{\shortstack{$R_Y$ \\ $\alpha x_{11}$}} &&    
        \end{quantikz}
        }
        \label{subfig: fs}
        }
        \\
        \subfigure[Feature Weighting]{
        \scalebox{0.8}{
        \begin{quantikz}[column sep=2mm]
            \lstick{$q_0$}& \gate[style={fill=red!20}]{\shortstack{$R_Y$ \\ $ w_0\alpha x_0$}} & \gate[style={fill=green!20}]{\shortstack{$R_X$ \\ $ w_2\alpha x_2$}} & \gate[style={fill=blue!20}]{\shortstack{$R_Z$ \\ $ w_4\alpha x_4$}} & \ctrl{1} & \gate[style={fill=red!20}]{\shortstack{$R_Y$ \\ $ w_6\alpha x_6$}} & \gate[style={fill=green!20}]{\shortstack{$R_X$ \\ $ w_8\alpha x_8$}} & \gate[style={fill=blue!20}]{\shortstack{$R_Z$ \\ $ w_{10}\alpha x_{10}$}}&&    \\
            \lstick{$q_1$}& \gate[style={fill=red!20}]{\shortstack{$R_Y$ \\ $ w_1\alpha x_1$}} & \gate[style={fill=green!20}]{\shortstack{$R_X$ \\ $ w_3\alpha x_3$}} & \gate[style={fill=blue!20}]{\shortstack{$R_Z$ \\ $ w_5\alpha x_5$}} & \targ[style={fill=blue!50,draw=black}]{} & \gate[style={fill=red!20}]{\shortstack{$R_Y$ \\ $ w_7\alpha x_7$}} & \gate[style={fill=green!20}]{\shortstack{$R_X$ \\ $ w_9\alpha x_9$}} & \gate[style={fill=blue!20}]{\shortstack{$R_Z$ \\ $ w_{11}\alpha x_{11}$}} &&    
        \end{quantikz}
        }
        \label{subfig: fw}
        }
        &
        \subfigure[Feature Ordering]{
        \scalebox{0.8}{
        \begin{quantikz}[column sep=2mm]
            \lstick{$q_0$}& \gate[style={fill=red!20}]{\shortstack{$R_Y$ \\ $\alpha x_4$}} & \gate[style={fill=green!20}]{\shortstack{$R_X$ \\ $\alpha x_8$}} & \gate[style={fill=blue!20}]{\shortstack{$R_Z$ \\ $\alpha x_1$}} & \ctrl{1} & \gate[style={fill=red!20}]{\shortstack{$R_Y$ \\ $\alpha x_7$}} & \gate[style={fill=green!20}]{\shortstack{$R_X$ \\ $\alpha x_{11}$}} & \gate[style={fill=blue!20}]{\shortstack{$R_Z$ \\ $\alpha x_{10}$}}&&    \\
            \lstick{$q_1$}& \gate[style={fill=red!20}]{\shortstack{$R_Y$ \\ $\alpha x_2$}} & \gate[style={fill=green!20}]{\shortstack{$R_X$ \\ $\alpha x_3$}} & \gate[style={fill=blue!20}]{\shortstack{$R_Z$ \\ $\alpha x_6$}} & \targ[style={fill=blue!50,draw=black}]{} & \gate[style={fill=red!20}]{\shortstack{$R_Y$ \\ $\alpha x_0$}} & \gate[style={fill=green!20}]{\shortstack{$R_X$ \\ $\alpha x_5$}} & \gate[style={fill=blue!20}]{\shortstack{$R_Z$ \\ $\alpha x_9$}} &&    
        \end{quantikz}
        }
        \label{subfig: fo}
        }
    \end{tabular}
    \caption{We report four examples to visualize different feature encoding manipulations considering a generic feature map. We define $\alpha$ as a multiplicative factor applied to all features (a common hyper parameter in QML methods). In these examples, we assume to have input data with 12 features $\{x_0, \dots, x_{11}\}$ to encode. In Feature Weighting, $w_i$ is the scaling factor applied to the i-th feature.  Note that combination of manipulations are also valid, e.g., feature weighting and ordering.}
    \label{fig: QFO}
\end{figure*}
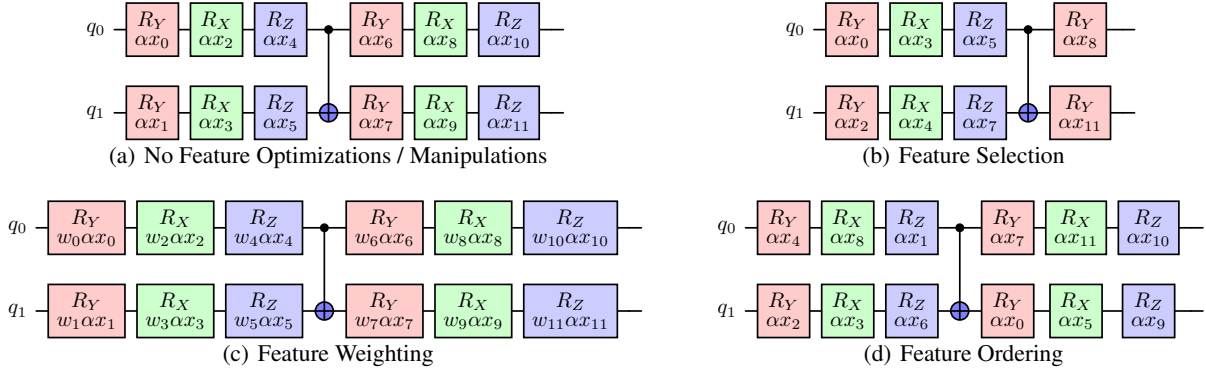

\comment{
\begin{figure*}
    \centering
    \subfigure[No Feature Optimizations]{
        \includegraphics[width=0.45\textwidth]{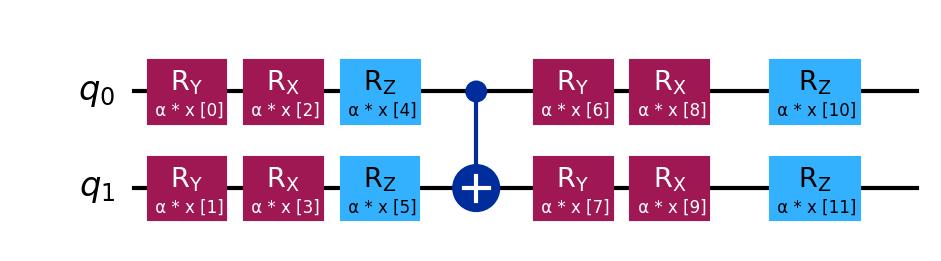}
        \label{subfig: nfo}
    }
    \subfigure[Feature Selection]{
        \includegraphics[width=0.45\textwidth]{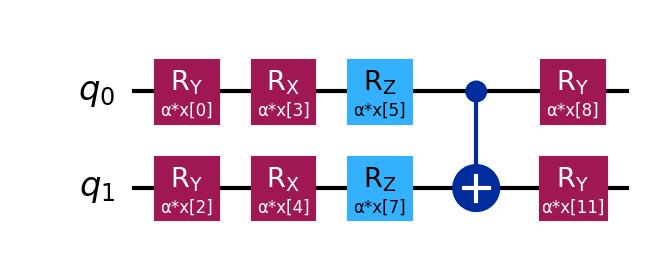}
        \label{subfig: fs}
    }\\
    \subfigure[Feature Weighting]{
        \includegraphics[width=0.45\textwidth]{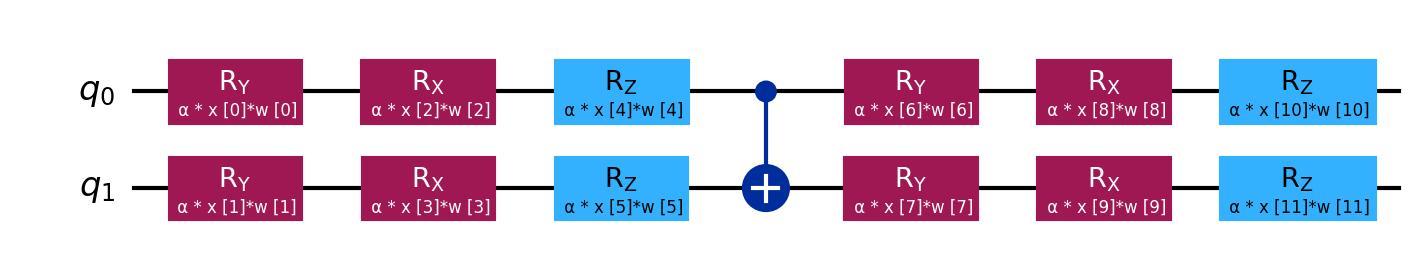}
        \label{subfig: fw}
    } 
    \subfigure[Feature Ordering]{
        \includegraphics[width=0.45\textwidth]{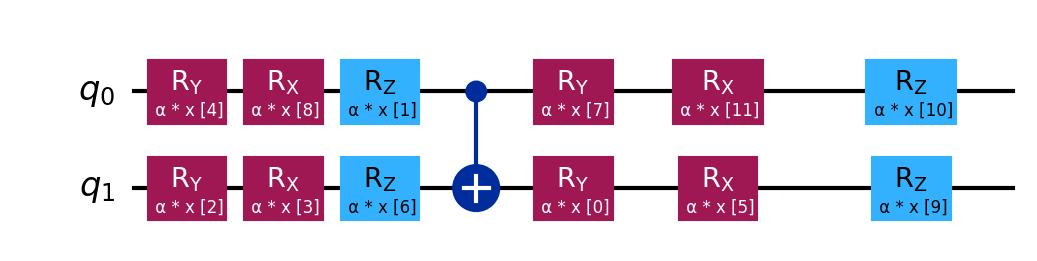}
        \label{subfig: fo}
    }
    \caption{We report four examples to visualize different feature encoding optimizations considering a generic ansatz. We assume to have input data with 12 features $\{x[0], \dots, x[11]\}$ to encode. In Figure~\ref{subfig: nfo}, we report the encoding of the whole input data without feature optimization; in Figure~\ref{subfig: fs}, we report an example of the Feature Selection optimization where we assume to select the best 8 features; in Figure~\ref{subfig: fw}, we provide Feature Weighting optimization which can be seen as a relaxation of Feature Selection where we encode all the 12 features and scale the rotation factor with different weights $w[i]$ based on the feature relevance; in Figure~\ref{subfig: fo}, we report the Feature Ordering optimization where the features are reordered before being encoded. We also combine these feature encoding optimizations to implement Feature Selection Ordering and Feature Weighting Ordering, as described in Section~\ref{sec: feat_selection_ordering} and in Section~\ref{sec: feat_weighting_ordering} respectively.}
    \label{fig: QFO}
\end{figure*}
}
\subsubsection{Feature Selection}
\label{par: feat_selection}
In general, a feature selection procedure reduces the dimensionality of the feature space of a given dataset $\mathcal{D}\in \mathbb{R}^{d\times p}$, by selecting the best $r$ features out of $p$, with $r<p$. In this work, we apply feature selection as a particular data manipulation technique to select the best $r$ features corresponding to the highest $r$ input weights generated in step 3 of the Algorithm \ref{alg:feat_opt}, reducing the dimensionality of the input data and thus the time to fit the QML model. In practice the goal is to find the set of $r$ features that makes the QML classifier outperform the No Feature Optimization (NFO) baseline, which is the QML model where we statically encode the input features without any kind of optimization.
\subsubsection{Feature Weighting}
\label{par: feat_weight}
The feature weighting is a relaxation of feature selection optimization where the less important features are weighted less than the more important ones but are not removed - which could influence the impact different features have on the quantum state encoding. We assume that the set of weights is composed of a real number between 0 and 1, $\mathcal{W} = [0,1]$.
I.e., a useless feature is weighted with a 0 value and, as the importance of the feature increases, the associated weight value also gets higher up to the value of 1, which is associated with the most important features. Therefore, in steps 5 and 10 of Algorithm~\ref{alg:feat_opt}, we use the input weights to scale the value of the corresponding feature and, therefore, the rotation angles used for the encoding in the feature map. The functionality is similar to feature selection but not removing any feature could lead to the creation of more unique and complex quantum feature maps. The goal is to find the set of weights $\mathcal{W}$ which scales the input features and allows to obtain better performance compared to the standard QML models.

\subsubsection{Feature Ordering}
\label{par: feat_ordering}
The goal of this data manipulation is to encode the features in an order that maximizes the QML model performance. Following Algorithm~\ref{alg:feat_opt}, we order the features based on the weights $\mathcal{W}$: the rule is that the features corresponding to the highest weight are encoded first in the feature map $\mathcal{F}$. In this way, we test whether encoding the features in a certain order matters for the feature map and thus for the performance of the whole QML model.

\subsubsection{Feature Selection Ordering}
\label{sec: feat_selection_ordering}
This is a combined data manipulation technique, where we join feature selection and feature ordering described in Section~\ref{par: feat_selection} and~\ref{par: feat_ordering} respectively. The aim is to see whether selecting and then ordering the features yields a greater advantage compared to only performing feature selection. Even though we perform two types of manipulations, we only leverage one set of weights for both techniques: we use input weights $\mathcal{W}$ to select $r$ out of $p$ features, where $r<p$, and then we order the selected features based on the corresponding weights. For example, suppose $r=3$ and an initial set of weights $\mathcal{W}=\{w_0=0.1, w_1=0.5, w_3=0.02, w_4=0.8, w_5=0.4\}$ related to features $\mathcal{X}=\{x_1, x_2, x_3, x_4, x_5\}$. We select features $x_1, x_4,$ and $x_5$ associated to highest weights $w_1, w_4,$ and $w_5$, encoding them in the order $x_4, x_5,$ and $x_1$.
\subsubsection{Feature Weighting Ordering}
\label{sec: feat_weighting_ordering}
As for the feature selection ordering described in Section~\ref{sec: feat_selection_ordering}, we combine two different manipulation procedures. In this case, we perform feature ordering after scaling their values with feature weighting described in Section~\ref{par: feat_weight}. The purpose is similar to the one of features selection ordering; we want to test if ordering features after scaling them brings better benefits than just scaling. We present two different approaches: one involving a single set of weights for both feature weighting and ordering (FWO), and the other employing two distinct sets of weights for the respective manipulations (FWOW). This is because we want to check whether two sets of weights are required to capture any possible weighting and ordering combination. Suppose $\mathcal{W}_w=\{ww_1, ww_2, ww_3, ww_4, ww_5\}$ and $\mathcal{W}_o=\{wo_1=0.1, wo_2=0.02, wo_3=0.6, wo_4=0.8, wo_5=0.4\}$ as initial sets of weights, related to features $\mathcal{X}=\{x_1, x_2, x_3, x_4, x_5\}$, for feature weighting and ordering optimizations respectively. We use weights $\mathcal{W}_w$ to scale features as $x_1\cdot ww_1, x_2\cdot ww_2, x_3\cdot ww_3, x_4\cdot ww_4, x_5\cdot ww_5$ and then order them as $x_4\cdot ww_4, x_3\cdot ww_3, x_5\cdot ww_5, x_1\cdot ww_1, x_2\cdot ww_2$ based on the values of $\mathcal{W}_o$. The procedure with just one set of weights for both the optimizations follows the one explained in Section~\ref{sec: feat_selection_ordering} with the difference that the weights are used to scale and not to select the features.
\comment{
\subsection{Circuit Expressibility}
\label{subsec: expressibility}
In this section, we explore how the data encoding strategies influence the expressiveness of a quantum circuit -- that is, its ability to generate a rich variety of quantum states. A circuit with higher expressiveness can explore a broader portion of the Hilbert space.
To quantify expressiveness, we begin by generating a set of random features sampled from a normal distribution. 
Then, we apply random permutations of the features (Feature Ordering), random weighting (Feature Weighting), or feature subset selection (Feature Selection) across multiple iterations to assess the circuit's sensitivity to different feature optimization strategies. Specifically, for each of the $T = 1000$ iterations, we apply a unique permutation, weighting, or subset selection to the input features, encode them into a quantum state, and extract the resulting statevector. This process yields a final statevector $S \in \mathbb{C}^{T\times2^{n}}$, where $n$ is the number of qubits.

To analyze the diversity of the resulting state space, we apply Principal Component Analysis (PCA) on the matrix $S$ by employing SVD. The number of principal components required to retain a specified level of variance serves as an indicator of expressiveness \gab{citation?} -- the more components needed, the more expressive the circuit. 
Additionally, we assess the statevector reconstruction error to further evaluate expressiveness. This is done by truncating the SVD to a rank $r$, reconstructing the statevector matrix $\hat{S}_{r}$, and computing the reconstruction error as $E_r = || S - \hat{S}_r||_2$, where $\hat{S}_r = U_r\Sigma_rV^{\dagger}_r$.
Higher reconstruction error for smaller ranks suggests greater circuit expressiveness since it indicates that the circuit explores a broader and more complex region of the Hilbert space that cannot be easily captured by a low-dimensional representation.
The entire procedure is repeated 30 times, each time with a newly generated set of input features sampled from the same distribution. The final metrics, such as the reconstruction error, are then averaged across all repetitions. This allows us to evaluate the expressiveness of the circuit under varying input conditions, ensuring the robustness of the results with respect to the input data.
Figures~\ref{subfig: l2_error_4qubits} and~\ref{subfig: l2_error_9qubits} show the statevector reconstruction error for Feature Ordering, Feature Selection, and Feature Weighting optimization strategies, applied to circuits with 4 and 9 qubits, respectively. In Figures~\ref{subfig: n_components_4_qubits} and~\ref{subfig: n_components_9_qubits}, we present the number of principal components retained after applying PCA in each case.
The most notable result is that Feature Ordering enables the circuit to explore a broader portion of the Hilbert space compared to Feature Weighting. This is evidenced by the consistently higher reconstruction error at lower ranks, as well as the greater number of principal components required to retain a given amount of variance. Additionally, it is observed that, as the number of qubits increases, the reconstruction error curves become smoother. In the Feature Ordering case, the number of retained components grows significantly with the number of qubits, whereas in the Feature Weighting case, it remains relatively stable, indicating a limited increase in expressiveness. Regarding Feature Selection, it is noteworthy that as the number of qubits increases, both the reconstruction error and the number of retained principal components gradually shift from resembling the Feature Ordering case to approaching the Feature Weighting case. However, in the high-variance regime, the number of retained components remains approximately twice as high as in the Feature Weighting case.

\begin{figure*}[t]
    \centering
    \subfigure[L2 norm reconstruction error retained 4 qubits]{
        \includegraphics[width=0.45\textwidth]{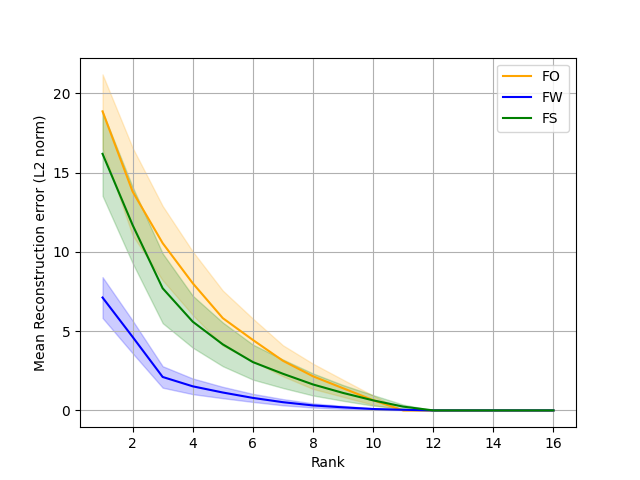}
        \label{subfig: l2_error_4qubits}
    }
    \subfigure[L2 norm reconstruction error retained 9 qubits]{
        \includegraphics[width=0.45\textwidth]{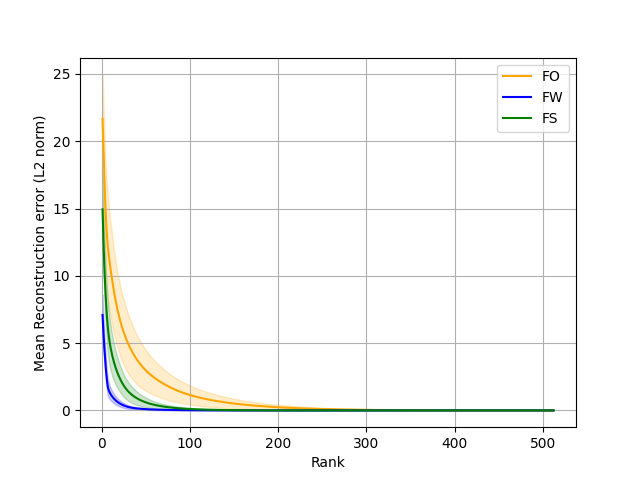}
        \label{subfig: l2_error_9qubits}
    }\\
    \subfigure[Principal components retained 4 qubits]{
        \includegraphics[width=0.45\textwidth]{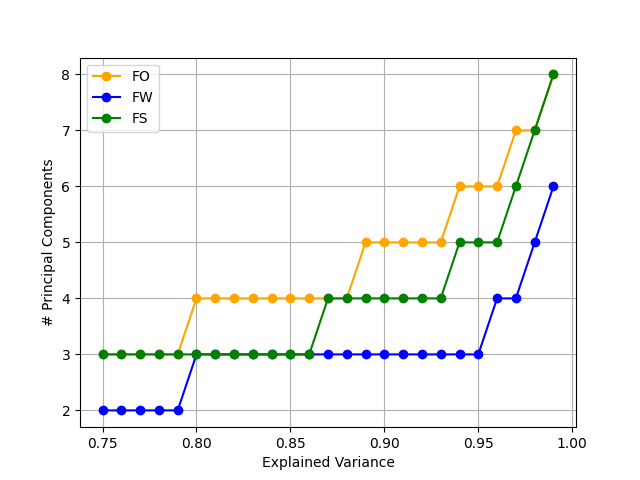}
        \label{subfig: n_components_4_qubits}
    } 
    \subfigure[Principal components retained 9 qubits]{
        \includegraphics[width=0.45\textwidth]{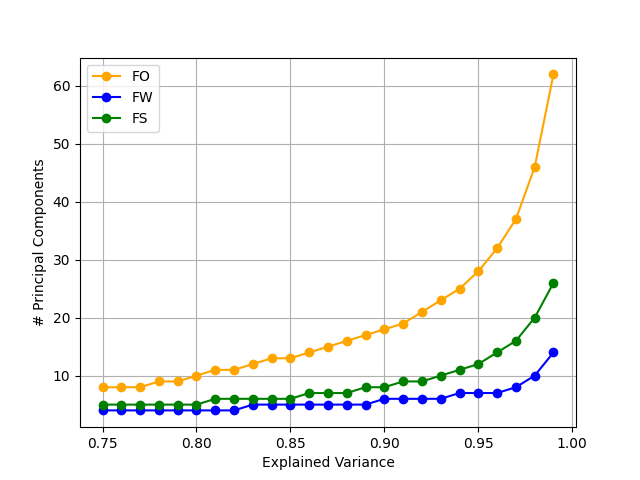}
        \label{subfig: n_components_9_qubits}
    }
    \caption{Reconstruction error (Figures~\ref{subfig: l2_error_4qubits},~\ref{subfig: l2_error_9qubits}) and number of retained principal components (Figures~\ref{subfig: n_components_4_qubits},~\ref{subfig: n_components_9_qubits}) as a function of the SVD rank and explained variance, respectively, for circuits with 4 and 9 qubits. The results compare Feature Ordering (FO), Feature Selection (FS), and Feature Weighting (FW) optimization strategies using the Heisenberg Hamiltonian as the feature map. Each curve represents the average over 30 independent runs with randomly sampled input features.}
    \label{fig: circuit expressibility}
\end{figure*}
}

\subsection{Projected Quantum Feature Map}
\label{subsec: pqfm}
\begin{figure}[h]
\centering
\includegraphics[width=0.48\textwidth]{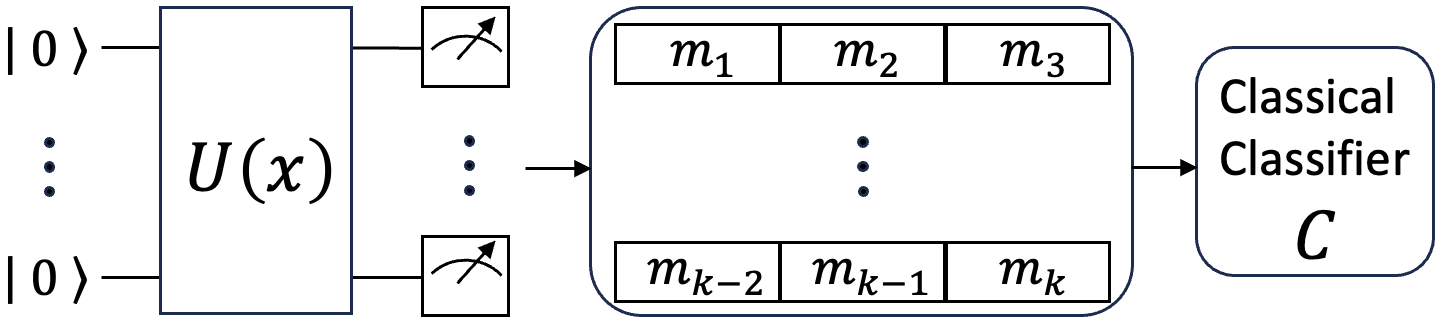}
\caption{Representative QML model schema. Feature map $U$ is used to encode input data $x$ into quantum registers. Then, measurements of expectation values for Pauli X, Y, and Z observables are performed to extract a new set of real-valued classical features that are used to fit a classifier $C$.}
\label{fig: QML model}
\end{figure}

Our framework is generic to any QML model. In this work, we consider as a representative QML model the one illustrated in Figure~\ref{fig: QML model}.
More precisely, we take into account a unitary $U(x)$ that acts as a dense quantum feature map encoding input data $x\in \mathbb{R}^{d\times p}$\footnote{In practice, depending on the feature map encoding, we restrict the value range.} into a $n$-qubit quantum register. Then, we measure expectations for Pauli X, Y, and Z observables per qubit obtaining three different real values per qubit, as $m_1, m_2,$ and $m_3$ for the first qubit. We stack real values $m_1, m_2, \dots, m_k$ as new features set for the input data $x$ which means projecting $x: \mathbb{R}^{d\times p}\rightarrow \mathbb{R}^{d\times 3n}$. Finally, we fit a classifier $C$ with the newly obtained feature set. We refer to this approach as the Projected Quantum Feature Map (PQFM) and which is an extension of the Projected Quantum Kernel (PQK) technique~\cite{Huang_2021} which approximates fidelity-based quantum kernels, and has also been referred to as a post-variational quantum neural net approach ~\cite{huang2023post,yogaraj2025post}, as it also approximates quantum neural nets.  Thus it is representative of both main stream near-term QML approaches - quantum kernel methods and QNNs. 

The projected quantum kernel approach itself may not handle large number of features effectively and is restricted to kernel-based models, which may not adequately fit the input data. This can be problematic in certain scenarios, for example, when irrelevant features are generated from the observables. 
Instead by using the measurements directly as transformed features we enable using any classical ML model / classifier instead of just kernel methods, including those implicitly incorporating feature selection.  By employing the projected quantum approach, the time complexity also reduces since we approximate the One-Particle Reduced Density Matrix (1-RDM) computation measuring expectations for Pauli X,Y, and Z observables per qubit which contain the same information as the 1-RDMs representation. If input data $x$ has large dimension $p$ and the number of qubits used to encode is significantly lower (so $n\ll p$), projected quantum approaches can also allow for a reduction to a low-dimensional classical feature space. For tailored problems, PQK can outperform standard Quantum Kernel (QK) in generalization~\cite{Huang_2021}; accordingly, PQFM, as an extension of PQK, may also achieve better generalization when the data are suitably structured for it. 
The performance of this type of QML model depends on both the feature map $U$ and the classifier $C$: the specific choices we used in this work regarding these two modules are defined in Section~\ref{subsub: ansatz} and~\ref{subsub:classifier} respectively.

\subsection{Framework Implementation}
\label{subsec: framework implementation}

\begin{figure}[h]
\centering
\includegraphics[width=0.48\textwidth]{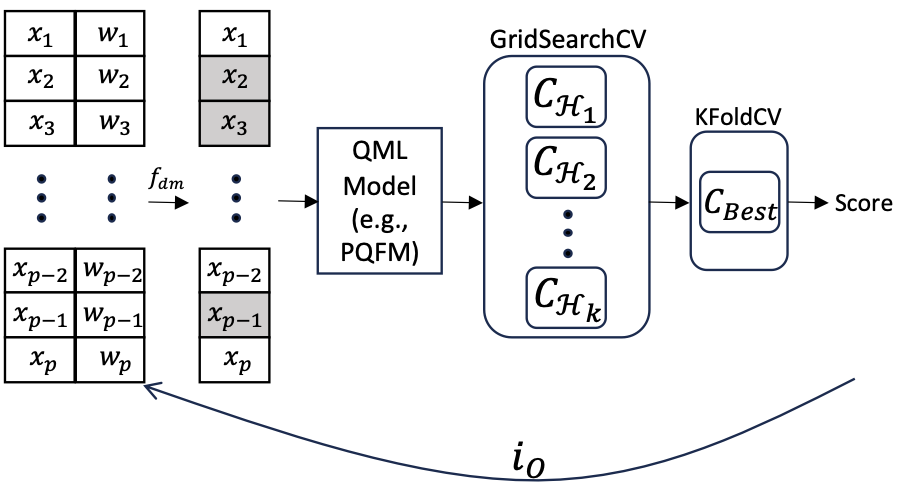}
\caption{QFEO framework phases overview. We illustrate feature selection as the data manipulation technique where the gray features are the ones that are used to fit the QML model, based on the input weights $w$. We apply grids search with cross-validation (GridSearchCV) to search for the best QML model classifier $C$ among $k$ different sets of hyperparameters $\mathcal{H}_k$, and estimate generalization / test performance with a different cross-validation evaluation (KFoldCV).
}
\label{fig: BO}
\end{figure}

Here we describe a concrete implementation of our framework for the goal of optimizing the feature encoding for a QML classification task, which will be the focus of our experimental study as well.  
In Figure~\ref{fig: BO}, we provide as overview of the main phases of our framework for this task with an illustrative feature encoding data manipulation being optimized.

 We start by generating initial (random) weights $w_i$ that drive the manipulation and, thus, the input data features to be encoded into the QML model. In practice, since we experiment with simple and interpretable data manipulation techniques, we generate a number of weights equal to the number of input features for most manipulation methods. 
 We use a classical data manipulation technique $f_{dm}$ to manipulate input features -- e.g., feature selection by selecting $k$ features, with $k<p$, based on the input weights, or feature ordering, where features are sorted based on the weights. Then, we fit a generic QML model with the manipulated feature set. Before encoding, we also scale the values of the features to a fixed interval (in our experiments, $[0.3, 2.8]$ using MinMaxScaler~\cite{scikit-learn-MinMaxScaler}) so that they do not lead to over-rotation with the rotational angle encoding and with the aim to reduce loss of information with dense encoding. By employing PQFM as a representative QML model, we project the feature set into a new classical feature set that we use to fit a classifier $\mathcal{C}$. Following this, we tune $\mathcal{C}$ given the current QML model encoding. In practice, we perform grid search with cross-validation -- GridSearchCV ~\cite{scikit-learn-GridSearchCV} -- (with the number of folds set to 5 for the cross-validation strategy in our experiments) to exhaustively test all $k$ possible combinations of hyperparameter values, defined in Frame~\ref{lst: hyperparams} in Appendix~\ref{app: hyperparams}, to find the best configuration which makes $\mathcal{C}$ achieve the highest score. By accurately reflecting the realistic modeling process, this approach enables direct optimization of the true performance -- the essential factor for ensuring consistent applicability and effectiveness when applied in practice / at test time. 
 
 Since the final score also depends on the hyperparameters of the feature map, in practice the ideal procedure to get the best QML model performance possible would include the PQFM hyperparameters in the GridSearchCV step as well, to search for the best model configuration not only among the hyperparameter values of the classifier but also among these of the feature map (such as the rotational scaling paramter $\alpha$ also sometimes referred to as the kernel bandwidth \cite{park2020practical,shaydulin2022importance}). This complete procedure would require a lot of computational power and execution time, and since the focus of this work is on improving the encoding given a fixed feature map, in our experiments here we fix the hyperparameters of a given feature map and evaluate how much our data manipulation framework alone can improve performance for each feature map, and perform GridSearchCV only among the hyperparameter values of the classifier.  In order to evaluate the ability to consistently improve performance across different possible feature maps and their hyperparameters, we apply our framework across a variety of multiple different feature maps and hyperparameters - as consistent improvement would imply benefit regardless of what particular feature map hyperparameters may work best for a given task.  The different feature maps and hyperparameters used in our experiments are reported in Frames~\ref{lst: SE_config} -~\ref{lst: RepeatedPauli_config} in Appendix~\ref{sub: ansatz_hyperparams}.
 
 Finally, we derive an estimate of the test performance by employing K-fold cross-validation -- KFoldCV ~\cite{scikit-learn-KFoldCV} -- (in this case with the number of folds K=10, and using a different random seed from the grid search) on the best classifier $\mathcal{C}_{best}$ found by the GridSearchCV to compute the final score (which corresponds to the cost function) as the average of the K different scores. We repeat this procedure $i_{\mathcal{O}}$ times, applying a suitable optimizer to adjust the initial weights at each iteration, driven by the objective of maximizing the final score.    At the end of the optimization procedure, we select the classifier which results in the best score over the $i_{\mathcal{O}}$ iterations to evaluate its performance over the projected test set. 

 In practice for our concrete implementation and our experiments we use Bayesian Optimization (BO) with a gaussian process surrogate model ~\cite{frazier2018tutorial,head_2021_5565057}.  We decided to use BO as it is an effective black box optimization approach that is targeted at quickly exploring a large, complex, and uncertain optimization landscape and in particular for cases when the evaluation cost is high - which matches QML modeling scenarios since applying the feature map could in general requiring executing multiple circuits on a quantum computer - so a limited number of evaluations is desirable or even necessary.  Additionally with the simple kind of manipulations we explore, the number of optimization variables (weights $\mathcal{W}$) is not too large (on the order of number of features) making BO still a suitable choice.  In practice, however, we could use any black box optimizer, and exploring alternatives such as including local search is an area of future work.  
Additionally, speculated and theoretical limitations of BO in higher dimensions prompted us to explore other optimization methods / extensions, such as SAAS-BO~\cite{eriksson2021highdimensionalbayesianoptimizationsparse} and LA-MCTS~\cite{song2022montecarlotreesearch}, which aim to improve black box optimization in high dimensions. However, we found traditional BO to still outperformed these alternatives in our case in preliminary experiments, leading us to select this approach, and demonstrating its capability of effective performance even with types of problem sizes coming from our feature encoding optimization scenarios. 

%% file: sections/experiments.tex
\section{Noiseless Simulator Experiments}
\label{sec:experiments}
\subsection{Datasets}
\label{subsec: dataset}
To validate the performance of the proposed techniques under different data structures, we test four benchmark datasets for classification, namely Churn~\cite{churn_dataset}, Virtual Screening~\cite{vs_dataset}, German Numeric~\cite{german_dataset}, and PLAsTiCC Astronomy~\cite{plastic_dataset} (details in Table~\ref{tab:training_test}). 

\comment{
\begin{table}[ht]
\centering
\caption{Dataset Characteristics (n=\# of data points, p=\# of features)}
\begin{tabular}{lccc}
\toprule
Dataset & n & p & \% pos. \\
\midrule
Churn & 4406 & 97 & 50.00 \\
Virtual Screening & 4700 & 1024 & 1.17 \\
German Numeric & 1000 & 20 & 30.00 \\
PLAsTiCC & 4863 & 67 & - \\
\bottomrule
\end{tabular}
\end{table}
}
\begin{table}[ht]
\centering
\small
\begin{tabular}{l|cc|c|c|c}
\hline
\multirow{2}{*}{Dataset} & \multicolumn{2}{c|}{n} & \multirow{2}{*}{p} & \multirow{2}{*}{\% pos.} \\ \cline{2-3}
        & Training & Test & &        \\ \hline
Churn & 4406 & 2938  & 97  & 50.0\%  \\ 
Virtual Screening & 13651 & 6725 & 47  & 26.4\%  \\ 
German Numeric & 670  & 330 & 24  & 30.0\%  \\ 
PLAsTiCC Astronomy & 2349 & 1157 & 67  & 66.0\%  \\ 
\hline
\end{tabular}
\caption{Dataset Characteristics (n=\# of data points, p=\# of features, \% pos.= percentage of positive labels)}
\label{tab:training_test}
\end{table}

For each dataset, we create 10 random train-test splits which refer to as dataset batches.  For each dataset batch, we randomly split the data into training and test sets reserving $33\%$ of the  samples for the test set. We use a stratified sampling procedure to ensure that the relative class frequencies is preserved on both the train and test sets. In this way, for each dataset, we train and test our QML models over ten different dataset batches to obtain more accurate and reliable estimates of the classifier performances, and characterize the variance in performance as well. Since the Churn dataset is highly imbalanced, we balance it by randomly under-sampling the majority class both in the training and in the test sets - to allow for more consistent comparison across datasets and application of standard metrics. The Virtual Screening, German Numeric, and PLAsTiCC Astronomy datasets were already sufficiently balanced / only minorly imbalanced. For the Virtual Screening dataset, we take the ALDH1 target~\cite{vs_dataset} because it showed the most promise in the prior work -- i.e., lower performance with classical models and more potential for quantum methods.

\subsection{Experiment setup}

\label{subsec: experiments_setup}
\subsubsection{Feature Maps}
\label{subsub: ansatz}
We consider three different dense encoding feature maps to encode input data: the Separate Entangled~\cite{peters2021machinelearninghighdimensional}, the Heisenberg Hamiltonian~\cite{Huang_2021}, and the Repeated Pauli, an extension of the Pauli feature map~\cite{Havl_ek_2019,qiskit_paulifeaturemap} to enable encoding more features than qubits. The Separate Entangled applies $R_X, R_Y,$ and $R_Z$ rotations together with CNOTs entanglement operations, as exemplified in Figure~\ref{fig: SE_ansatz}.
\comment{
\begin{figure*}[h]
    \centering
    \begin{tabular}{ll}
        \subfigure[No Feature Optimizations]{
        \footnotesize
        \begin{quantikz}[column sep=2mm]
            \lstick{$q_0$}& \gate[style={fill=red!20}]{\shortstack{$R_Y$ \\ $ \theta_0$}} & \gate[style={fill=green!20}]{\shortstack{$R_X$ \\ $\theta_2$}} & \gate[style={fill=blue!20}]{\shortstack{$R_Z$ \\ $ \theta_4$}} & \ctrl{1} & \gate[style={fill=red!20}]{\shortstack{$R_Y$ \\ $ \theta_6$}} & \gate[style={fill=green!20}]{\shortstack{$R_X$ \\ $ \theta_8$}} & \gate[style={fill=blue!20}]{\shortstack{$R_Z$ \\ $ \theta_{10}$}}&&    \\
            \lstick{$q_1$}& \gate[style={fill=red!20}]{\shortstack{$R_Y$ \\ $ \theta_1$}} & \gate[style={fill=green!20}]{\shortstack{$R_X$ \\ $ \theta_3$}} & \gate[style={fill=blue!20}]{\shortstack{$R_Z$ \\ $ \theta_5$}} & \targ[style={fill=blue!50,draw=black}]{} & \gate[style={fill=red!20}]{\shortstack{$R_Y$ \\ $ \theta_7$}} & \gate[style={fill=green!20}]{\shortstack{$R_X$ \\ $ \theta_9$}} & \gate[style={fill=blue!20}]{\shortstack{$R_Z$ \\ $ \theta_{11}$}} &&    
        \end{quantikz}
        \label{fig: SE_ansatz}
        }
        &
        \subfigure[Feature Selection]{
        \footnotesize
        \begin{quantikz}[column sep=1mm]
            \lstick{$q_0$}& 
            \gate[style={fill=gray!20}]{U_3} & 
            \gate[2, style={fill=cyan!20}]{\shortstack{$R_{ZZ}$ \\ $\theta_0$}} & 
            \gate[2, style={fill=brown!20}]{\shortstack{$R_{YY}$ \\ $ \theta_0$}} & 
            \gate[2, style={fill=yellow!20}]{\shortstack{$R_{XX}$ \\ $ \theta_0$}} & 
            {} & 
            {} & 
            {} & 
            \gate[2, style={fill=cyan!20}]{\shortstack{$R_{ZZ}$ \\ $\theta_3$}} & 
            \gate[2, style={fill=brown!20}]{\shortstack{$R_{YY}$ \\ $ \theta_3$}} & 
            \gate[2, style={fill=yellow!20}]{\shortstack{$R_{XX}$ \\ $ \theta_3$}}
            & \\
            \lstick{$q_1$} & 
            \gate[style={fill=gray!20}]{U_3} &
            {} & 
            {}& 
            {} & 
            \gate[2, style={fill=cyan!20}]{\shortstack{$R_{ZZ}$ \\ $\theta_2$}} & 
            \gate[2, style={fill=brown!20}]{\shortstack{$R_{YY}$ \\ $ \theta_2$}} & 
            \gate[2, style={fill=yellow!20}]{\shortstack{$R_{XX}$ \\ $ \theta_2$}} && &&  
            \\
            \lstick{$q_2$}& 
             \gate[style={fill=gray!20}]{U_3} & 
             \gate[2, style={fill=cyan!20}]{\shortstack{$R_{ZZ}$ \\ $\theta_1$}} & 
            \gate[2, style={fill=brown!20}]{\shortstack{$R_{YY}$ \\ $ \theta_1$}} & 
            \gate[2, style={fill=yellow!20}]{\shortstack{$R_{XX}$ \\ $ \theta_1$}}& 
            {} &
            {} & 
            {}
            && &&  
            \\
            \lstick{$q_3$}& 
            \gate[style={fill=gray!20}]{U_3}
            & {}
            & {}
            & {}
            & {}
            & {}
            & {}
            & {}
            & {}
            & {}
            &
        \end{quantikz}
        \label{fig: HH_ansatz}
        }
        \\
        \subfigure[Feature Weighting]{
        \footnotesize
        \begin{quantikz}[column sep=2mm]
            \lstick{$q_0$}& \gate[style={fill=orange!20}]{H} & \gate[style={fill=green!20}]{\shortstack{$R_X$ \\ $\pi/2$}} & \gate[style={fill=violet!20}]{\shortstack{$P$ \\ $ \theta_0$}} & \gate[style={fill=green!20}]{\shortstack{$R_X$ \\ $-\pi/2$}} & {} & \ctrl{1} & {} & \ctrl{1} & \gate[style={fill=orange!20}]{H} & \gate[style={fill=green!20}]{\shortstack{$R_X$ \\ $\pi/2$}} & \gate[style={fill=violet!20}]{\shortstack{$P$ \\ $ \theta_2$}} & \gate[style={fill=green!20}]{\shortstack{$R_X$ \\ $-\pi/2$}} & {} & {} & \ctrl{1} & {} & \ctrl{1} & {}
            && \\
            \lstick{$q_1$}& 
            \gate[style={fill=orange!20}]{H} & \gate[style={fill=green!20}]{\shortstack{$R_X$ \\ $\pi/2$}} &\gate[style={fill=violet!20}]{\shortstack{$P$ \\ $ \theta_1$}} & 
            \gate[style={fill=green!20}]{\shortstack{$R_X$ \\ $-\pi/2$}} & 
            \gate[style={fill=orange!20}]{H} & 
            \targ[style={fill=blue!50,draw=black}]{} & 
            \gate[style={fill=violet!20}]{\shortstack{$P$ \\ $ \frac{1}{\alpha}\theta_0\theta_1$}} & 
            \targ[style={fill=blue!50,draw=black}]{} & 
             \gate[style={fill=orange!20}]{H} & 
              \gate[style={fill=orange!20}]{H} &
              \gate[style={fill=green!20}]{\shortstack{$R_X$ \\ $\pi/2$}} &\gate[style={fill=violet!20}]{\shortstack{$P$ \\ $ \theta_3$}} & 
            \gate[style={fill=green!20}]{\shortstack{$R_X$ \\ $-\pi/2$}} & 
            \gate[style={fill=orange!20}]{H} & 
            \targ[style={fill=blue!50,draw=black}]{} & 
            \gate[style={fill=violet!20}]{\shortstack{$P$ \\ $\frac{1}{\alpha} \theta_2\theta_3$}}& 
            \targ[style={fill=blue!50,draw=black}]{} 
            &
            \gate[style={fill=orange!20}]{H}
            &&    
        \end{quantikz}
        \label{fig: HH_ansatz}
        }
    \end{tabular}
    \caption{}
\end{figure*}
}
\begin{figure}[h]
    \centering
    \scalebox{0.9}{
    \begin{quantikz}[column sep=2mm]
            \lstick{$q_0$}& \gate[style={fill=red!20}]{\shortstack{$R_Y$ \\ $ \alpha x_0$}} & \gate[style={fill=green!20}]{\shortstack{$R_X$ \\ $\alpha x_2$}} & \gate[style={fill=blue!20}]{\shortstack{$R_Z$ \\ $ \alpha x_4$}} & \ctrl{1} & \gate[style={fill=red!20}]{\shortstack{$R_Y$ \\ $ \alpha x_6$}} & \gate[style={fill=green!20}]{\shortstack{$R_X$ \\ $ \alpha x_8$}} & \gate[style={fill=blue!20}]{\shortstack{$R_Z$ \\ $ \alpha x_{10}$}}&&    \\
            \lstick{$q_1$}& \gate[style={fill=red!20}]{\shortstack{$R_Y$ \\ $ \alpha x_1$}} & \gate[style={fill=green!20}]{\shortstack{$R_X$ \\ $ \alpha x_3$}} & \gate[style={fill=blue!20}]{\shortstack{$R_Z$ \\ $ \alpha x_5$}} & \targ[style={fill=blue!50,draw=black}]{} & \gate[style={fill=red!20}]{\shortstack{$R_Y$ \\ $ \alpha x_7$}} & \gate[style={fill=green!20}]{\shortstack{$R_X$ \\ $ \alpha x_9$}} & \gate[style={fill=blue!20}]{\shortstack{$R_Z$ \\ $ \alpha x_{11}$}} &&    
        \end{quantikz}
        }
    \caption{Example of 2-qubits Separate Entangled feature map. We define $\alpha$ as the Pauli rotation factor and $x_i$ is the $i$-th feature, $i=0, ..., 11$. \comment{We use the two-qubit controlled-X ($CX$) gate as the entangling operation.}}
    \label{fig: SE_ansatz}
\end{figure}

\comment{\begin{figure}[h]
\centering
\includegraphics[width=0.5\textwidth]{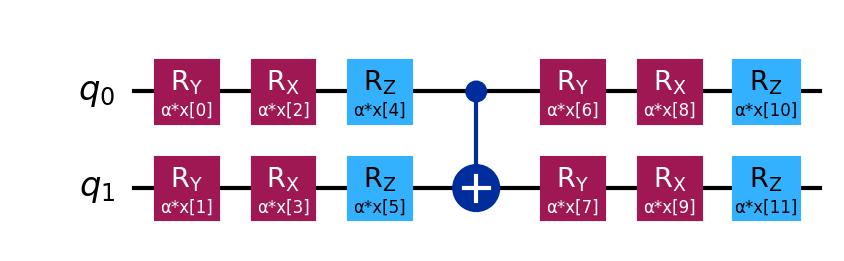}
\caption{Example of Separate Entangled ansatz. We assume to encode 12 features $\{x[0], x[1], \dots, x[11]\}$ into two qubits $q_0$ and $q_1$. We use the two-qubits controlled not-gate $C_X$ as entanglement operation.}
\label{fig: SE_ansatz}
\end{figure}
}
The Heisenberg Hamiltonian instead applies the unitary random gates $U3$ on all qubits, followed by entangling two-qubit rotations $R_{zz}, R_{xx}, R_{yy}$  to encode the input data, as illustrated in Figure~\ref{fig: HH_ansatz}. 
Lastly, the Repeated Pauli feature map is a modified version of the Pauli Feature Map~\cite{Havl_ek_2019,qiskit_paulifeaturemap}. In the original version, the number of qubits equals the number of feature, whereas we encode more features in the same number of qubits by repeating multiple blocks of Pauli Feature Maps, as depicted in Figure~\ref{fig: RepeatedPauli_ansatz}. The rotation angle alpha used in this feature map satisfies the condition $\alpha x_i x_j \in [0, 2\pi)$.

\begin{figure}[h]
    \centering
    \scalebox{0.7}{
    \begin{quantikz}[column sep=1mm]
            \lstick{$q_0$}& 
            \gate[style={fill=gray!20}]{U3} & 
            \gate[2, style={fill=cyan!20}]{\shortstack{$R_{ZZ}$ \\ $\alpha x_0$}} & 
            \gate[2, style={fill=brown!20}]{\shortstack{$R_{YY}$ \\ $ \alpha x_0$}} & 
            \gate[2, style={fill=yellow!20}]{\shortstack{$R_{XX}$ \\ $ \alpha x_0$}} & 
            {} & 
            {} & 
            {} & 
            \gate[2, style={fill=cyan!20}]{\shortstack{$R_{ZZ}$ \\ $\alpha x_3$}} & 
            \gate[2, style={fill=brown!20}]{\shortstack{$R_{YY}$ \\ $ \alpha x_3$}} & 
            \gate[2, style={fill=yellow!20}]{\shortstack{$R_{XX}$ \\ $ \alpha x_3$}}
            & \\
            \lstick{$q_1$} & 
            \gate[style={fill=gray!20}]{U3} &
            {} & 
            {}& 
            {} & 
            \gate[2, style={fill=cyan!20}]{\shortstack{$R_{ZZ}$ \\ $\alpha x_2$}} & 
            \gate[2, style={fill=brown!20}]{\shortstack{$R_{YY}$ \\ $ \alpha x_2$}} & 
            \gate[2, style={fill=yellow!20}]{\shortstack{$R_{XX}$ \\ $ \alpha x_2$}} && &&  
            \\
            \lstick{$q_2$}& 
             \gate[style={fill=gray!20}]{U3} & 
             \gate[2, style={fill=cyan!20}]{\shortstack{$R_{ZZ}$ \\ $\alpha x_1$}} & 
            \gate[2, style={fill=brown!20}]{\shortstack{$R_{YY}$ \\ $ \alpha x_1$}} & 
            \gate[2, style={fill=yellow!20}]{\shortstack{$R_{XX}$ \\ $ \alpha x_1$}}& 
            {} &
            {} & 
            {}
            && &&  
            \\
            \lstick{$q_3$}& 
            \gate[style={fill=gray!20}]{U3}
            & {}
            & {}
            & {}
            & {}
            & {}
            & {}
            & {}
            & {}
            & {}
            &
        \end{quantikz}
        }
    \caption{Example of Heisenberg Hamiltonian feature map. We define $\alpha$ as the Pauli rotation factor and $x_i$ is the i-th data feature. We assume to have input data with 4 features $\{x_0, x_1, x_2, x_3\}$ to encode into four qubits $q_0, q_1, q_2,$ and $q_3$.\comment{We apply random unitary gates at the beginning, followed by entanglement and encoding operations through $R_{ZZ}$, $R_{XX}$, and $R_{YY}$ rotation gates.}}
    \label{fig: HH_ansatz}
\end{figure}

\comment{in the sense that, in the phase gates between the entanglement operators, we apply a phase by directly performing multiplication between feature values instead of performing $\alpha*\prod_{i}(\pi-x[i])$.}

\comment{
\begin{figure}[h]
\centering
\includegraphics[width=0.5\textwidth]{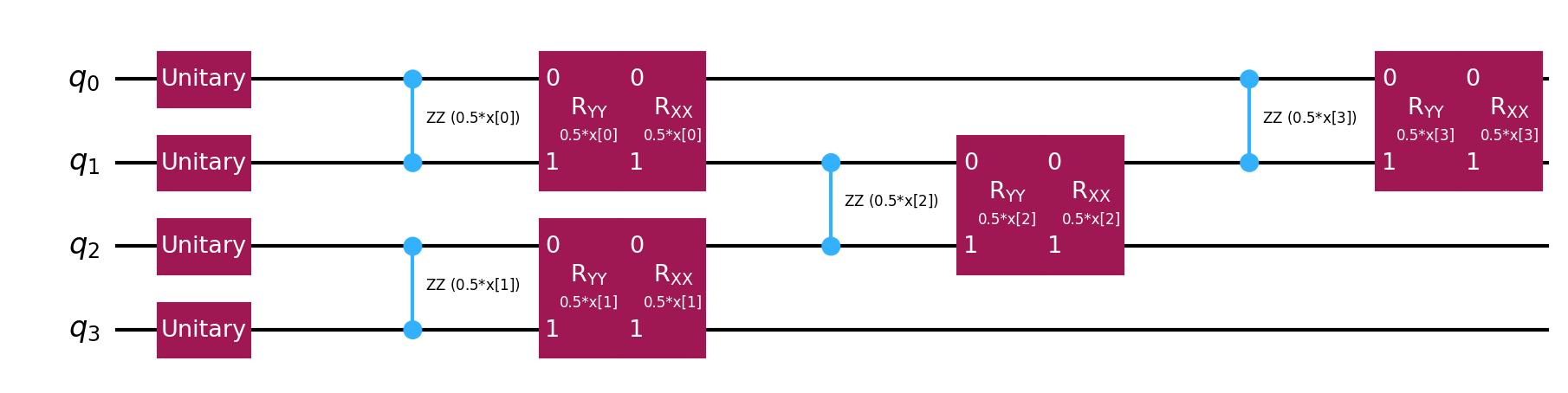}
\caption{Example of Heisenberg Hamiltonian ansatz. We assume to encode 4 features $\{x[0], x[1], x[2], x[3]\}$ into four qubits $q_0, q_1, q_2,$ and $q_3$. We apply random unitary gates at the beginning, followed by entanglement and encoding operations through ZZ-gates and $R_{xx}, R_{yy}$ rotation gates.}
\label{fig: HH_ansatz}
\end{figure}
}

\comment{
\begin{figure}[h]
    \centering
    \footnotesize
    \begin{quantikz}[column sep=2mm]
            \lstick{$q_0$}& \gate[style={fill=orange!20}]{H} & \gate[style={fill=red!20}]{\shortstack{$R_Y$ \\ $\theta_0$}} & \ctrl{1} & {} & \ctrl{1} & {} & \gate[style={fill=orange!20}]{H} & \gate[style={fill=red!20}]{\shortstack{$R_Y$ \\ $\theta_2$}} & {} & {} & \ctrl{1} & {} & \ctrl{1} & {}
            & \\
            \lstick{$q_1$}& 
             \gate[style={fill=red!20}]{\shortstack{$R_Y$ \\ $-\theta_1$}} & {} &
            \targ[style={fill=blue!50,draw=black}]{} & 
            \gate[style={fill=violet!20}]{\shortstack{$P$ \\ $\theta_{01}$}} & 
            \targ[style={fill=blue!50,draw=black}]{} & 
             \gate[style={fill=orange!20}]{H} & 
              &
              \gate[style={fill=red!20}]{\shortstack{$R_Y$ \\ $-\theta_3$}} &
             & 
            \targ[style={fill=blue!50,draw=black}]{} & 
            \gate[style={fill=violet!20}]{\shortstack{$P$ \\ $\theta_{23}$}}& 
            \targ[style={fill=blue!50,draw=black}]{} 
            &
            \gate[style={fill=orange!20}]{H}
            &&    
        \end{quantikz}
    \caption{Example of Repeated Pauli ansatz. We define $\theta_i=\alpha\cdot x_i$, where $\alpha$ is the Pauli rotation factor and $x_i$ is the i-th data feature. We assume to have input data with 4 features $\{x_0, x_1, x_2, x_3\}$ to encode into two qubits $q_0$ and $q_1$. \comment{The application of Hadamard and $R_x$ gates depends on the Pauli operators which in this example are [Y, XZ] as reported in Frame~\ref{lst: RepeatedPauli_config} in Appendix~\ref{app: hyperparams}.}}
    \label{fig: RepeatedPauli_ansatz}
\end{figure}
}

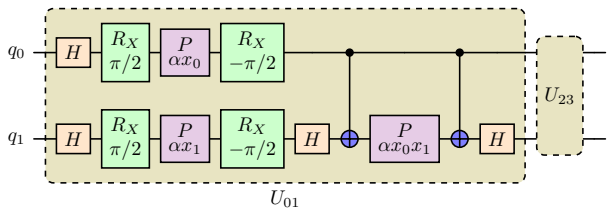
\begin{figure}[h]
    \centering
    \scalebox{0.75}{
    \begin{quantikz}[column sep=2mm]
            \lstick{$q_0$}&{}&\gate[style={fill=orange!20}]{H} \gategroup[2,steps=9,style={dashed,rounded
                        corners,fill=olive!20, inner
                        xsep=2pt},background,label style={label position=below,anchor=north,yshift=-0.2cm}]{{\sc
                        $U_{01}$}}
            & \gate[style={fill=green!20}]{\shortstack{$R_X$ \\ $\pi/2$}} & \gate[style={fill=violet!20}]{\shortstack{$P$ \\ $ \alpha x_0$}} & \gate[style={fill=green!20}]{\shortstack{$R_X$ \\ $-\pi/2$}} & {} & \ctrl{1} & {} & \ctrl{1} & {} & {} & \gate[2, style={dashed,rounded corners,fill=olive!20}]{U_{23}} & {}
            & \\
            \lstick{$q_1$}& {} &
            \gate[style={fill=orange!20}]{H} & \gate[style={fill=green!20}]{\shortstack{$R_X$ \\ $\pi/2$}} &\gate[style={fill=violet!20}]{\shortstack{$P$ \\ $ \alpha x_1$}} & 
            \gate[style={fill=green!20}]{\shortstack{$R_X$ \\ $-\pi/2$}} & 
            \gate[style={fill=orange!20}]{H} & 
            \targ[style={fill=blue!50,draw=black}]{} & 
            \gate[style={fill=violet!20}]{\shortstack{$P$ \\ $ \alpha x_0x_1$}} & 
            \targ[style={fill=blue!50,draw=black}]{} & 
             \gate[style={fill=orange!20}]{H} & {} 
            & &&    
        \end{quantikz}
        }
    \caption{Example of Repeated Pauli feature map. We define $\alpha$ as the Pauli rotation factor and $x_i$ as the i-th data feature. We assume to have input data with 4 features $\{x_0, x_1, x_2, x_3\}$ to encode into two qubits $q_0$ and $q_1$. Unitary $U_{23}$ represents the repeated initial block but for features $x_2$ and $x_3$.}
    \label{fig: RepeatedPauli_ansatz}
\end{figure}

\comment{
\begin{quantikz}[column sep=2mm]
            \lstick{$q_0$}& \gate[style={fill=orange!20}]{H} & \gate[style={fill=green!20}]{\shortstack{$R_X$ \\ $\pi/2$}} & \gate[style={fill=violet!20}]{\shortstack{$P$ \\ $ \alpha x_0$}} & \gate[style={fill=green!20}]{\shortstack{$R_X$ \\ $-\pi/2$}} & {} & \ctrl{1} & {} & \ctrl{1} & {} & \gate[style={fill=orange!20}]{H} & \gate[style={fill=green!20}]{\shortstack{$R_X$ \\ $\pi/2$}} & \gate[style={fill=violet!20}]{\shortstack{$P$ \\ $ \alpha x_2$}} & \gate[style={fill=green!20}]{\shortstack{$R_X$ \\ $-\pi/2$}} & {} & \ctrl{1} & {} & \ctrl{1} & {}
            & \\
            \lstick{$q_1$}& 
            \gate[style={fill=orange!20}]{H} & \gate[style={fill=green!20}]{\shortstack{$R_X$ \\ $\pi/2$}} &\gate[style={fill=violet!20}]{\shortstack{$P$ \\ $ \alpha x_1$}} & 
            \gate[style={fill=green!20}]{\shortstack{$R_X$ \\ $-\pi/2$}} & 
            \gate[style={fill=orange!20}]{H} & 
            \targ[style={fill=blue!50,draw=black}]{} & 
            \gate[style={fill=violet!20}]{\shortstack{$P$ \\ $ \alpha x_0x_1$}} & 
            \targ[style={fill=blue!50,draw=black}]{} & 
             \gate[style={fill=orange!20}]{H} & 
              \gate[style={fill=orange!20}]{H} &
              \gate[style={fill=green!20}]{\shortstack{$R_X$ \\ $\pi/2$}} &\gate[style={fill=violet!20}]{\shortstack{$P$ \\ $ \alpha x_3$}} & 
            \gate[style={fill=green!20}]{\shortstack{$R_X$ \\ $-\pi/2$}} & 
            \gate[style={fill=orange!20}]{H} & 
            \targ[style={fill=blue!50,draw=black}]{} & 
            \gate[style={fill=violet!20}]{\shortstack{$P$ \\ $\alpha x_2 x_3$}}& 
            \targ[style={fill=blue!50,draw=black}]{} 
            &
            \gate[style={fill=orange!20}]{H}
            &&    
        \end{quantikz}
\begin{figure}[h]
\centering
\includegraphics[width=0.5\textwidth]{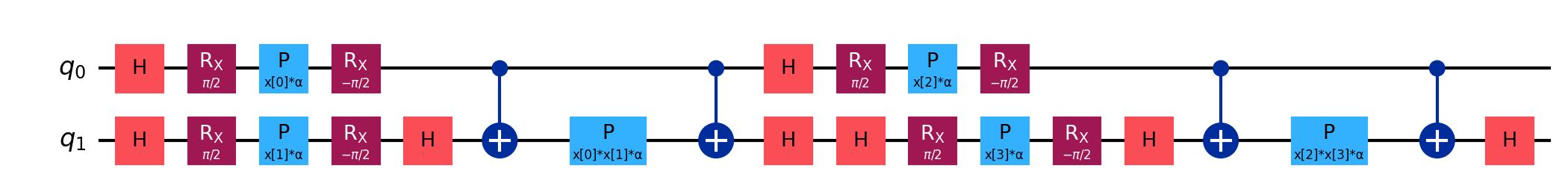}
\caption{Example of Repeated Pauli ansatz. We assume to encode 4 features $\{x[0], x[1], x[2], x[3]\}$ into two qubits $q_0$ and $q_1$. The application of Hadamard and $R_x$ gates depends on the Pauli operators which in this example are [Y, XZ] as reported in Frame~\ref{lst: RepeatedPauli_config} in Appendix~\ref{app: hyperparams}.}
\label{fig: RepeatedPauli_ansatz}
\end{figure}
}

The entanglement structure of the feature map and the value of $\alpha$ used are reported in Appendix~\ref{sub: ansatz_hyperparams}. 
They were kept fixed across experiments, while varying only the number of qubits between 9 and 15, given our goal of analysing the performance when the number of qubits increases. We limit our experiments to a maximum of 15 qubits since the classical simulation of generic quantum circuits with 16 qubits requires hours of execution on a M1 MacBook Pro with 32GB of RAM.

\subsubsection{Classifier}
\label{subsub:classifier}
The features, processed according to our method, are fed to a classifier $C$, and scored according to the Area Under Receiver Operating Characteristic (ROC AUC or simply AUC), a standard metric for classification ~\cite{bradley1997use,fawcett2006introduction}. In our experiments, the classifier is XGBoost, that is a non-kernel based model. Additional experiments with a Support Vector Classifier (SVC) are reported in Appendix~\ref{app: svc}. Multiple sets of classifier  hyperparameters were tested, as documented in Appendix~\ref{sub: classifier_hyperparams}. The best set is selected via grid search with cross-validation (GridSearchCV) ~\cite{scikit-learn-GridSearchCV}, see Figure~\ref{fig: BO}.

\subsection{Results}
\label{sec:results}
\begin{figure*}
    \centering
    \subfigure[Churn results]{
        \includegraphics[width=0.45\textwidth]{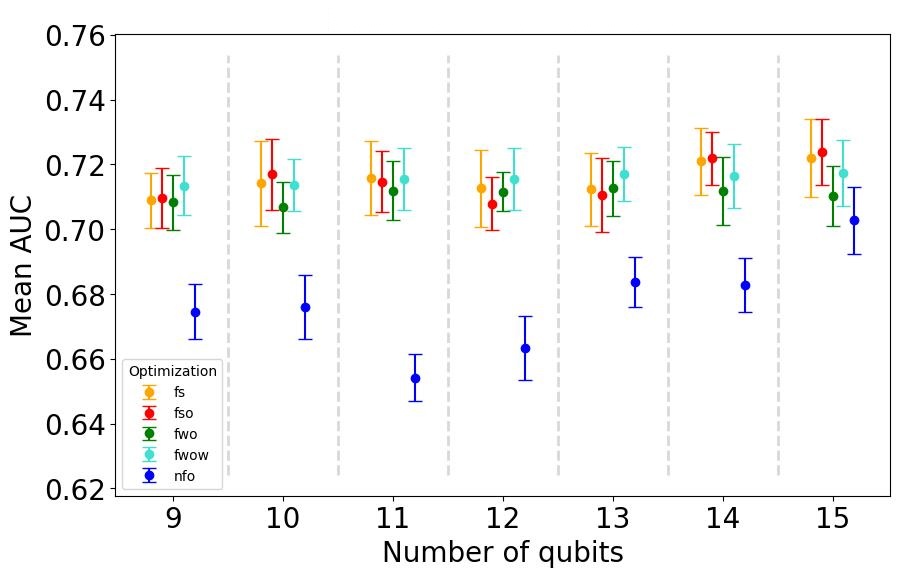}
        \label{subfig: SE1_churn}
    }
    \subfigure[Virtual Screening results]{
        \includegraphics[width=0.45\textwidth]{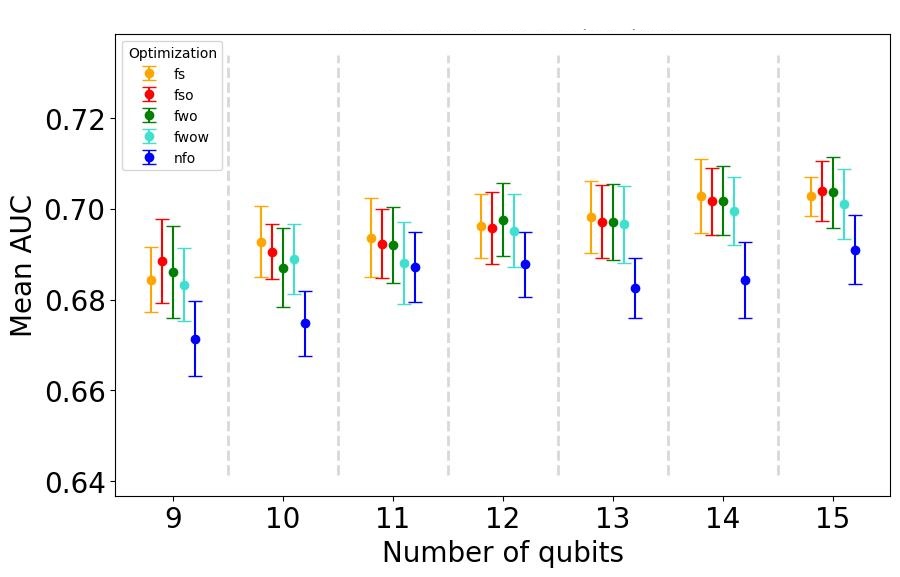}
        \label{subfig: HH1_virtual}
    }
    \caption{Mean ± Std Dev AUC Test Scores for QML Method with Different Scoring Methods and Qubits. In Figure~\ref{subfig: SE1_churn}, we visualize the results obtained on the Churn dataset with QML model exploiting configuration 1 of the Separate Entangled feature map, reported in Frame~\ref{lst: SE_config} in Appendix~\ref{sub: ansatz_hyperparams}. In Figure~\ref{subfig: HH1_virtual}, we report the results related to the Virtual Screening dataset obtained with QML model exploiting configuration 1 of the Heisenberg Hamiltonian feature map, reported in Frame~\ref{lst: HH_config} in Appendix~\ref{sub: ansatz_hyperparams}. We remember that the score reported for each number of qubits is the average over the 10 different dataset splits together with the corresponding standard deviation.}
    \label{fig: barplot_main}
\end{figure*}

\comment{
\gab{Merge this subsec with the Analysis} Tables~\ref{tab: pct_diff_results_SE1},~\ref{tab: pct_diff_results_HH1}, and~\ref{tab: pct_diff_results_RP0} present the most representative results for each ansatz, with XGBoost as the classifier. \gab{before introducing a new table, comment on the previous ones.} In Table~\ref{tab: pct_diff_overall}, we also provide the overall average performance for each different ansatz configuration (the interested reader could find the definition of each ansatz configuration in Appendix~\ref{sub: ansatz_hyperparams}) to evaluate the overall performance of the quantum feature encoding optimization regardless of the dataset.
We report the results in terms of AUC percentage change of each quantum feature encoding optimization with respect to the NFO baseline: a positive percentage highlights the benefit of the specific quantum feature encoding optimization while a negative percentage indicates a better NFO baseline performance. In Appendix~\ref{app: XGB experiments appendix} and~\ref{app: svc}, the interested reader can also find the detailed results for every configuration of each different ansatz exploiting XGBoost and SVC as classifier. 
In this first analysis, we report the Feature Selection (FS), Feature Selection Ordering (FSO), Feature Weighting Ordering (FWO), and Feature Weighting Ordering Weighting (FWOW) results compared with the baseline (NFO) since they are the optimization methods that have proven to be the best on average. In Figure~\ref{fig: barplot_main}, we also depict the granular results for two significant cases for the Separate Entangled and Heisenberg Hamiltonian ansatz, so that we can visualize how the results vary with the number of qubits. In Section~\ref{sub: ablation}, we also report the results for the Feature Weighting and Feature Ordering optimization methodologies. For the FS and FSO experiments, we select different number of features depending on the dataset: 40 for the Churn, 30 for the Virtual screening, and 18 for the German Numeric.
}

\begin{table*}[h]
\centering
\begin{tabular}{|c|c|c|c|c|}
\hline
\multirow{2}{*}{Dataset} & \multicolumn{4}{c|}{Scoring Method} \\ \cline{2-5} 
                                           & FS      & FSO & FWO & FWOW  \\ \Xhline{3\arrayrulewidth}
\multirow{1}{*}{Churn} 
&$+5.73\%$ \scriptsize$(2.17\%)$&$+5.70\%$ \scriptsize$(2.01\%)$&$+5.03\%$ \scriptsize$(2.47\%)$& {\boldmath$+5.77\%$} \scriptsize$(2.32\%)$\\ \cline{2-5}
\hline
\multirow{1}{*}{Virtual Screening}
&$+2.15\%$ \scriptsize$(1.92\%)$&$+2.29\%$ \scriptsize$(1.54\%)$&{\boldmath $+2.47\%$} \scriptsize$(1.90\%)$& $+2.26\%$ \scriptsize$(1.84\%)$\\ \cline{2-5}
\hline
\multirow{1}{*}{German Numeric} &$-1.43\%$ \scriptsize$(0.92\%)$&$ -0.19\%$ \scriptsize$(0.95\%)$&$+0.10\%$ \scriptsize$(1.02\%)$ &{\boldmath$+0.33\%$} \scriptsize$(0.82\%)$ \\ \cline{2-4}
\hline
\multirow{1}{*}{PLAsTiCC Astronomy} &$+4.76\%$ \scriptsize$(3.01\%)$&{\boldmath$+6.79\%$} \scriptsize$(2.15\%)$&$+5.65\%$ \scriptsize$(2.32\%)$ &$+5.82\%$ \scriptsize$(2.11\%)$ \\ \cline{2-4}
\Xhline{3\arrayrulewidth}
\multicolumn{1}{|c|}{\textbf{Overall Average}}  & $+2.80\%$ \scriptsize$(3.20\%)$& {\boldmath$+3.64\%$} \scriptsize$(3.19\%)$ & $+3.31\%$ \scriptsize$(2.54\%)$& $+3.54\%$ \scriptsize$(2.71\%)$ \\ \hline
\end{tabular}
\caption{AUC Percentage Change Test Scores (and its Percentage Std Dev) with respect to NFO for QML model using configuration 1 of Separate Entangled feature map reported in Frame~\ref{lst: SE_config} in Appendix~\ref{sub: ansatz_hyperparams}. Each score is computed as the average percentage variation of the specific optimization exploiting Separate Entangled feature map, with qubits ranging from 9 to 15, relative to the baseline. In Figure~\ref{subfig: SE1_churn}, we report the granular results for the Churn case.}
\label{tab: pct_diff_results_SE1}
\end{table*}

\begin{table*}[h]
\centering
\begin{tabular}{|c|c|c|c|c|}
\hline
\multirow{2}{*}{Dataset} & \multicolumn{4}{c|}{Scoring Method} \\ \cline{2-5} 
                                           & FS      & FSO & FWO & FWOW  \\ 
                                           \Xhline{3\arrayrulewidth}
\multirow{1}{*}{Churn} 
&$+8.68\%$ \scriptsize$(2.63\%)$&{\boldmath$+8.71\%$} \scriptsize$(2.42\%)$&$+7.69\%$ \scriptsize$(2.23\%)$& $+7.53\%$ \scriptsize$(2.49\%)$\\ \cline{2-5}
\hline
\multirow{1}{*}{Virtual Screening}
&{\boldmath$+1.92\%$} \scriptsize$(0.68\%)$&$+1.91\%$ \scriptsize$(0.70\%)$&$+1.81\%$ \scriptsize$(0.60\%)$& $+1.54\%$ \scriptsize$(0.74\%)$\\ \cline{2-5}
\hline
\multirow{1}{*}{German Numeric} 
&$+1.68\%$ \scriptsize$(1.86\%)$&$+1.97\%$ \scriptsize$(2.82\%)$&{\boldmath$+1.99\%$} \scriptsize$(2.83\%)$ & $+1.65\%$ \scriptsize$(2.69\%)$\\ \cline{2-5}
\hline
\multirow{1}{*}{PLAsTiCC Astronomy} 
&$+10.55\%$ \scriptsize$(1.73\%)$&{\boldmath$+11.05\%$} \scriptsize$(2.24\%)$&$+9.32\%$ \scriptsize$(2.45\%)$& $+8.14\%$ \scriptsize$(2.33\%)$\\ \cline{2-5}
\Xhline{3\arrayrulewidth}
\multicolumn{1}{|c|}{\textbf{Overall Average}}
&$+5.71\%$ \scriptsize$(4.58\%)$& {\boldmath$+5.91\%$} \scriptsize$(4.68\%)$& $+5.20\%$ \scriptsize$(3.87\%)$&  $+4.71\%$ \scriptsize$(3.61\%)$\\ \hline
\end{tabular}
\caption{AUC Percentage Change Test Scores (and its Percentage Std Dev) with respect to NFO for QML model using configuration 1 of Heisenberg Hamilton feature map reported in Frame~\ref{lst: HH_config} in Appendix~\ref{sub: ansatz_hyperparams}. Each score is computed as the average percentage variation of the specific optimization exploiting Heisenberg Hamiltonian feature map, with qubits ranging from 9 to 15, relative to the baseline. In Figure~\ref{subfig: HH1_virtual}, we report the granular results for the Virtual Screening case.}
\label{tab: pct_diff_results_HH1}
\end{table*}

\begin{table*}[h]
\centering
\begin{tabular}{|c|c|c|c|c|}
\hline
\multirow{2}{*}{Dataset} & \multicolumn{4}{c|}{Scoring Method} \\ \cline{2-5} 
                                           & FS      & FSO & FWO & FWOW  \\ 
                                           \Xhline{3\arrayrulewidth}
\multirow{1}{*}{Churn} 
&{\boldmath$+4.75\%$} \scriptsize$(1.2\%)$&$+4.69\%$ \scriptsize$(1.32\%)$&$+4.03\%$ \scriptsize$(1.41\%)$&$+4.02\%$ \scriptsize$(1.28\%)$\\ \cline{2-5}
\hline
\multirow{1}{*}{Virtual Screening}
&$+2.26\%$ \scriptsize$(0.50\%)$&$+2.59\%$ \scriptsize$(0.62\%)$&{\boldmath$+2.76\%$} \scriptsize$(0.87\%)$&$+2.34\%$ \scriptsize$(0.88\%)$\\ \cline{2-5}
\hline
\multirow{1}{*}{German Numeric}
&$+0.22\%$ \scriptsize$(2.45\%)$&$+0.01\%$ \scriptsize$(2.09\%)$&{\boldmath$+0.78\%$} \scriptsize$(2.4\%)$&$-0.08\%$ \scriptsize$(2.73\%)$\\ \cline{2-5}
\hline
\multirow{1}{*}{PLAsTiCC Astronomy}
&$+9.64\%$ \scriptsize$(1.57\%)$&{\boldmath$+10.00\%$} \scriptsize$(1.71\%)$&$+9.32\%$ \scriptsize$(2.01\%)$&$+8.92\%$ \scriptsize$(1.99\%)$\\ \cline{2-5}
\Xhline{3\arrayrulewidth}
\multicolumn{1}{|c|}{\textbf{Overall Average}}  
&$+4.22\%$ \scriptsize$(4.06\%)$&{\boldmath$+4.32\%$} \scriptsize$(4.24\%)$&$+4.22\%$ \scriptsize$(3.65\%)$&$+3.80\%$ \scriptsize$(3.81\%)$\\ \hline
\end{tabular}
\caption{AUC Percentage Change Test Scores (and its Percentage Std Dev) with respect to NFO for QML model using configuration 0 of Repeated Pauli feature map reported in Frame~\ref{lst: RepeatedPauli_config} in Appendix~\ref{sub: ansatz_hyperparams}. Each score is computed as the average percentage variation of the specific optimization exploiting Repeated Pauli feature map, with qubits ranging from 9 to 15, relative to the baseline.}
\label{tab: pct_diff_results_RP0}
\end{table*}

\begin{table*}[h]
\centering
\begin{tabular}{|c|c|c|c|c|}
\hline
\multirow{2}{*}{Feature Map} & \multicolumn{4}{c|}{Scoring Method} \\ \cline{2-5} & FS  & FSO & FWO & FWOW  \\ \Xhline{3\arrayrulewidth}
\multirow{1}{*}{SE$_0$} 
& $+2.97\%$ \scriptsize$(3.80\%)$& $+3.95\%$ \scriptsize$(3.22\%)$& $+4.02\%$ \scriptsize$(2.67\%)$& {\boldmath$+4.30\%$} \scriptsize$(2.56\%)$\\ \cline{2-5}
\hline
\multirow{1}{*}{SE$_1$}
& $+2.80\%$ \scriptsize$(3.20\%)$& {\boldmath$+3.64\%$} \scriptsize$(3.19\%)$& $+3.31\%$ \scriptsize$(2.54\%)$& $+3.54\%$ \scriptsize$(2.71\%)$ \\ \cline{2-5}
\hline
\multirow{1}{*}{SE$_2$} 
& $+3.13\%$ \scriptsize$(3.8\%)$& {\boldmath$+3.64\%$} \scriptsize$(4.02\%)$& $+2.97\%$ \scriptsize$(2.91\%)$& $+3.37\%$ \scriptsize$(2.96\%)$\\ \cline{2-5}
\hline
\multirow{1}{*}{HH$_0$} 
 &$+5.28\%$ \scriptsize$(4.61\%)$& {\boldmath$+5.57\%$} \scriptsize$(4.88\%)$& $+4.99\%$ \scriptsize$(4.03\%)$ &  $+4.53\%$ \scriptsize$(3.80\%)$\\ \cline{2-5}
\hline
\multirow{1}{*}{HH$_1$} 
&$+5.71\%$ \scriptsize$(4.58\%)$& {\boldmath$+5.91\%$} \scriptsize$(4.68\%)$& $+5.20\%$ \scriptsize$(3.87\%)$&  $+4.71\%$ \scriptsize$(3.61\%)$\\ \cline{2-5}
\hline
\multirow{1}{*}{RP$_0$} 
&$+4.22\%$ \scriptsize$(4.06\%)$&{\boldmath$+4.32\%$} \scriptsize$(4.24\%)$&$+4.22\%$ \scriptsize$(3.65\%)$&$+3.80\%$ \scriptsize$(3.81\%)$\\ \cline{2-5}
\hline
\end{tabular}
\caption{Overall AUC Percentage Change Test Scores with respect to NFO for all the different feature map configurations reported in Appendix~\ref{sub: ansatz_hyperparams}. Each score is the overall average among the corresponding three different datasets scores.}
\label{tab: pct_diff_overall}
\end{table*}

\begin{figure*}[h]
    \centering
    \subfigure[Separate Entangled ansatz]{
        \includegraphics[width=0.45\textwidth]{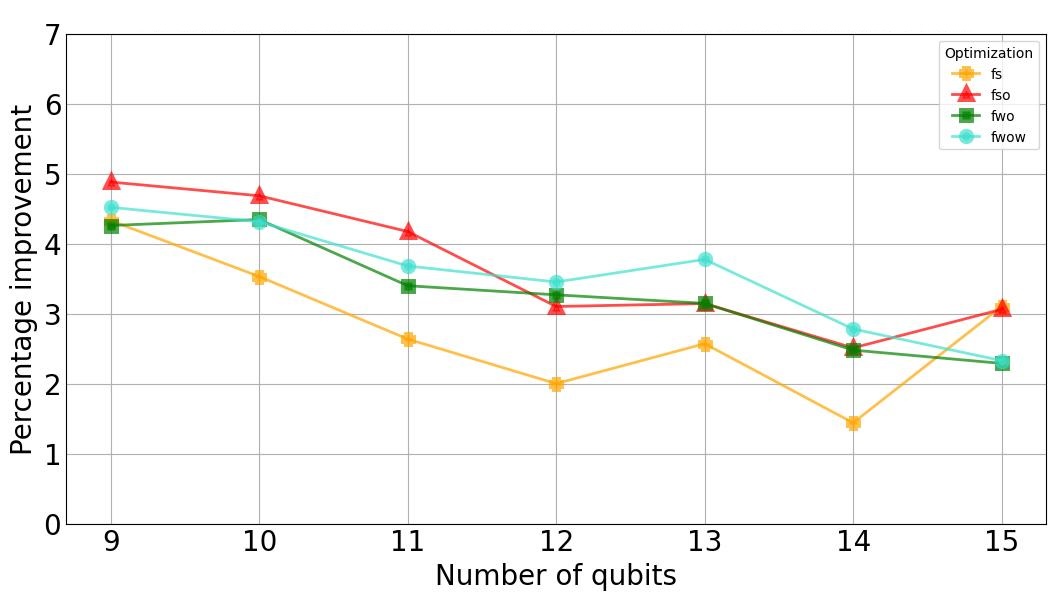}
        \label{subfig: ptc_improvements_per_qubit_SE1}
    }
    \subfigure[Heisenberg Hamiltonian ansatz]{
        \includegraphics[width=0.45\textwidth]{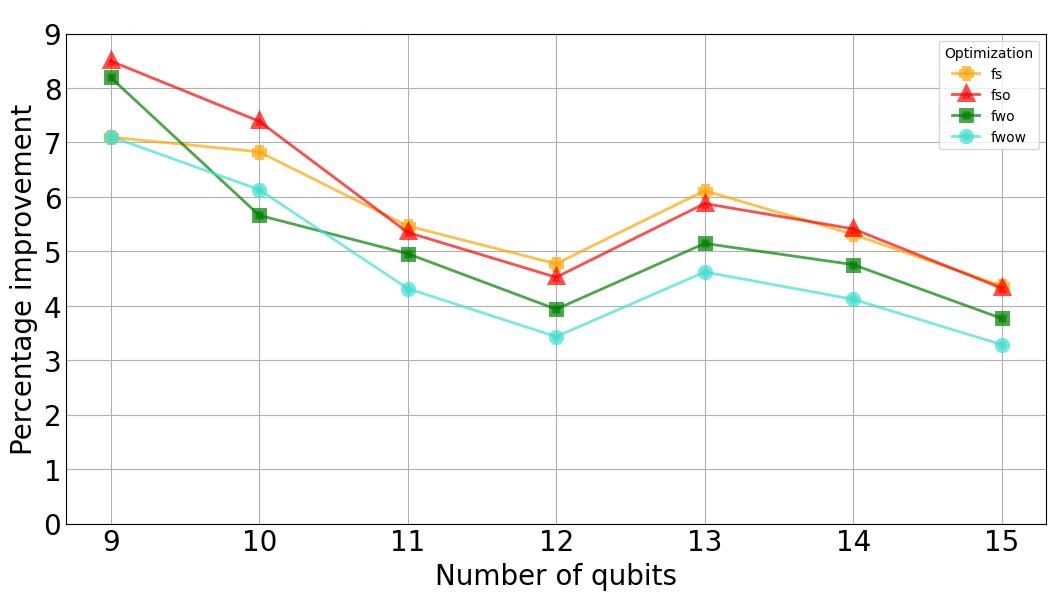}
        \label{subfig: ptc_improvements_per_qubit_HH1}
    }
    \caption{AUC percentage change improvement per qubit averaged across all the datasets used. In Figures~\ref{subfig: ptc_improvements_per_qubit_SE1} and~\ref{subfig: ptc_improvements_per_qubit_HH1}, we depict this kind of information exploiting configuration 1 of Separate Entangled and Heisenberg Hamiltonian feature maps respectively.}
    \label{fig: ptc_improvements_per_qubit}
\end{figure*}

\comment{
\subsection{Analyses}
\label{sub: analyses}
\gab{I'd remove the following paragraph}We start by analyzing the numerical results together with an in-depth analysis on the training phase of the QML models. Then, we propose a study about feature importance regarding the quantum feature encoding optimizations tested.
\gab{The titles of the subsubsecs are not very clear. What about `results of the trained models' and `analysis of the training process'?}
}
\subsubsection{Numerical results analyses}
\label{subsub: numerical_analyses}
In this first analysis, we report the feature selection (FS), feature selection ordering (FSO), feature weighting ordering (FWO), and feature weighting ordering weighting (FWOW) results compared with the baseline (NFO) since they are the optimization methods that have proven to be the best on average. In Section~\ref{sub: ablation}, we also report the results for the feature weighting and feature ordering optimization methodologies. For the FS and FSO experiments, we select different number of features depending on the dataset: 40 for  Churn, 30 for Virtual Screening and PLAsTiCC Astronomy, and 18 for German Numeric, to account for the varying feature sizes of the datasets while still enabling evaluating different fractions of selected features via the difference across datasets.

Overall, we can identify two principal findings. The first is that by using QFEO, we achieve competitive or better performance with respect to NFO on all four datasets with notably better performance on Churn, Virtual Screening, and PLAsTiCC Astronomy datasets (Tables~\ref{tab: pct_diff_results_SE1}-\ref{tab: pct_diff_results_RP0} and~\ref{tab: XGB_SE0}-\ref{tab: pct_diff_results_german_svc_ablation}; Figures~\ref{subfig: SE1_churn}-\ref{subfig: HH1_virtual}). The second is that by looking at the overall average results (Table \ref{tab: pct_diff_overall}), we would generally expect a significant improvement with the usage of QFEO: of course, this claim depends and is limited to the four datasets tested, but it is certainly a good indicator of the quality of these methodologies. 

In Table~\ref{tab: pct_diff_results_SE1}, we report the Separate Entangled feature map results. It is interesting to see that FWOW optimization yields the best performance increase over the Churn and the German Numeric dataset; however, the overall average score of FSO is the highest, owing to the significant improvement over the PLAsTiCC Astronomy dataset and the comparable performance in relation to the other optimizations across the remaining datasets. In the case of Virtual Screening dataset, we can appreciate that FWO optimization produces the highest performance enhancement. This is an interesting result since it means that by optimizing just one set of weights instead of two, as FWOW does, we can achieve comparable or even better performance depending on the data. Another important insight that we highlight is the fact that there is no ideal data manipulation function that is always advantageous but, depending on the data we have available and the chosen feature map, one optimization can be more effective than another. In fact, we can notice that for Churn and Virtual Screening datasets we obtain the best results with FWO(W), with competitive performances even in the case of FSO and FS; this is not true in the German Numeric dataset, where both FS and FSO are not only not comparable with FWO but are actually inferior to the NFO baseline. This is an interesting finding which demonstrates the importance of a dense encoding feature map in our problem: when the number of data features is small, as the number of qubits increases, we encode fewer features per qubit, which in turn reduces the entanglement and simplifies the complexity of the wave function. This behavior is most evident in the German Numeric dataset, which has only 24 features. It becomes even more apparent when using the feature selection manipulation, where reducing the number of features to 18 leads to a noticeable performance degradation.

By examining at Tables~\ref{tab: pct_diff_results_HH1} and~\ref{tab: pct_diff_results_RP0}, a similar analysis could be conducted on the performance of the QML model exploiting both the Heisenberg Hamiltonian and the Repeated Pauli feature map respectively. In the case of Heisenberg Hamiltonian feature map, we can notice how feature selection ordering and feature selection achieve noteworthy performance, resulting in the best optimizations over the Churn, Virtual Screening, and PLAsTiCC Astronomy datasets. Again, we observe that FWO performs well, peaking on the German Numeric dataset and outperforming the other optimization methods, while FSO achieves comparable performance.
Overall, we can observe that FSO results in the highest performance improvement across the datasets; one possible hypothesis for this behavior could be related to the fact that Heisenberg feature map is a more complex circuit / involving more operations per feature as it uses multiple 2-qubit rotations for each, therefore FS(O), by removing features, may have a bigger impact in simplifying the circuit if key features for inclusion are found through optimization.

Furthermore, to confirm the impact that the choice of the feature map has on the final results, we highlight that, by exploiting the Heisenberg Hamiltonian one, we always obtain an improvement in performance with respect to NFO, even in the case of German Numeric dataset differently from the Separate Entangled and Repeated Pauli feature map cases. The effectiveness of the FWO optimization is also confirmed by the Repeated Pauli results, with FS(O) and FWO achieving comparable overall performance, whereas FWOW shows a greater deviation from the others. Finally, it is also worth highlighting the overall results presented in Table~\ref{tab: pct_diff_overall}. It is noteworthy that FSO emerges as the optimization with the highest average percentage improvement for all the feature maps, with the exception of the first configuration of the Separate Entangled, where FWOW takes the lead. These overall results appear to also highlight the impact of the entanglement typology: indeed, only the first Separate Entangled configuration exhibits full entanglement, whereas all other feature map configurations involve pairwise entanglement (note: illustrations of the entanglement patterns are provided in Appendix \ref{app: hyperparams}). 
This may be a point of further investigation in subsequent studies. Related to the overall results, it is interesting to also consider the outcomes reported in Figure~\ref{fig: ptc_improvements_per_qubit} that allow to visualize the AUC percentage change improvement general trend per qubit for the Separate Entangled and Heisenberg Hamiltonian feature maps. The most evident point is that we obtain the highest improvement using feature map that involve the lowest number of qubits (except for FWO optimization on the Separate Entangled, where we get it for 10 qubits which can still be considered a low number). This result emphasizes the importance of dense feature encoding: as previously said, for a fixed number of features, increasing the number of qubits makes the feature map less dense, leading to fewer features encoded per qubit. This reduction limits entanglement layers and decreases the complexity of the wave function. Focusing on the individual subfigures~\ref{subfig: ptc_improvements_per_qubit_SE1} and~\ref{subfig: ptc_improvements_per_qubit_HH1}, it is interesting to see that, for the Separate Entangled case, the improvement trend is the same both for FS and FSO optimizations while it is not exactly the case for FWO and FWOW. This result makes sense since FS and FSO are from the same data manipulation class, while for the case of FWOW, optimizing two sets of weights instead of one can lead to different results. Directly related to this, it is worth noting how in the case of Heisenberg Hamiltonian the improvement trend is the same for all optimizations, confirming even more the impact that different feature map have on QML model performance.
\subsubsection{Training analysis}
\label{subsub: bo_convergence_main}
\begin{figure}[h]
\centering
\includegraphics[width=0.4\textwidth]{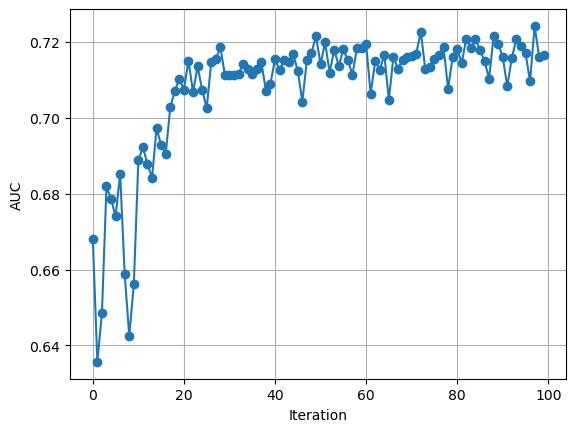}
\caption{Bayesian Optimization convergence trend during feature selection ordering on the Churn dataset, using the 15-qubit Separate Entangled feature map. The $y$-axis shows the average training AUC score computed over 10 batches at each iteration, reported on the $x$-axis.}
\label{fig: BO_convergence_SE_main}
\end{figure}

We also propose a case study about the training phase of our QML models. The objective is to observe the trend of the Bayesian Optimization procedure (that we described in Figure~\ref{fig: BO}) across the different iterations to see if the specific QML model is actually learning the optimal set of weights that determines the features to be encoded in the feature map. We set the number of iterations to 100 to balance result quality and execution time, since classical simulation of QML models is computationally intensive. Figure~\ref{fig: BO_convergence_SE_main} shows the convergence trend of Bayesian Optimization for the QML model that uses a 15-qubit Separate Entangled feature map to encode the Churn dataset features, combined with XGBoost as the classifier. We notice that, already with 100 iterations of Bayesian Optimization procedure, FSO optimization procedure shows a trend of convergence. This is a significant finding demonstrating that, for the specific QML setting described, FSO optimization enables the model to learn the optimal set of weights, allowing us to select the best set of features and order them in the most effective way. In Section~\ref{app:BO_convergence} of the Appendix, we report a more exhaustive set of examples concerning this type of study.
\subsubsection{Feature importance}
\label{subsub: feat_importance}

This analysis aims to demonstrate that, by employing QFEO framework, different data manipulation methods can assign differing importance to features, thereby highlighting how each approach offers a unique perspective on the feature space. The importance depends on the type of optimization: for example, 
while, for feature weighting, it is related to the magnitude of the weight that is used to scale the rotation in the encoding of the corresponding feature.
For example, in the case of Churn dataset encoded with Separate Entangled feature map -- with the configuration 1 reported in Frame~\ref{lst: SE_config} -- we find that the 28th feature is the one that is weighted the most in feature weighting optimization and its importance is confirmed by the
feature selection and feature selection ordering manipulations. Indeed, feature 28 is the most selected one and, also considering the ordering procedure, it is the highest-ranked one. This means that, on average over the 10 dataset batches, we always select feature 28 and, in the case of feature selection ordering, we encode it as first feature in the Separate Entangled feature map. Also in the case of feature weighting ordering, we find that feature 28 is one
of the most important one both in terms of weighting and ordering. This is not true for the feature weighting ordering weighting technique, where other features are considered more important. For example, feature 83 is very important in FWOW as well as in FSO but not so much in the case of FS. The heatmaps reported in Figure~\ref{fig:feature_scoring_churn} and Appendix Figure~\ref{app: noiseless_feature_importance} show additional cases, highlighting that, depending on the data we are working on, we could exploit QFEO framework with different data manipulation approaches. In support of this analysis, in Figure~\ref{fig: feat_importance_fs_per_ansatz} in Appendix~\ref{app: feature_importance}, we also demonstrate how the importance of features may be shared or differentiated across different feature maps.

\comment{

The objective of this analysis is to demonstrate that different quantum feature encoding optimization methods may or may not consider the same features important, highlighting how each procedure provides a different view of the feature space. The importance depends on the type of optimization: for example, in case of Feature Selection, it refers to the frequency with which a feature is selected and encoded in the ansatz while, for Feature Weighting, it is related to the magnitude of the weight that is used to scale the rotation in the encoding of the corresponding feature. The heatmap, in Figure~\ref{fig: feat_importance_subset}, 
depicts exactly this type of analysis for the Churn dataset allowing to visualize the importance of each feature in relation to the type of optimization used to encode it in the Separate Entangled ansatz. We report just a subset of features since it is enough to show this analysis; the interested reader could find the complete plot in Figure~\ref{fig:feature_scoring_churn} in Appendix~\ref{app: feature_importance}. 
Starting from the feature 28, we can notice that it is the one that is weighted the most in Feature Weighting optimization. The importance of this feature is confirmed by the
Feature Selection and Feature Selection Ordering optimization; as we can see, this feature is the most selected one and, also considering the ordering procedure, it is the highest-ranked one. This means that, on average
over the 10 dataset splits, we always select feature 28 and, in the case of Feature Selection Ordering, we encode it as first
feature in the Separate Entangled ansatz. Also in the case of Feature Weighting Ordering, we can see that feature 28 is one
of the most important one both in terms of weighting and ordering. This is not true for the Feature Weighting Ordering
Weighting optimization, where other features are considered more important: for example, feature 83 is very important in FWOW as well as in the FSO optimization but not so much in the case of FS. By looking at the reported heatmap, we can discover more of these cases which highlights the fact that, depending on the data we are working on, we could exploit QFEO framework with different data manipulation approaches. In support of this analysis, in Figure~\ref{fig: feat_importance_fs_per_ansatz} in Appendix~\ref{app: feature_importance}, we also demonstrate how the importance of features may be shared or differentiated across different feature map. 

\begin{figure}[h]
\centering
\includegraphics[width=0.48\textwidth]{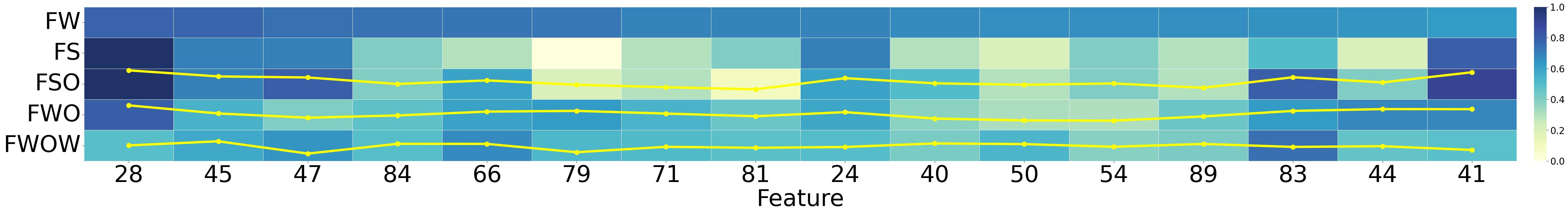}
\caption{Feature scoring analysis on Churn dataset corresponding to the experiments regarding Separate Entangled ansatz whose results are reported in Table~\ref{tab: pct_diff_results_SE1}. The color represents the importance of each feature: the darker, the greater importance. Features are sorted by their importance according to the Feature Weighting optimization. For the feature encoding optimizations involving ordering, we report the yellow line which represents the ordering of each feature for the corresponding optimization. Results are the average over the 10 different dataset splits.}
\label{fig: feat_importance_subset}
\end{figure}

\begin{figure*}[h]
\centering
\includegraphics[width=\textwidth]{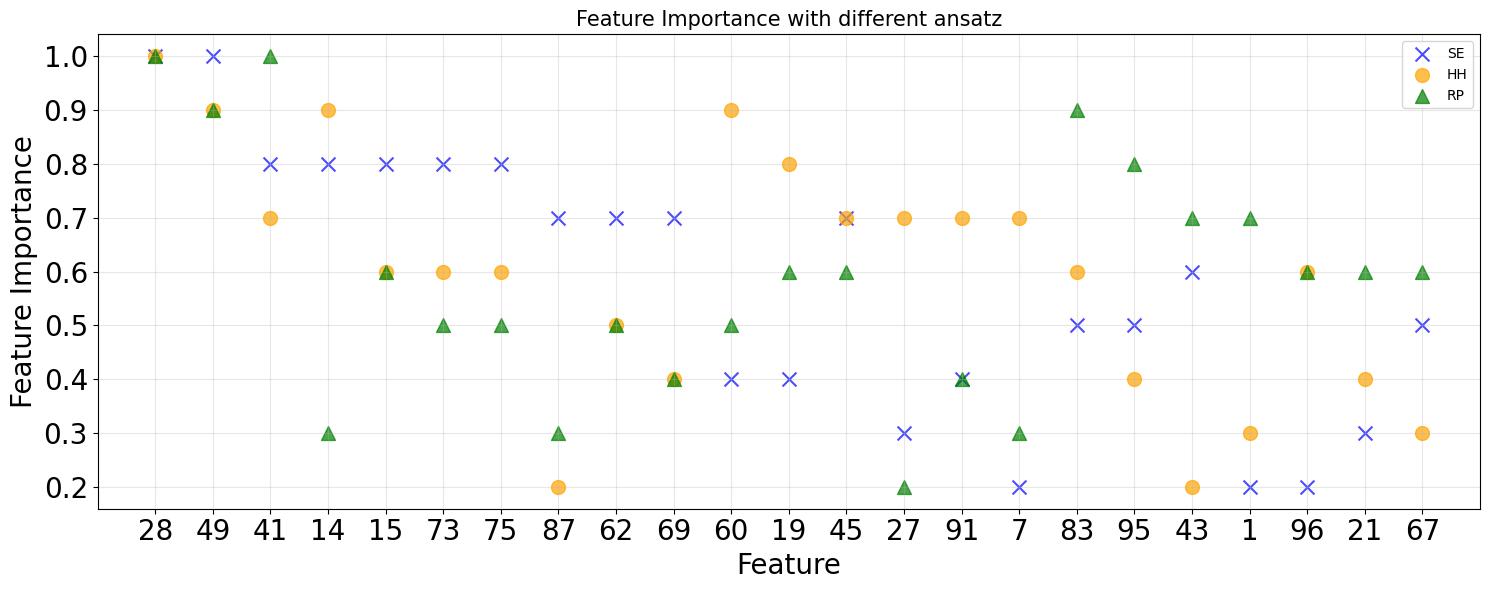}
\caption{Feature scoring analysis on Churn dataset per ansatz exploiting Feature Selection optimization. The represented features are the union of the top 10 selected features across the three different ansatz.}
\label{fig: feat_importance_fs_per_ansatz}
\end{figure*}

}

\subsection{Ablation study}
\label{sub: ablation}
As additional experiments, we perform feature weighting (FW) and feature ordering (FO) optimizations on the Churn and German Numeric datasets only, which are the bigger and the smallest datasets, and only considering the Separated Entangled and Heisenberg Hamiltonian feature maps\comment{\andres{only considering the Separated Entangled and Heisenberg Hamiltonian feature maps (This phrase gives the impression that there are other ansatzes apart from the Separate Entangled and the Heisenberg. Are there? \tom{Yes, we also have Repeated Pauli ansatz, figure \ref{fig: RepeatedPauli_ansatz}})}} since it is enough to demonstrate the usefulness of FW and FO optimizations. The corresponding complete set of results can be found in Appendix~\ref{app: XGB experiments appendix} in Tables~\ref{tab: XGB_SE1_ablation_numerical} and~\ref{tab: XGB_HH1_numerical_ablation}.
We also keep the FWOW results here just for comparison with the simpler FW technique. The considerations do not differ much from the analysis made in Section~\ref{subsub: numerical_analyses}. Indeed, in Table~\ref{tab: pct_diff_results_SE_1_ablation}, we can notice that, in the case of Separate Entangled feature map, FWOW performs better also with respect to FO and FW. A noteworthy observation is that the performance of FO is highly comparable to that of FWOW on the Churn dataset. For the Heisenberg Hamiltonian feature map\comment{reported in Table~\ref{tab: pct_diff_results_HH_1_ablation}}, the FO method does not perform as well as it does for the Separate Entangled case, but this outcome was expected. In fact, the higher complexity of the Heisenberg Hamiltonian relative to the Separate Entangled feature map, combined with the factorial computational cost\footnote{For $n$ features, there are potentially $n!$ ordering combinations to explore. As $n$ increases, the complexity grows rapidly; for the Churn dataset, this results in $O(97!)$.} of the feature ordering manipulation, renders this type of optimization particularly challenging for this feature map. This behavior is much clearer on the Churn dataset, which has a significantly higher dimensionality compared to the German Numeric dataset. Specifically, we observe that for both the Churn and German Numeric datasets, FW achieves the best score.\footnote{It is important, however, to consider the results reported in Section~\ref{subsub: numerical_analyses}. For instance, although the $+8.31\%$ score of FW represents an excellent result, it is not the best for the Churn dataset when compared to FS(O).} The observation that, for the German Numeric dataset, the Heisenberg Hamiltonian feature map consistently outperforms the Separate Entangled feature map when applying data manipulation techniques can be connected to the insights from the analysis in Section~\ref{subsub: numerical_analyses}. Indeed, in the Separate Entangled case, as for the FS(O) optimization, we do not observe an improvement in the usage of FW and FO. An improvement that instead becomes concrete in the case of the Heisenberg Hamiltonian feature map. As for the comparison between FWOW and FW results, we can observe that, in the case of Separate Entangled feature map, we always obtain better results with FWOW while, in the Heisenberg Hamiltonian case, we obtain more similar performances with the FW outperforming both FO and FWOW. As previously mentioned, these are interesting points as they highlight the diversity of these data manipulation techniques and the potential to achieve either very similar or significantly different results depending on the data and feature map employed. 

\comment{
\begin{table}[t]
\centering
\scalebox{0.72}{
\begin{tabular}{|c|c|c|c|}
\hline
\multirow{2}{*}{Dataset} & \multicolumn{3}{c|}{Scoring Method} \\ \cline{2-4} 
                                           & FO      & FW & FWOW   \\ 
                                           \Xhline{3\arrayrulewidth}
\multirow{1}{*}{Churn} 
&$+5.48\%$ \scriptsize$(2.28\%)$&$+2.39\%$ \scriptsize$(1.83\%)$&{\boldmath$+5.77\%$} \scriptsize$(2.32\%)$\\ \cline{2-4}
\hline
\multirow{1}{*}{German Numeric} &$-0.69\%$ \scriptsize$(1.05\%)$&$-0.37\%$ \scriptsize$(0.90\%)$&{\boldmath$+0.33\%$} \scriptsize$(0.82\%)$\\ \cline{2-4}
\Xhline{3\arrayrulewidth}
\multicolumn{1}{|c|}{\textbf{Overall Average}}  &$+2.39\%$ \scriptsize$(4.36\%)$&$+1.01\%$ \scriptsize$(1.95\%)$&{\boldmath$+3.06\%$} \scriptsize$(3.84\%)$\\ \hline
\end{tabular}
}
\caption{AUC Percentage Change Test Scores with respect to NFO for QML model using configuration 1 of Separate Entangled ansatz reported in Frame~\ref{lst: SE_config} in Appendix~\ref{sub: ansatz_hyperparams}. Each score is computed as the average percentage variation of the specific optimization exploiting Separate Entangled ansatz, with qubits ranging from 9 to 15, relative to the baseline.}
\label{tab: pct_diff_results_SE_1_ablation}
\end{table}

\begin{table}[t]
\centering
\scalebox{0.72}{
\begin{tabular}{|c|c|c|c|}
\hline
\multirow{2}{*}{Dataset} & \multicolumn{3}{c|}{Scoring Method} \\ \cline{2-4} 
                                           & FO      & FW & FWOW   \\ 
                                           \Xhline{3\arrayrulewidth}
\multirow{1}{*}{Churn} 
&$+4.30\%$ \scriptsize$(2.32\%)$&{\boldmath$+8.31\%$} \scriptsize$(3.01\%)$&$+7.53\%$ \scriptsize$(2.49\%)$\\ \cline{2-4}
\hline
\multirow{1}{*}{German Numeric} &$+0.96\%$ \scriptsize$(2.49\%)$&{\boldmath$+1.81\%$} \scriptsize$(2.89\%)$&$+1.65\%$ \scriptsize$(2.69\%)$\\ \cline{2-4}
\Xhline{3\arrayrulewidth}
\multicolumn{1}{|c|}{\textbf{Overall Average}}  &$+2.63\%$ \scriptsize$(2.36\%)$&{\boldmath$+5.06\%$} \scriptsize$(4.59\%)$&$+4.59\%$ \scriptsize$(4.16\%)$\\ \hline
\end{tabular}
}
\caption{AUC Percentage Change Test Scores with respect to NFO for QML model using configuration 1 of Heisenberg Hamiltonian ansatz reported in Frame~\ref{lst: HH_config} in Appendix~\ref{sub: ansatz_hyperparams}. Each score is computed as the average percentage variation of the specific optimization exploiting Separate Entangled ansatz, with qubits ranging from 9 to 15, relative to the baseline.}
\label{tab: pct_diff_results_HH_1_ablation}
\end{table}
}
\begin{table}[t]
\centering
\scalebox{0.6}{
\begin{tabular}{|c|c|c|c|c|}
\hline
\multirow{2}{*}{Feature Map} &\multirow{2}{*}{Dataset} & \multicolumn{3}{c|}{Scoring Method} \\ \cline{3-5} 
                                         &  & FO      & FW & FWOW   \\ 
                                           \Xhline{3\arrayrulewidth}
{SE1}&\multirow{1}{*}{Churn} 
&$+5.48\%$ \scriptsize$(2.28\%)$&$+2.39\%$ \scriptsize$(1.83\%)$&{\boldmath$+5.77\%$} \scriptsize$(2.32\%)$\\ \cline{2-4}
\hline
{SE1}&\multirow{1}{*}{German Numeric} &$-0.69\%$ \scriptsize$(1.05\%)$&$-0.37\%$ \scriptsize$(0.90\%)$&{\boldmath$+0.33\%$} \scriptsize$(0.82\%)$\\ \cline{3-5}
\Xhline{3\arrayrulewidth}
{}&\multicolumn{1}{|c|}{\textbf{Overall Average}}  &$+2.39\%$ \scriptsize$(4.36\%)$&$+1.01\%$ \scriptsize$(1.95\%)$&{\boldmath$+3.06\%$} \scriptsize$(3.84\%)$\\ \hline
\Xhline{3\arrayrulewidth}
{HH1}&\multirow{1}{*}{Churn} 
&$+4.30\%$ \scriptsize$(2.32\%)$&{\boldmath$+8.31\%$} \scriptsize$(3.01\%)$&$+7.53\%$ \scriptsize$(2.49\%)$\\ \cline{3-5}
\hline
{HH1}&\multirow{1}{*}{German Numeric} &$+0.96\%$ \scriptsize$(2.49\%)$&{\boldmath$+1.81\%$} \scriptsize$(2.89\%)$&$+1.65\%$ \scriptsize$(2.69\%)$\\ \cline{3-5}
\Xhline{3\arrayrulewidth}
{}&\multicolumn{1}{|c|}{\textbf{Overall Average}}  &$+2.63\%$ \scriptsize$(2.36\%)$&{\boldmath$+5.06\%$} \scriptsize$(4.59\%)$&$+4.59\%$ \scriptsize$(4.16\%)$\\ \hline
\end{tabular}
}
\caption{AUC Percentage Change Test Scores with respect to NFO for QML model using configuration 1 of Separate Entangled and Heisenberg Hamiltonian feature maps reported in Frames~\ref{lst: SE_config} and~\ref{lst: HH_config} in Appendix~\ref{sub: ansatz_hyperparams}. Each score is computed as the average percentage variation of the specific optimization, across all qubits ranging from 9 to 15, relative to the baseline.}
\label{tab: pct_diff_results_SE_1_ablation}
\end{table}

\section{Real Hardware Experiments}
\label{sec: real_hw_experiments}
\begin{figure}[h]
\centering
\includegraphics[width=0.5\textwidth]{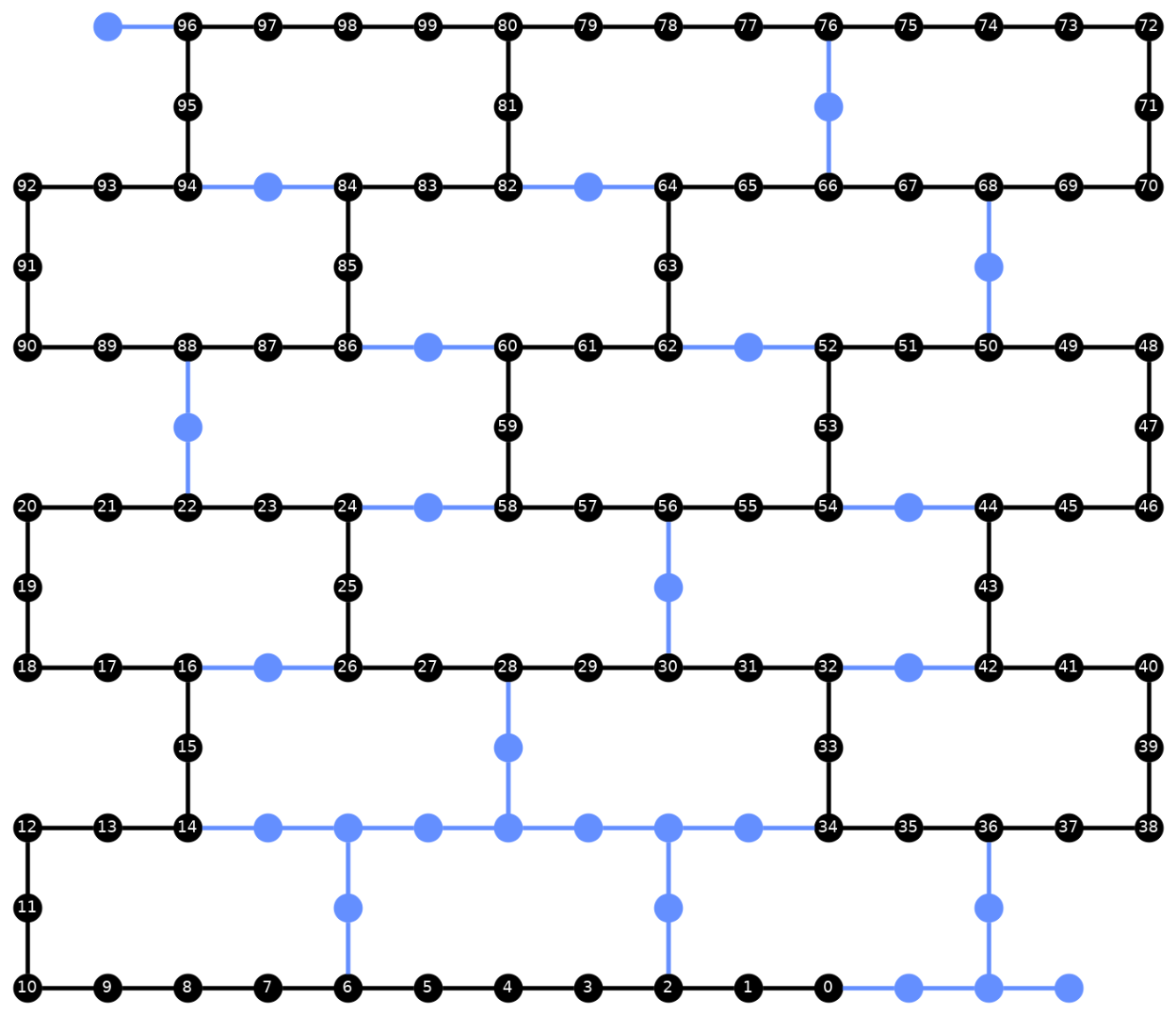}
\caption{ibm\_brisbane layout considering 100-qubits Heisenberg Hamiltonian feature map fitted with NFO technique.}
\label{fig: real_HW_layout}
\end{figure}

To assess whether our QFEO framework has a tangible impact on QML model performance in the presence of noise inherent to current QPUs, we conducted a reduced set of experiments on the 127-qubit IBM Eagle r3 processor (\texttt{ibm\_brisbane}). This QPU exhibits an average 2Q error rate of $3.74 \times 10^{-3}$, an Error Per Layer Gate (EPLG) of $2.07 \times 10^{-2}$, and a Circuit Layer Operations Per Second (CLOPS) of 180K. Since we were unable to precisely replicate all the experiments conducted with the simulator due to queue times and restricted access to the QPUs, we randomly subsample half of the samples of the PLAsTiCC Astronomy dataset, just for one single seed, resulting in a training and test sets of 1174 and 1157 records respectively. Also, we selected Heisenberg Hamiltonian feature map with configuration 1, as reported in Frame~\ref{lst: HH_config} in Appendix~\ref{sub: ansatz_hyperparams}, with 100 qubits to fully leverage the structure and capabilities of the selected QPU. Considering the previously mentioned limitations for QPU experiments, we only perform FSO, as well as the NFO baseline, using XGboost as the sole classifier. Since PLAsTiCC Astronomy consists of 67 features and our feature map comprises 100 qubits, we applied the data reloading technique~\cite{P_rez_Salinas_2020} by tiling the same input features next to the original ones, as if we were repeating the same feature map twice,
 fully entangling all the qubits and enabling the generation of a more complex wave function -- we report an example illustration in Figure~\ref{fig: data_reuploading circuit} in Appendix~\ref{app: data_reloading_setup}. In the case of NFO, applying data reloading leads to encoding twice the number of initial features into the circuit, resulting in a total of 134 features. Conversely, for FSO, this approach not only doubles the initial feature set to 134 but also proportionally increases the BO weights. However, only the 99 most relevant features, as determined by the optimal weights after 100 Bayesian Optimization iterations, are ultimately selected and encoded into the circuit. A key aspect to note is that for the reloaded features, both for NFO and FSO, we applied an alpha scaling factor that is twice the one reported in configuration 1 in Frame~\ref{lst: HH_config} in Appendix~\ref{sub: ansatz_hyperparams} for the original features. In quantum machine learning, feature scaling plays a significant role, as different scaling choices can lead to distinct rotations and ultimately result in different models. By weighting the reloaded features differently, we aim to enable more possible variation in the feature map that can result from optimization via our framework to potentially obtain a better performing QML model.  
Moreover, we use basic error mitigation and twirling~\cite{Wallman_2016}, to avoid increasing QPU run times. In Figure~\ref{fig: real_HW_layout}, we present the ibm\_brisbane layout obtained from the quantum circuit implementing NFO.  

\subsection{Results}
\label{sub: real_hw_results}

\comment{
\subsection{Analyses}
\label{sub: real_hw_analyses}
}
\subsubsection{Numerical results analyses}
\label{subsub: numerical_real_hw_analyses}

In Table~\ref{tab: real_HW_results}, we report the NFO and FSO scores obtained with the ibm\_brisbane QPU. Additionally, we apply the same feature optimization to the reduced PLAsTiCC astronomy dataset on the noiseless simulator, leveraging a 15-qubits Heisenberg Hamiltonian feature map. The corresponding results are presented in a smaller font. The comparison with real hardware results is particularly relevant as it allows for an evaluation of the benefits of utilizing a larger number of qubits, even in the presence of noise, compared to a smaller number of noiseless qubits. We do not report standard deviation in the results, as the experiments were conducted using only a single seed of the PLAsTiCC astronomy dataset due to the high cost of the real hardware experiments, as previously mentioned. In the following analysis sections, when we refer to the noiseless experiments in comparison with the real hardware experiments, we are specifically referring to this 15-qubits Heisenberg Hamiltonian noiseless experiment.
\begin{table}[h]
\centering
\begin{tabular}{|c|c|c|}
        \hline
        Metric & NFO & FSO \\
        \hline
        \multirow{2}{*}\centering{AUC} 
        & \makecell{$0.824$ \\ \scriptsize $0.620^*$}  
        & \makecell{{\boldmath$0.843$} \\ \scriptsize {\boldmath$0.656^*$}} \\
        \hline
        \multirow{2}{*}\centering{AUC percentage change} 
        & \makecell{- \\ \scriptsize -}  
        & \makecell{$+2.4\%$ \\ \scriptsize $+5.8\%^*$} \\
        \hline
    \end{tabular}
\caption{AUC (Percentage Change) Test Scores on the PLAsTiCC Astronomy reduced dataset. We report both scores obtained on ibm\_brisbane QPU (in larger font) and from the noiseless simulator (in smaller font and marked with *) with a 15-qubit Heisenberg Hamiltonian feature map. Results are not directly comparable to Section~\ref{sec:results}, as the data reloading technique increases the number of features and the dataset size was reduced for execution on real hardware.}
\label{tab: real_HW_results}
\end{table}

The first key result to highlight is the confirmation that FSO manipulation provides improvements over NFO. This finding is consistent with the results presented in Section~\ref{sec:experiments}, as it demonstrates that optimizing the way we fed features for encoding is beneficial not only on an ideal noiseless quantum computer but also on currently available noisy quantum hardware.
The second important point to note is the comparison between results obtained from real hardware and those from noiseless$^*$ simulations.  A significant AUC improvement is observed, highlighting the value and utility of using a larger number of qubits. 
Despite the presence of noise, these qubits produce better results than a smaller number of noiseless qubits. However, the AUC percentage change shows a greater improvement in the noiseless experiment compared to the real hardware one. While this can be partly attributed to the presence of noise, it also aligns with the analysis discussed in Figure~\ref{fig: ptc_improvements_per_qubit}. There, we highlighted that, with a fixed feature map scheme, increasing only the number of qubits leads to a reduction in the percentage change, primarily due to the decreased encoding density of the feature map.
Also, during the real hardware execution, we observed that the noise profile of ibm\_brisbane QPU changed, which ultimately impacted performance, particularly in the case of FSO, where the most execution time was spent running 100 Bayesian Optimization iterations. For this reason, after completing the training phase, we re-run the final circuit fitted with the best set of weights found in the 100 training iterations.
We performed this step to evaluate the test set performance under the same hardware error profile that was present during training of the final model with the optimal weights. Of course, we did it also for NFO in order to compare the results obtained with FSO using the same hardware error profile. We also tried re-transpiling the final circuit to adapt it to the current hardware error configuration. However, we found that re-running the circuit while maintaining the same hardware configuration that was used to obtain the optimal weights yielded the best results.
Given the variation in the hardware error profile, we believe that our results are not overly optimistic and may, in fact, underestimate the actual final AUC value. As quantum processors continue to advance alongside improvements in error mitigation techniques, adopting more sophisticated methods beyond twirling, such as Zero-Noise Extrapolation (ZNE), would likely yield further benefits. However, we were unable to test these approaches due to the increased execution time and associated costs. Similarly, leveraging the latest Heron processors, which exhibit lower error rates compared to the Eagle series, could further enhance performance, but we were unable to test them due to their significantly longer queue times compared to Eagle.

\subsubsection{Training Analysis}
\label{subsub: real_hw_training_analysis}
\begin{figure}[h]
\centering
\includegraphics[width=0.4\textwidth]{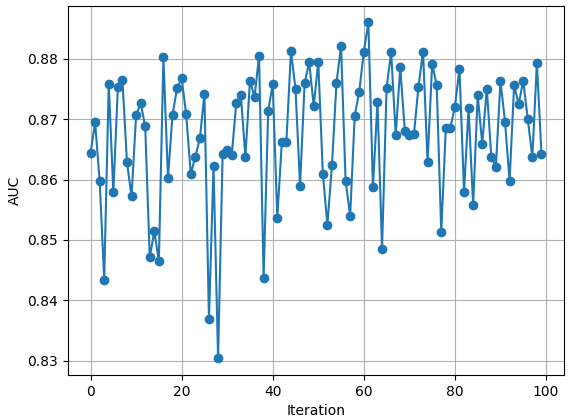}
\caption{Bayesian Optimization convergence trend observed executing Feature Selection Ordering optimization on the reduced PLAsTiCC Astronomy dataset employing imb\_brisbane QPU for 100-qubits Heisenberg Hamiltonian feature map. On the $y-$axis, we report the average training AUC score over the 10 different splits at each iteration, reported on the $x-$axis.}
\label{fig: BO_convergence_HH_brisbane}
\end{figure}

As in Section~\ref{subsub: bo_convergence_main} for the noiseless experiments, we also show the trend of the Bayesian Optimization iterations in Figure \ref{fig: BO_convergence_HH_brisbane} to illustrate the learning process of our QML model fitted with FSO manipulation on real hardware. It is worth noting that, despite being executed on a noisy processor, the optimization process exhibits a convergence trend. Particular attention should be given to two specific points in the plot, around the 27th and 64th iterations. In the first instance, a significant drop in training performance is observed, followed by a rapid recovery. In the second case, however, performance declines and stabilizes at values lower than the previously attained optimum. These fluctuation are of interest as they coincide with time intervals in which changes in the hardware error profile were detected. Nevertheless, this behavior may not be solely attributed to hardware-induced noise but could also result from natural exploration dynamics of the Bayesian Optimization feature space. Further investigations will be conducted to better understand these effects.
In general, Bayesian Optimization demonstrates strong performance, effectively handling the presence of noise. A crucial factor contributing to this robustness is likely the application of twirling, which, in this specific case study, appears sufficient to mitigate error disturbances. However, as previously mentioned, more advanced error mitigation techniques beyond twirling could further enhance both test and training performance.

\subsubsection{Feature Importance}
\label{subsub: real_hw_feat_importance}
In this section, we present a slightly different approach compared to the one reported in Section~\ref{subsub: feat_importance}. Specifically, since feature reloading was applied during real hardware execution, this analysis focuses on identifying which features with the FSO manipulation are selected in both their original and reloaded forms, as well as those that are never selected or are selected only in their original or reloaded form. We observe cases of features which are never selected, cases where a feature is selected among the top 99 features in both its original and reloaded forms, and cases where a feature is selected in noiseless execution but not in the real hardware one. In Figure~\ref{fig: real_hw_feature_importance_comparison} in Appendix~\ref{app: realHW_feature_importance}, we report a complete overview of this analysis for real hardware and noiseless execution. Given the relevance of ordering in FSO, we extend our analysis to examine the ranking of the 99 selected features, identifying the most optimal encoding order. Out of the 99 selected features, 14 (or 13 in the noiseless simulator case) originated from the original features, 11 (or 12 in the noiseless simulator case) from the re-uploaded (scaled) features, and 37 were chosen from both versions. Among these 37 features, where both the original and scaled versions were included, the order was nearly balanced, with the scaled version preceding the original in 20 cases (about $54.05\%$ of the whole set of features) and the original appearing first in 17 cases (about $45.95\%$ of the whole set of features). In Figure~\ref{fig: real_hw_feature_ordering_comparison} in Appendix ~\ref{app: realHW_feature_importance}, we present the top 20 features ranked by their Bayesian Optimization-assigned weights.

\comment{In this section, we present a slightly different approach compared to the one reported in Section~\ref{subsub: feat_importance}. Specifically, since for real hardware execution we applied feature reloading, this analysis focuses on identifying which features in the FSO optimization are selected both in their original and reloaded forms, as well as those that are never selected or only selected in their original or reloaded form. In Figure~\ref{fig: real_hw_feature_importance_comparison}, we compare this analysis for real hardware and noiseless execution. Notably, it is interesting to observe how there are cases, such as feature 57, which are never selected, feature 48, which is selected among the top 99 features in both its original and reloaded forms, and feature 2, which is selected in noiseless execution but not in the real hardware one. Given the relevance of ordering in FSO optimization, we extend our analysis to examine the ranking of the 99 selected features, identifying the most optimal encoding order. Figure~\ref{fig: real_hw_feature_ordering_comparison} presents the top 20 features ranked by their BO-assigned weights. Notably, feature 44 in real hardware execution and feature 1 in the noiseless case are both selected in original and reloaded versions and stand out among the most important in the ranking. Out of the 99 selected features, 14 (or 13 in the noiseless simulator case) originated from the original features, 11 (or 12 in the noiseless simulator case) from the reuploaded (scaled) features, and 37 were chosen from both versions. Among these 37 features, where both the original and scaled versions were included, the order was nearly balanced, with the scaled version preceding the original in 20 cases (about $54.05\%$ of the whole set of features) and the original appearing first in 17 cases (about $45.95\%$ of the whole set of features).
}
 
\comment{
\begin{figure*}
    \centering
    \subfigure{
        \includegraphics[width=\textwidth]{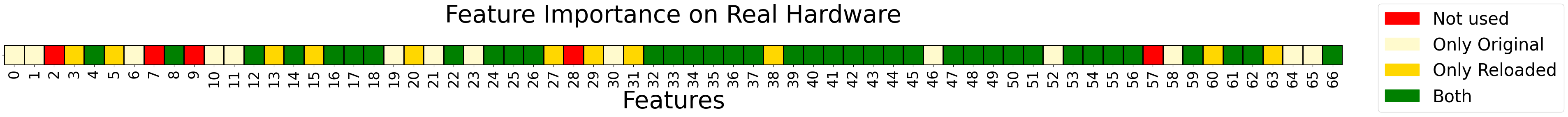}
        \label{subfig: real_hw_feature_scoring}
    }
    \\
    \subfigure{
        \includegraphics[width=\textwidth]{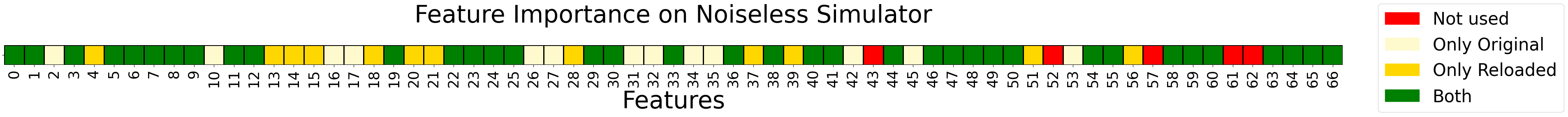}
        \label{subfig: noiseless_feature_scoring}
    } 
    \caption{Feature Importance Comparison between Hardware and Noiseless Simulator execution for FSO optimization. We can categorize the features into four groups: those that were never selected (red), those that were selected only from the original set of 67 features (light yellow), those selected only from the duplicated set -- i.e., those linked only to the different assigned weights for the reuploading setup -- (strong yellow), and those used twice in the circuit (green).}
    \label{fig: real_hw_feature_importance_comparison}
\end{figure*}

\begin{figure*}
    \centering
    \subfigure[Top 20 Features Ordering on Real Hardware]{
        \includegraphics[width=0.4\textwidth]{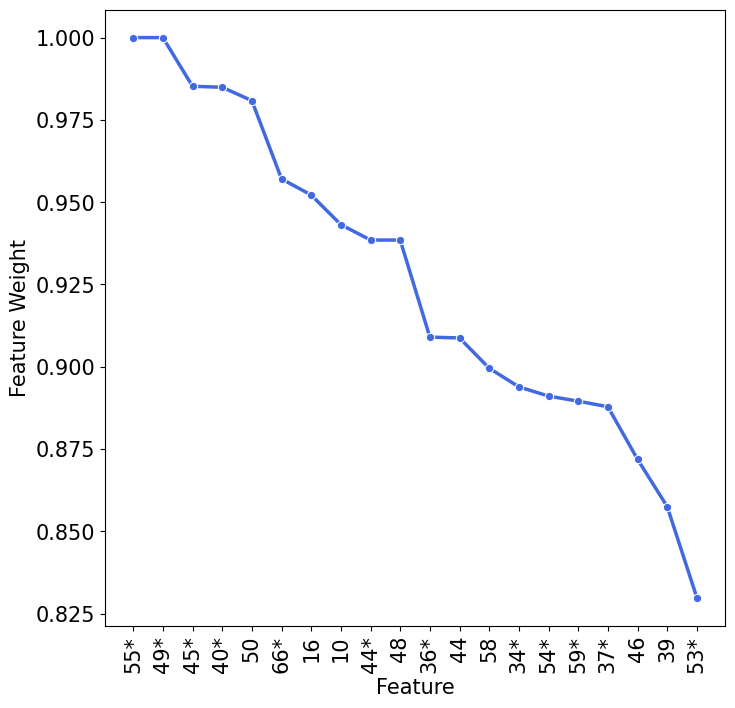}
        \label{subfig: real_hw_feature_ordering}
    }
    \subfigure[Top 20 Features Ordering on Noiseless Simulator]{
        \includegraphics[width=0.4\textwidth]{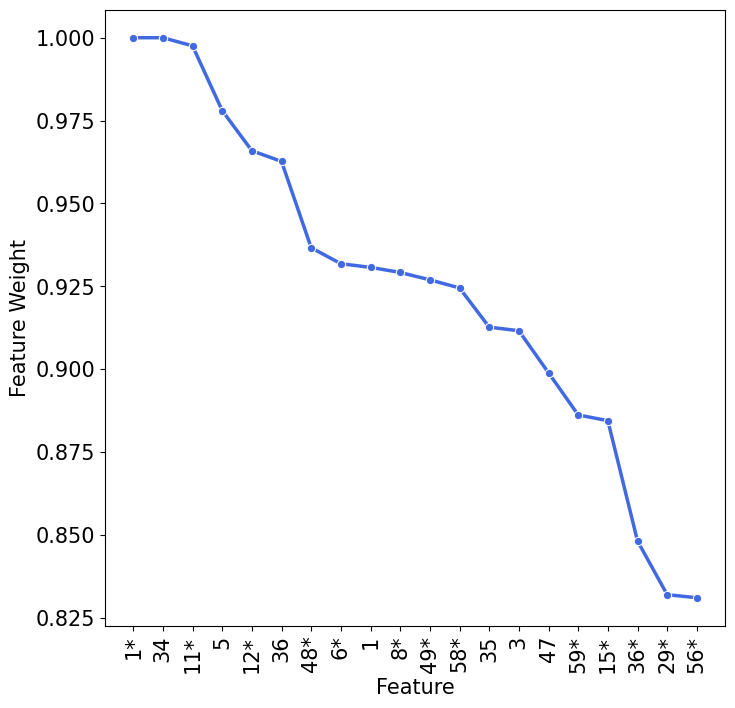}
        \label{subfig: noiseless_feature_ordering}
    } 
    \caption{Feature Ordering Comparison between Hardware and Noiseless Simulator execution for FSO optimization. We report the first 20 features out of the 99 selected, ranked by importance (and thus encoding), as measured by the magnitude of their weights. The features marked with an asterisk (*) are the reloaded features.}
    \label{fig: real_hw_feature_ordering_comparison}
\end{figure*}
}

\comment{
\begin{figure*}[h]
\centering
\includegraphics[width=\textwidth]{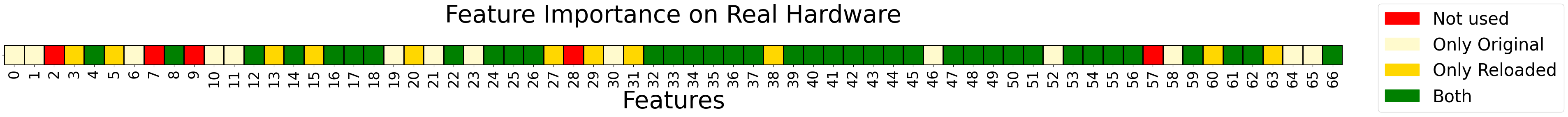}
\caption{Bayesian Optimization convergence trend observed executing Feature Selection Ordering optimization on the reduced PLAsTiCC Astronomy dataset employing imb\_brisbane QPU for 100-qubits Heisenberg Hamiltonian ansatz. On the $y-$axis, we report the average training AUC score over the 10 different splits at each iteration, reported on the $x-$axis.}
\label{fig: trial}
\end{figure*}
}

%% file: sections/conclusion.tex
\section{Conclusion}
\label{sec:conclusion}
We have proposed and implemented an innovative framework -- referred to as Quantum Feature Encoding Optimization (QFEO). This is a breakthrough from previous works where the feature encoding of a QML model, i.e., how data is preprocessed and fed into a given quantum feature map, remains fixed – and identifies and provides an effective way exploit unique-to-quantum aspects of manipulating input features to improve QML modeling.  Importantly, the framework is generic with respect to the QML model and the feature map used for the data encoding, the data manipulation techniques, and the input data. Moreover, QFEO leverages multiple rounds of cross-validation to accurately reflecting the realistic modeling process and most directly optimize the true expected model performance -- this is key for the approach to work consistently in practice.

We initially evaluated the feasibility and effectiveness of the procedure of optimizing the way we fed data for encoding with feature selection, feature ordering, feature weighting, and/or combinations of these techniques using noiseless simulation. This experimentation was conducted on a variety of datasets and varying number of qubits, and by selecting three specific feature maps with multiple different configurations, confirming that, on an ideal fault-tolerant quantum computer, our approach would consistently bring concrete improvement - demonstrating up to a $5.9\%$ average improvement across datasets and number of qubits, and up to $11\%$ average improvement for specific datasets and feature maps.

With experiments with an IBM Eagle QPU, we demonstrated that we can achieve improvement in QML model performance with our QFEO framework  even on modern noisy quantum processors. We manipulated the input features before encoding by applying feature selection ordering (FSO) achieving a $+2.4\%$ improvement in the AUC score compared to standard fixed feature encoding.

As already highlighted in the precedent analyses, the obtained results also depend on the data used. We conduct experiments on 4 datasets of different nature and dimensionality, using three different feature map with different configurations and varying the number of qubits from 9 to 15, reaching up to 100 in the case of experiments on the ibm\_brisbane QPU, which suggests the generality / general applicability of our approach.
As stated in Section \ref{subsec: data_manipulation}, for the sake of understanding, we report a subset of simple and interpretable data manipulation techniques that are interesting as being mostly unique to QML models and not yet explored in the literature. Nevertheless, the QFEO framework remains general and could support any variety of data manipulations. A possible direction for future research might be to go a step further than simple weighting / scaling by taking scaling into account and adding an offset, or a linear combination of features -- i.e., weight parameters form a weight matrix that maps the input features to linear combinations of them to then feed into the circuit. In general, depending on the weight parameters, we might take a different optimization approach to do it best and most efficiently.

As we move towards achieving quantum utility, the need for deeper and more complex circuits will only grow. Our analysis, as shown in Figure~\ref{fig: ptc_improvements_per_qubit}, indicates that circuits become increasingly complex as their density increases. This highlights the importance of exploring additional feature maps beyond those already used, as they could offer valuable insights for further advancements in circuit design.
\comment{A further point of investigation is a more detailed verification of how a dense encoding results in better performances compared to a less dense encoding, as highlighted in Figure~\ref{fig: ptc_improvements_per_qubit} \tom{add more fm}.}Furthermore, given the continuous improvement of quantum processors, it would be interesting to use the latest generation IBM Heron processors to understand if and how they would further improve performance since they have much lower error levels (2Q error rate of $3.74\times 10^{-3}$ and EPLG of $3.74\times 10^{-3}$) than Eagle processors. However, the results we obtained are good indicators of high reliability of Eagle processors.

Finally, while not the focus of this study, an interesting direction for future work is to investigate whether quantum ML models employing QFEO could improve over classical ML models, in combination with additional tuning of the quantum feature maps used. In Appendix~\ref{app: classical_ML results}, we report the performance of classical models on the same datasets used in the quantum experiments and discuss the results. Note that as the goal here was to improve a given QML model for a specified quantum feature map, we generally would not expect to have a high chance to improve over classical models without also further tuning or optimizing the particular feature map circuit selected as well.


%% file: sections/appendix.tex
\appendix
\onecolumn
\section{Feature importance analyses}
\label{app: feature_importance}

\subsection{Noiseless analyses}
\label{app: noiseless_feature_importance}
In this section, we report the complete overview regarding the feature importance for the dataset with the largest and smallest number of features, that are the Churn and the German Numeric dataset respectively. As already described in Section~\ref{subsub: feat_importance}, the objective is to illustrate that a feature considered crucial for one type of encoding optimization might be either equally important or entirely irrelevant for another.
\subsubsection{Churn feature importance}
\label{subsec: churn_importance_general}
In Section~\ref{subsub: feat_importance}, we briefly analyze some feature cases, such as feature 28. Here, we add other insights about other interesting features such as 41, 75, 49, and 15. If it is true that feature 41 is considered important in feature weighting (it is in fact the 16th most important feature) and this is also reflected in feature selection and feature selection ordering, it is not as true for the other features. In fact, features 75, 49, and 15 are very important for feature selection and feature selection ordering but much less so in the case of feature weighting. This highlights the fact that while feature weighting may place more focus on certain features, optimizations concerning feature selection may select others that are considered more promising to encode. It is also interesting to focus on feature 13: this is the least important feature for feature weighting and it is also considered not so important by all other optimization methods except FS. This is an interesting result since we can notice that FSO, although being a FS-based method, does not consider feature 13 as important as FS does. Always considering the four features mentioned above, we notice that feature 41 seems to be relevant both in terms of weight and order for the feature weighting ordering technique. In contrast, features 75, 49, and 15 seem not be so relevant for FWO, following the FW importance trend. However, there are cases such as feature 19 that results in high relevance for feature weighting ordering but not for feature weighting. For what concern FWOW importance results, by exploiting two different sets of weights, the weighting process is not so correlated with the ordering one, as expected. Indeed, we can notice examples of feature, such as the 27th and the 95th, where a darker color does not also correspond to higher ordering. Correlation that we instead find in the FSO and FWO processes since the ordering is optimized on the same sets of weights used for the selection and weighting respectively.

\begin{figure}[h]
\centering
\makebox[\textwidth]{\includegraphics[width=1.\textwidth]{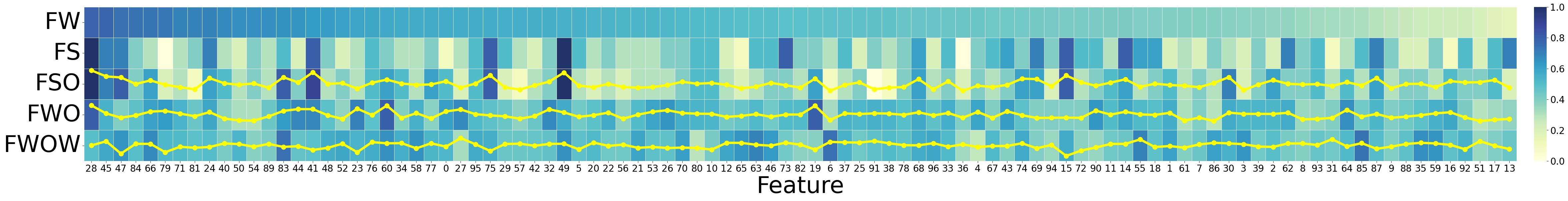}}
\caption{Feature importance analysis on Churn dataset leveraging Separate Entangled feature map, with the configuration 1 reported in Frame~\ref{lst: SE_config}, to encode features. The color represents the importance of each feature, labeled on the $x$-axis, concerning the optimization reported on the $y$-axis: the darker it is, the greater the importance. Feature weighting gives the order of the feature importance, starting from the most important one to the least one. For the manipulations involving ordering, we report the yellow line which represents the ordering of each feature for the corresponding optimization. The reported results, both in terms of colors and dotted line, are the average results over the 10 different dataset batches.}
\label{fig:feature_scoring_churn}
\end{figure}

To support the analyses, we report the top 10 selected features for the three different feature maps tested to assess whether these features are equally important across different feature maps. For example, it is noteworthy that feature 28 is consistently selected by the feature selection optimization to be encoded across all three different feature maps. Another notable case is feature 14 which is frequently selected for encoding in Separate Entangled and Heisenberg Hamiltonian, but not in the Repeated Pauli feature map. This figure aims to further emphasize how the feature map operates across different feature spaces, leading to the different performance outcomes as reported in Section \ref{sec:results}.

\begin{figure*}[h]
\centering
\includegraphics[width=\textwidth]{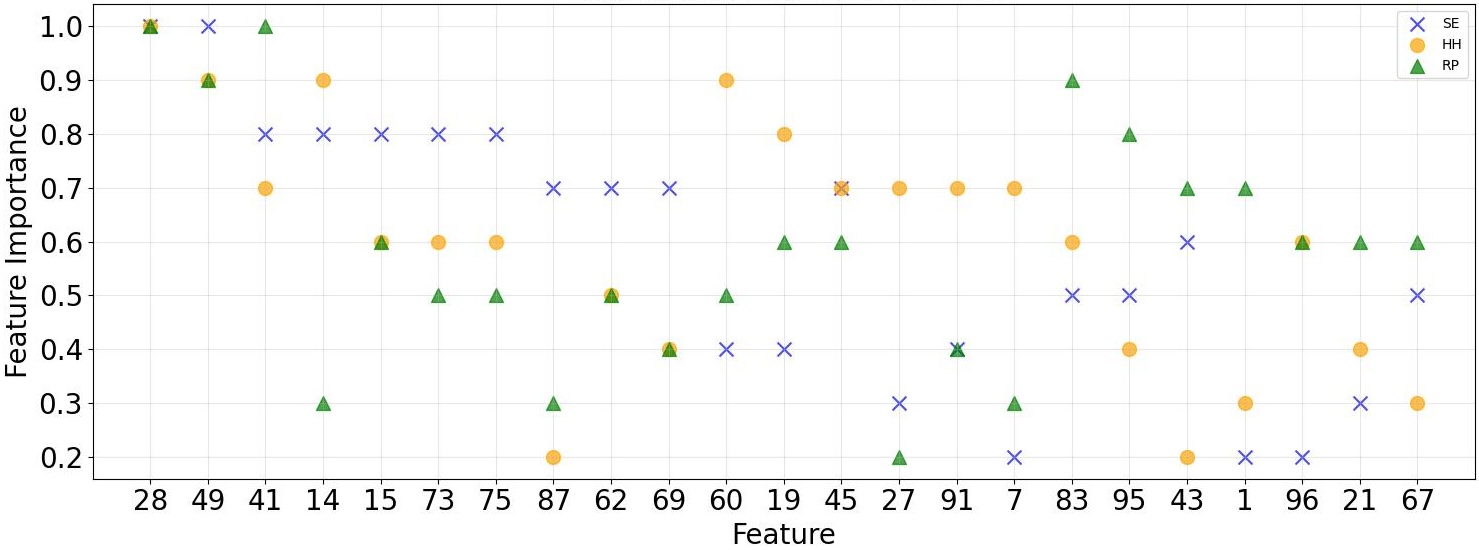}
\caption{Feature importance analysis on Churn dataset per feature map exploiting feature selection manipulation. The represented features are the union of the top 10 selected features across the three different feature maps.}
\label{fig: feat_importance_fs_per_ansatz}
\end{figure*}
\clearpage
\subsubsection{German Numeric feature importance}
\label{subsec: german_importance_general}
The analysis of the feature importance for the German Numeric dataset strictly follows the one of the Churn dataset. For example, we can notice cases of features, such as feature 10 or feature 0, which are considered important both in case of FS and FSO but not in the case of FWO and FWOW, and viceversa. One interesting insight that can be appreciated with this dataset is that in the optimizations that engage feature weighting, which means the optimizations that do not discard any feature, we can notice a more uniform feature importance trend, without features considered much more important than others. This fact is also reflected by the almost constant yellow ordering line in the FWO and FWOW cases.

\begin{figure}[h]
\centering
\makebox[\textwidth]{\includegraphics[width=1.\textwidth]{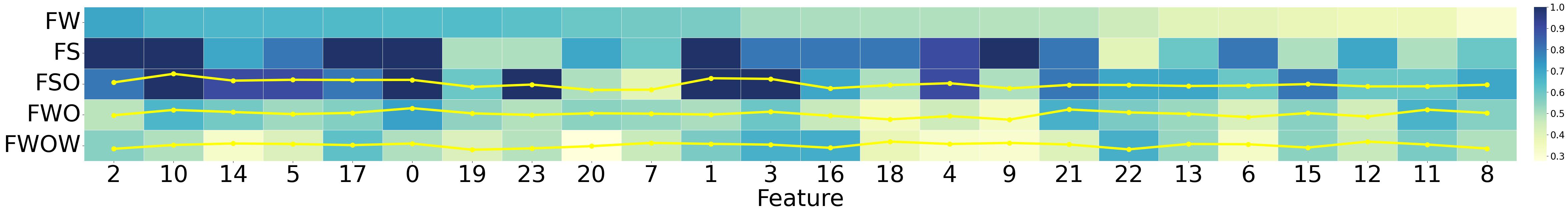}}
\caption{Feature importance analysis on German Numeric dataset leveraging Separate Entangled feature map, with the configuration 1 reported in Frame~\ref{lst: SE_config}, to encode features. The color represents the importance of each feature, labeled on the $x$-axis, concerning the optimization reported on the $y$-axis: the darker it is, the greater the importance. Feature weighting gives the order of the feature importance, starting from the most important one to the least one. For the optimizations involving ordering, we report the yellow line which represents the ordering of each feature for the corresponding optimization. The reported results, both in terms of colors and dotted line, are the average results over the 10 different dataset batches.}
\label{fig:feature_scoring_german}
\end{figure}

\subsection{Real Hardware analyses}
\label{app: realHW_feature_importance}

\begin{figure*}
    \centering
    \subfigure{
        \includegraphics[width=\textwidth]{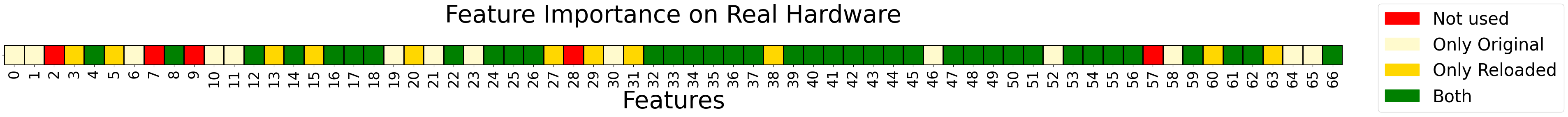}
        \label{subfig: real_hw_feature_scoring}
    }
    \\
    \subfigure{
        \includegraphics[width=\textwidth]{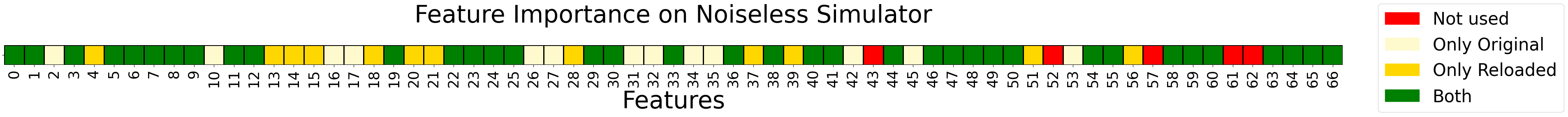}
        \label{subfig: noiseless_feature_scoring}
    } 
    \caption{Feature Importance Comparison between Hardware and Noiseless Simulator execution for FSO manipulation. We can categorize the features into four groups: those that were never selected (red), those that were selected only from the original set of 67 features (light yellow), those selected only from the duplicated set -- i.e., those linked only to the different assigned weights for the reloading setup -- (strong yellow), and those used twice in the circuit (green).}
    \label{fig: real_hw_feature_importance_comparison}
\end{figure*}

In this section, we report a complete analyses regarding feature importance mentioned in Section~\ref{subsub: real_hw_feat_importance}. Since for real hardware execution we applied feature reloading and we only test feature selection ordering (FSO) as data manipulation technique, this analysis focuses on identifying which features in the FSO optimization are selected both in their original and reloaded forms, as well as those that are never selected or only selected in their original (or reloaded) form. In Figure~\ref{fig: real_hw_feature_importance_comparison}, we report a comparison between real hardware and noiseless execution. Notably, it is interesting to observe cases, such as feature 57, which are never selected, feature 48, which is selected among the top 99 features in both its original and reloaded forms, and feature 2, which is selected in noiseless execution but not in the real hardware one. 

\begin{figure*}
    \centering
    \subfigure[Top 20 Features Ordering on Real Hardware]{
        \includegraphics[width=0.4\textwidth]{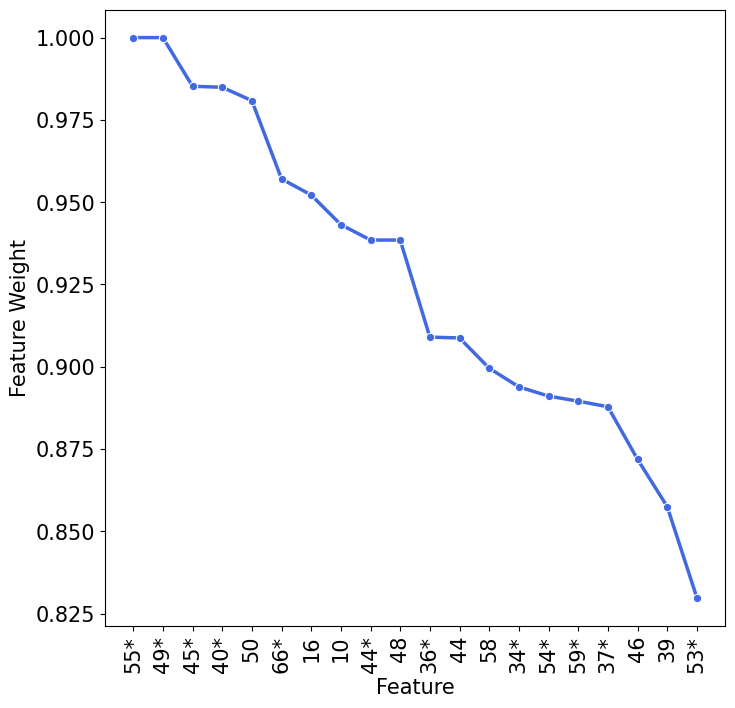}
        \label{subfig: real_hw_feature_ordering}
    }
    \subfigure[Top 20 Features Ordering on Noiseless Simulator]{
        \includegraphics[width=0.4\textwidth]{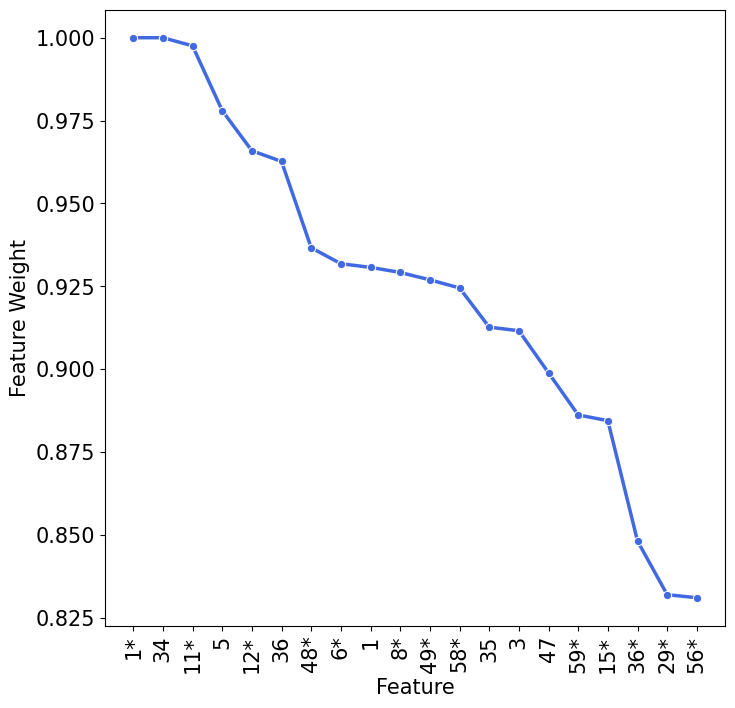}
        \label{subfig: noiseless_feature_ordering}
    } 
    \caption{Feature Ordering Comparison between Hardware and Noiseless Simulator execution for FSO manipulation. We report the first 20 features out of the 99 selected, ranked by importance (and thus encoding), as measured by the magnitude of their weights. The features marked with an asterisk (*) are the reloaded features.}
    \label{fig: real_hw_feature_ordering_comparison}
\end{figure*}

Given the relevance of ordering in FSO, we extend our analysis to examine the ranking of the 99 selected features, identifying the most optimal encoding order. Figure~\ref{fig: real_hw_feature_ordering_comparison} presents the top 20 features ranked by their Bayesian Optimization-assigned weights. Notably, feature 44 in real hardware execution and feature 1 in the noiseless case are both selected in original and reloaded versions and stand out among the most important in the ranking.

\newpage

\section{Circuit Expressibility}
\label{sec: expressibility}
In this section, we explore how the data encoding strategies influence the expressiveness of a quantum circuit -- that is, its ability to generate a rich variety of quantum states. A circuit with higher expressiveness can explore a broader portion of the Hilbert space (the entire space of quantum states).  A common approach looking at analyzing expressibility of a parameterized ansatz is to compare the distribution of states it generates when varying its parameters to a uniform (Haar) distribution over all states ~\cite{Cerezo_2021,sim2019expressibility}.  We follow a similar approach, but instead look from a linear algebra perspective and analyze the intrinsic dimension with the lower dimension projection needed to closely approximate the set of states the circuit is capable of generating with the given manipulations.  This gives an interpretable way to characterize the richness of possible quantum states that can be created, following widely used linear algebra ideas  ~\cite{jolliffe2002pca,jolliffe2016principal,kirby2000geometric}. 

To quantify expressiveness, we begin by generating a set of random features sampled from a normal distribution, as it is common in machine learning to assume normally distributed features or to standardize them in order to approximate this assumption. Then, we apply random permutations of the features (feature ordering), random weighting (feature weighting), or feature subset selection (feature selection) across multiple iterations to assess the circuit's sensitivity to different feature optimization strategies. Specifically, for each of the $T = 1000$ iterations, we apply a unique permutation, weighting, or subset selection to the input features, encode them into a quantum state, and extract the resulting statevector. This process yields a final statevector matrix $S \in \mathbb{C}^{T\times2^{n}}$, where $n$ is the number of qubits. \comment{\andres{[Interesting idea. However, I know that for random sampling you get concentration of states in certain parts of the Hilbert space (see \url{https://pennylane.ai/qml/demos/tutorial_haar_measure}) I'm curious about what was the reasoning behind to choose gaussian distributions and why would it be better than taking into account the Haar measure for sampling.] \tom{here we are not trying to get some uniform distribution across the random states. Instead, we are trying to model what are the typical classical features that you have and it is generally assumed that they come from the Gaussian/Normal distribution. Added a brief sentence to clarify this point when we specified the chosen normal distribution.}}}

To analyze the diversity of the resulting state space, we apply Principal Component Analysis (PCA) ~\cite{jolliffe2002pca,jolliffe2016principal} on the matrix $S$ by employing Singular Value Decomposition (SVD). The number of principal components required to retain a specified level of explained variance serves as an indicator of expressiveness -- the more components needed, the more expressive the circuit.  This is similar to the previous expressibility approaches, because a uniform distribution over states would require all dimensions to approximate well, whereas if states are mostly concentrated in a lower dimensional subspace then a good approximation could be had with few dimensions.  
Additionally, we assess the statevector reconstruction error to further evaluate expressiveness. This is done by truncating the SVD to a rank $r$, reconstructing the statevector matrix $\hat{S}_{r}$, and computing the reconstruction error as $E_r = || S - \hat{S}_r||_2$, where $\hat{S}_r = U_r\Sigma_rV^{\dagger}_r$. Basically, we strive to create all possible quantum states by performing a specific data manipulation (e.g., feature ordering) and then observe what the intrinsic dimensionality of the resulting statevector is. We restrict the intrinsic dimensionality and quantify the extent to which it can be recovered. High recoverability indicates that the data manipulation initially applied does not substantially increase the intrinsic dimensionality of the feature space. That is, higher reconstruction error $E_r$ for smaller ranks suggests greater circuit expressiveness since it indicates that the circuit explores a broader and more complex region of the Hilbert space that cannot be easily captured by a low-dimensional representation.

The entire procedure is repeated 30 times, each time with a newly generated set of input features sampled from the same distribution. The final metrics, such as the reconstruction error, are then averaged across all repetitions. This allows us to evaluate the expressiveness of the circuit under varying input conditions, ensuring the robustness of the results with respect to the input data.

\begin{figure*}[t]
    \centering
    \subfigure[L2 norm reconstruction error retained 4 qubits]{
        \includegraphics[width=0.45\textwidth]{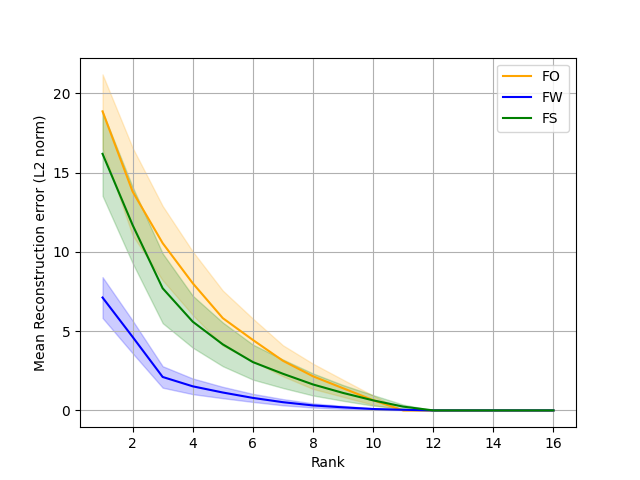}
        \label{subfig: l2_error_4qubits}
    }
    \subfigure[L2 norm reconstruction error retained 9 qubits]{
        \includegraphics[width=0.45\textwidth]{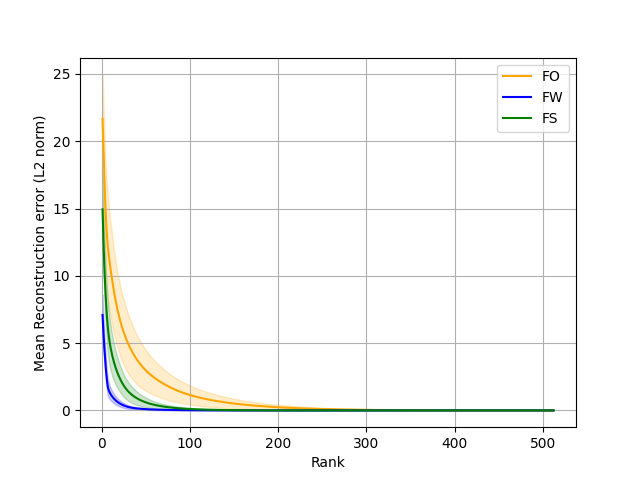}
        \label{subfig: l2_error_9qubits}
    }\\
    \subfigure[Principal components retained 4 qubits]{
        \includegraphics[width=0.45\textwidth]{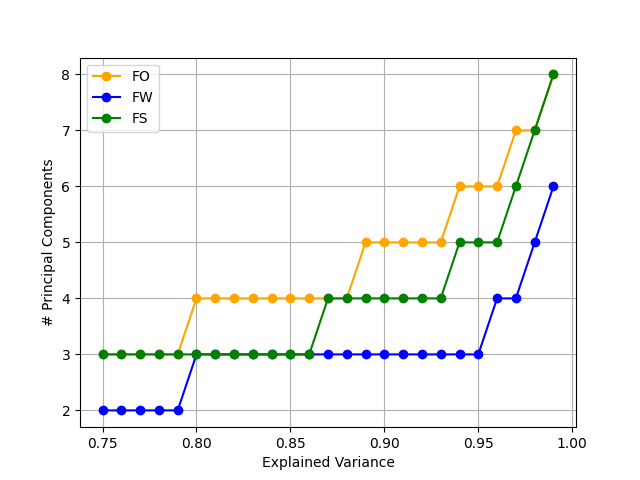}
        \label{subfig: n_components_4_qubits}
    } 
    \subfigure[Principal components retained 9 qubits]{
        \includegraphics[width=0.45\textwidth]{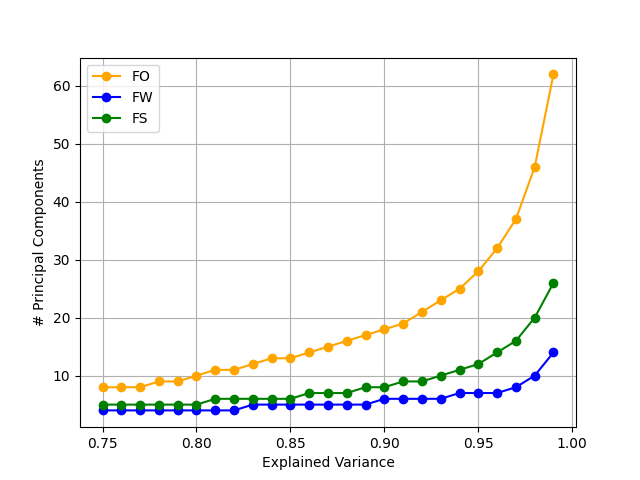}
        \label{subfig: n_components_9_qubits}
    }
    \caption{Reconstruction error (Figures~\ref{subfig: l2_error_4qubits},~\ref{subfig: l2_error_9qubits}) and number of retained principal components (Figures~\ref{subfig: n_components_4_qubits},~\ref{subfig: n_components_9_qubits}) as a function of the SVD rank and explained variance, respectively, for circuits with 4 and 9 qubits. The results compare feature ordering (FO), feature selection (FS), and feature weighting (FW) strategies using the Heisenberg Hamiltonian feature map. Each curve represents the average over 30 independent runs with randomly sampled input features.}
    \label{fig: circuit expressibility}
\end{figure*}

Figures~\ref{subfig: l2_error_4qubits} and~\ref{subfig: l2_error_9qubits} show the statevector reconstruction error for feature ordering, feature selection, and feature weighting strategies, applied to circuits with 4 and 9 qubits, respectively. In Figures~\ref{subfig: n_components_4_qubits} and~\ref{subfig: n_components_9_qubits}, we present the number of principal components retained after applying PCA in each case.
The most notable result is that feature ordering enables the circuit to explore a broader portion of the Hilbert space compared to feature weighting. This is evidenced by the consistently higher reconstruction error at lower ranks, as well as the greater number of principal components required to retain a given amount of variance. Additionally, it is observed that, as the number of qubits increases, the reconstruction error curves become smoother. In the feature ordering case, the number of retained components grows significantly with the number of qubits, whereas in the feature weighting case, it remains relatively stable, indicating a limited increase in expressiveness. Regarding feature selection, it is noteworthy that as the number of qubits increases, both the reconstruction error and the number of retained principal components gradually shift from resembling the feature ordering case to approaching the feature weighting case. However, in the high-variance regime, the number of retained components remains approximately twice as high as in the feature weighting case.

\newpage

\section{Bayesian Optimization convergence}
\label{app:BO_convergence}
In this section, we resume and complete the analysis started in Section~\ref{subsub: bo_convergence_main} on the training phase of our QML models. The goal is to observe the trend of the Bayesian Optimization procedure over all the iterations to see if our approach shows convergence patterns. 

\begin{figure}
    \centering
    \subfigure[FSO on Churn dataset]{
        \includegraphics[width=0.3\textwidth]{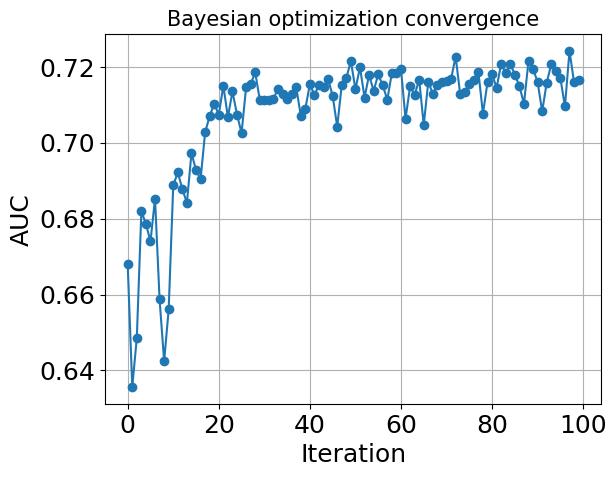}
        \label{subfig: fso1_churn}
    }
    \subfigure[FWO on Churn dataset]{
        \includegraphics[width=0.3\textwidth]{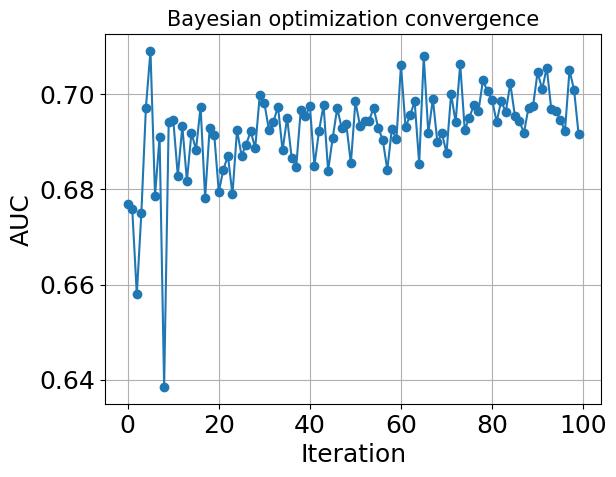}
        \label{subfig: fwo1_churn}
    }
    \subfigure[FWOW on Churn dataset]{
        \includegraphics[width=0.3\textwidth]{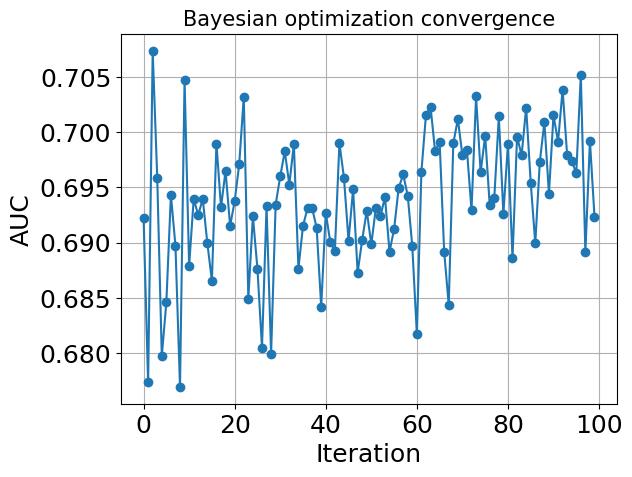}
        \label{subfig: fwow1_churn}
    }
    \\
    \subfigure[FSO on Virtual Screening dataset]{
        \includegraphics[width=0.3\textwidth]{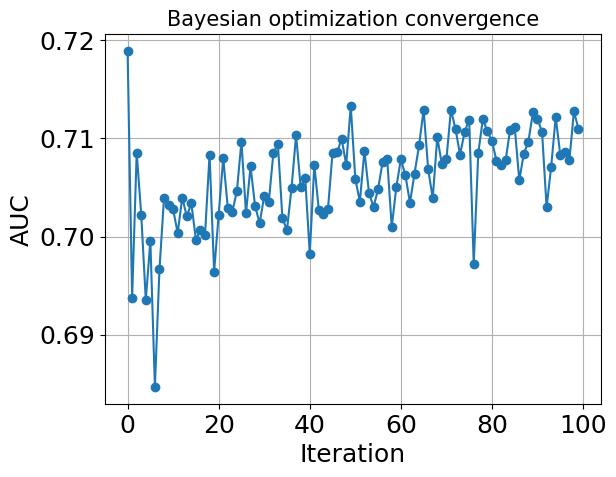}
        \label{subfig: fso1_virtual}
    } 
    \subfigure[FWO on Virtual Screening dataset]{
        \includegraphics[width=0.3\textwidth]{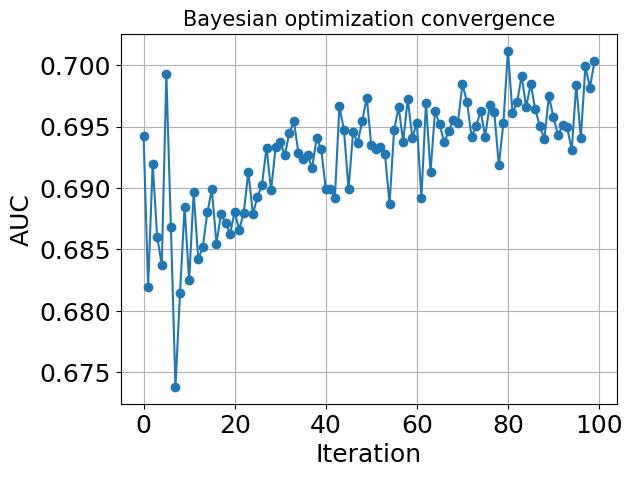}
        \label{subfig: fwo1_virtual}
    }
    \subfigure[FWOW on Virtual Screening dataset]{
        \includegraphics[width=0.3\textwidth]{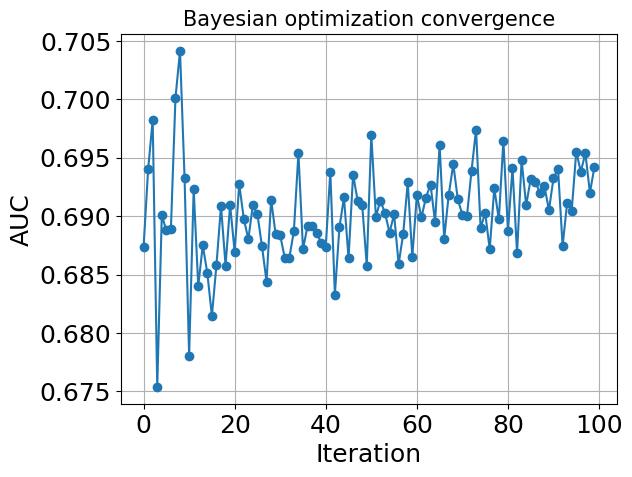}
        \label{subfig: fwow1_virtual}
    }
    \\
    \subfigure[FSO on German Numeric dataset]{
        \includegraphics[width=0.3\textwidth]{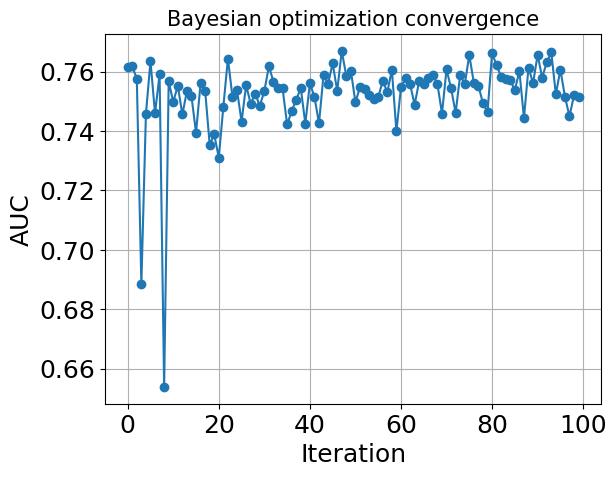}
        \label{subfig: fso1_german}
    } 
    \subfigure[FWO on German Numeric dataset]{
        \includegraphics[width=0.3\textwidth]{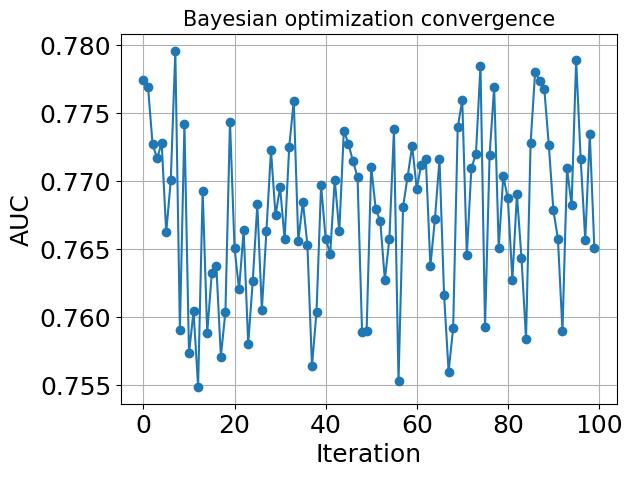}
        \label{subfig: fwo1_german}
    }
    \subfigure[FWOW on German Numeric dataset]{
        \includegraphics[width=0.3\textwidth]{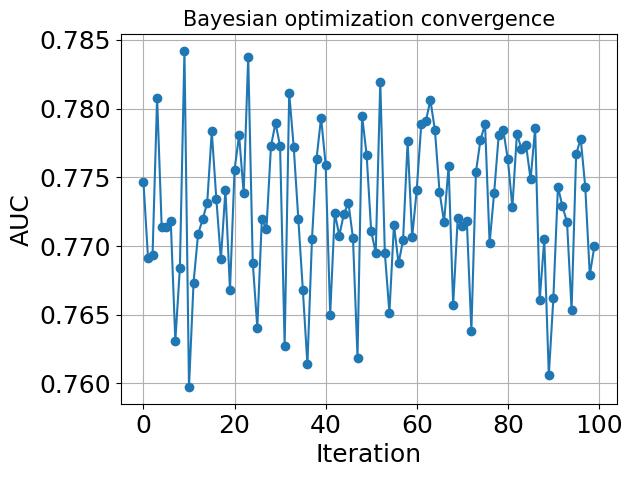}
        \label{subfig: fwow1_german}
    }
    \\
    \subfigure[FSO on Plastic Astronomy dataset]{
        \includegraphics[width=0.3\textwidth]{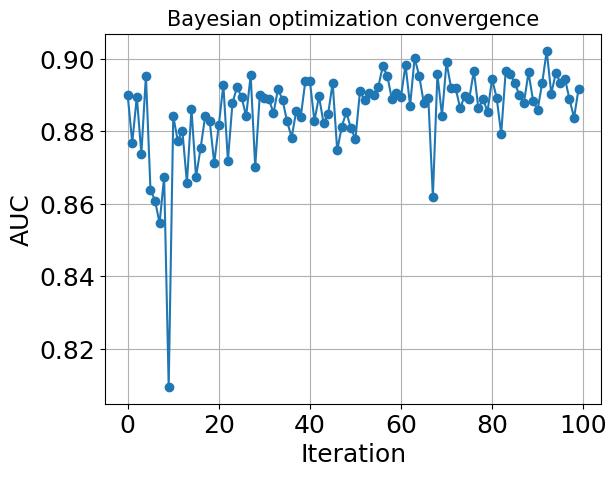}
        \label{subfig: fso1_plastic}
    } 
    \subfigure[FWO on Plastic Astronomy dataset]{
        \includegraphics[width=0.3\textwidth]{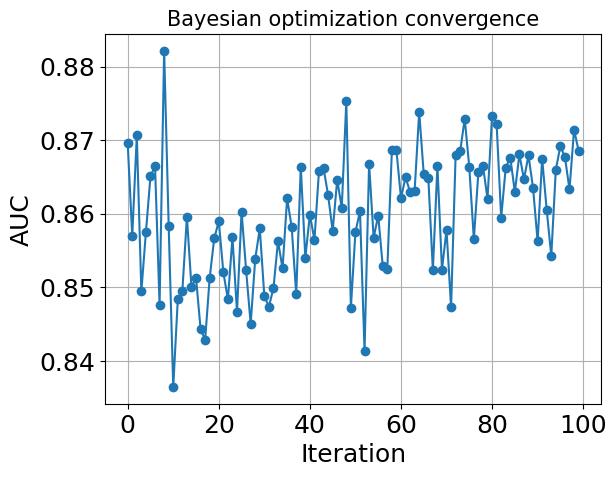}
        \label{subfig: fwo1_plastic}
    }
    \subfigure[FWOW on Plastic Astronomy dataset]{
        \includegraphics[width=0.3\textwidth]{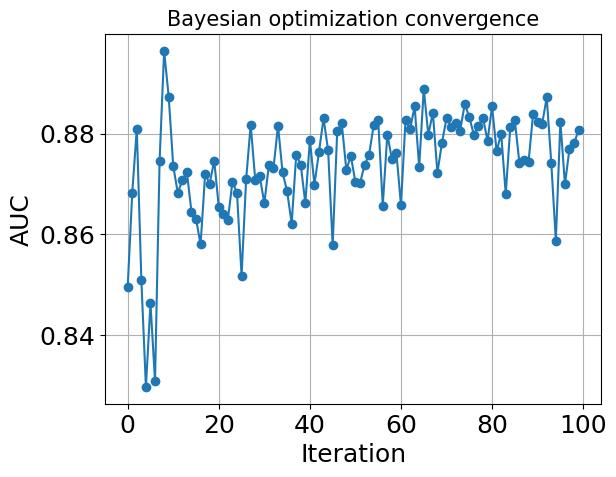}
        \label{subfig: fwow1_plastic}
    }
    \caption{Bayesian Optimization convergence trend of the QML model exploiting the 15-qubits Separate Entangled feature map. In Figure~\ref{subfig: fso1_churn},~\ref{subfig: fwo1_churn}, and~\ref{subfig: fwow1_churn}, we report the convergence trend of the Bayesian Optimization procedure for feature selection ordering, feature weighting, and feature weighting ordering weighting over the Churn dataset. In Figure~\ref{subfig: fso1_virtual},~\ref{subfig: fwo1_virtual}, and~\ref{subfig: fwow1_virtual}, we report the same study for the Virtual Screening dataset and, in Figures~\ref{subfig: fso1_german},~\ref{subfig: fwo1_german}, and~\ref{subfig: fwow1_german}, we present the same analysis but for the German Numeric dataset. Lastly, in Figures~\ref{subfig: fso1_plastic},~\ref{subfig: fwo1_plastic}, and~\ref{subfig: fwow1_plastic}, we report the BO convergence regarding Plastic Astronomy dataset.}
    \label{fig: BO convergence SE}
\end{figure}

\begin{figure}
    \centering
    \subfigure[FSO on Churn dataset]{
        \includegraphics[width=0.3\textwidth]{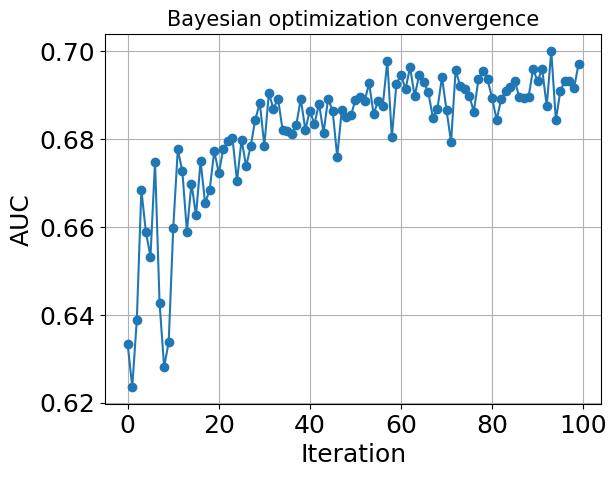}
        \label{subfig: fso4_churn}
    }
    \subfigure[FWO on Churn dataset]{
        \includegraphics[width=0.3\textwidth]{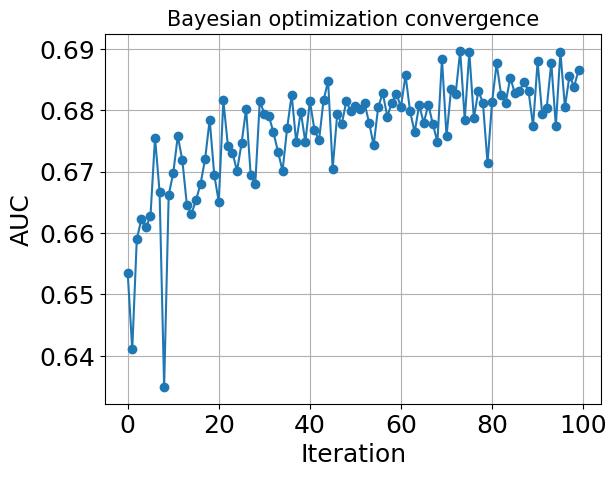}
        \label{subfig: fwo4_churn}
    }
    \subfigure[FWOW on Churn dataset]{
        \includegraphics[width=0.3\textwidth]{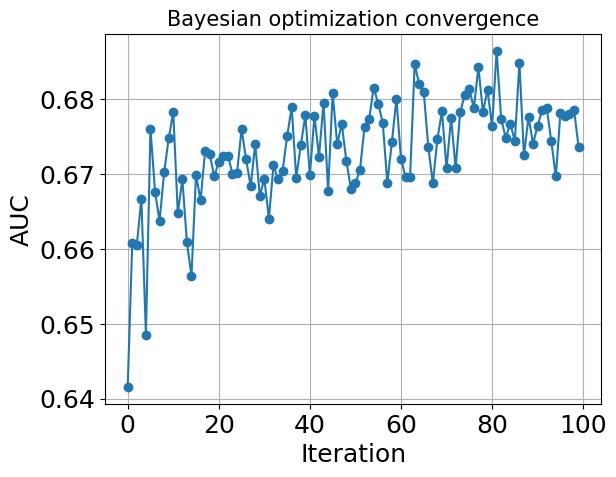}
        \label{subfig: fwow4_churn}
    }
    \\
    \subfigure[FSO on Virtual Screening dataset]{
        \includegraphics[width=0.3\textwidth]{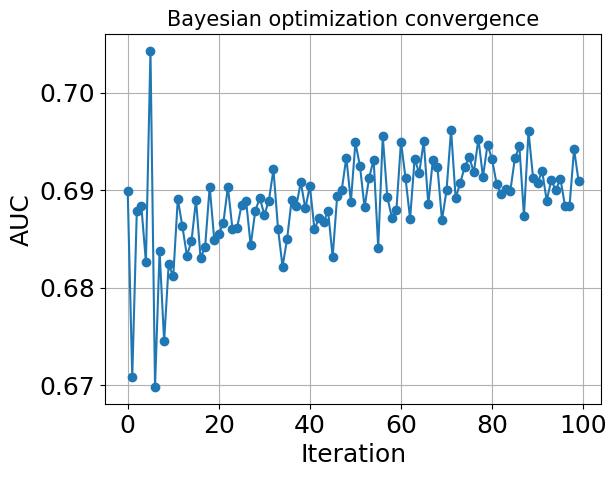}
        \label{subfig: fso4_virtual}
    } 
    \subfigure[FWO on Virtual Screening dataset]{
        \includegraphics[width=0.3\textwidth]{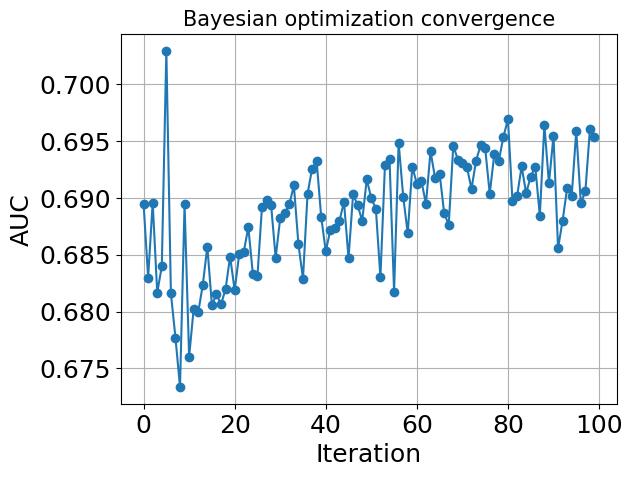}
        \label{subfig: fwo4_virtual}
    }
    \subfigure[FWOW on Virtual Screening dataset]{
        \includegraphics[width=0.3\textwidth]{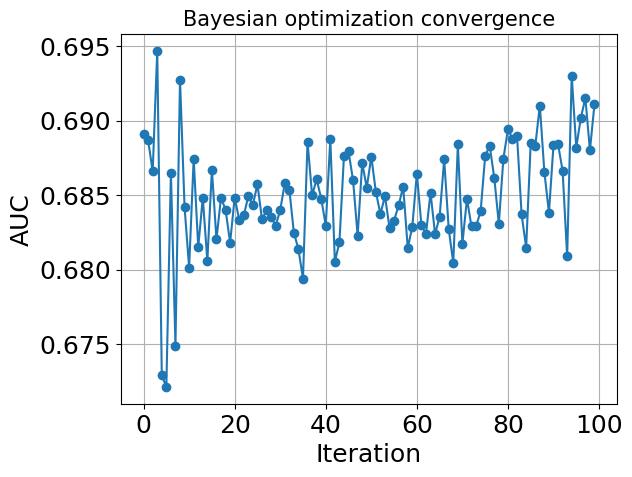}
        \label{subfig: fwow4_virtual}
    }
    \\
    \subfigure[FSO on German Numeric dataset]{
        \includegraphics[width=0.3\textwidth]{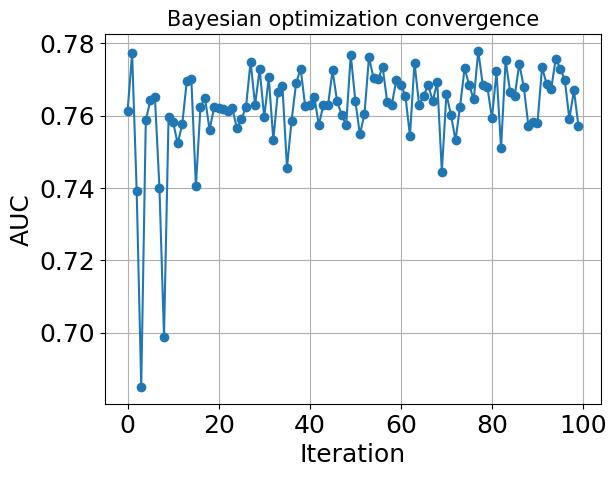}
        \label{subfig: fso4_german}
    } 
    \subfigure[FWO on German Numeric dataset]{
        \includegraphics[width=0.3\textwidth]{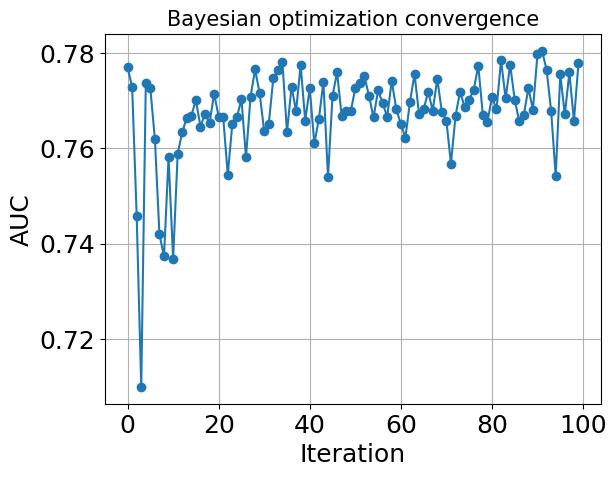}
        \label{subfig: fwo4_german}
    }
    \subfigure[FWOW on German Numeric dataset]{
        \includegraphics[width=0.3\textwidth]{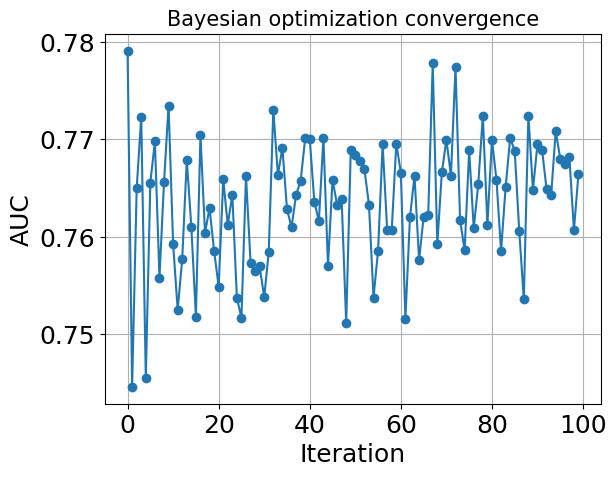}
        \label{subfig: fwow4_german}
    }
    \\
    \subfigure[FSO on Plastic Astronomy dataset]{
        \includegraphics[width=0.3\textwidth]{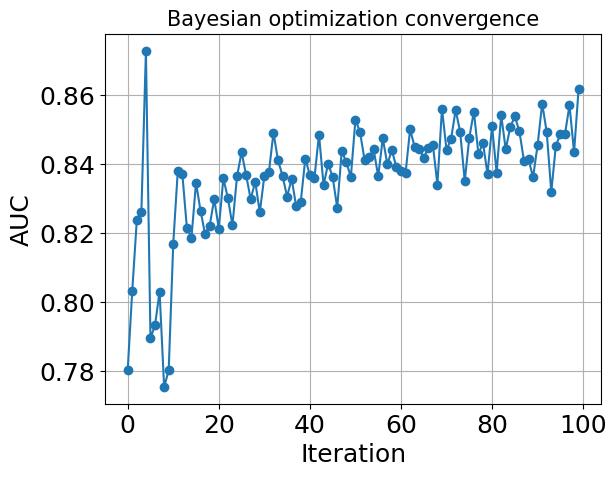}
        \label{subfig: fso4_plastic}
    } 
    \subfigure[FWO on Plastic Astronomy dataset]{
        \includegraphics[width=0.3\textwidth]{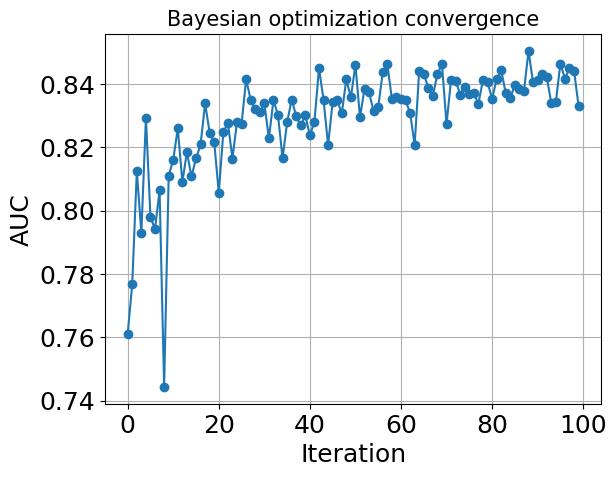}
        \label{subfig: fwo4_plastic}
    }
    \subfigure[FWOW on Plastic Astronomy dataset]{
        \includegraphics[width=0.3\textwidth]{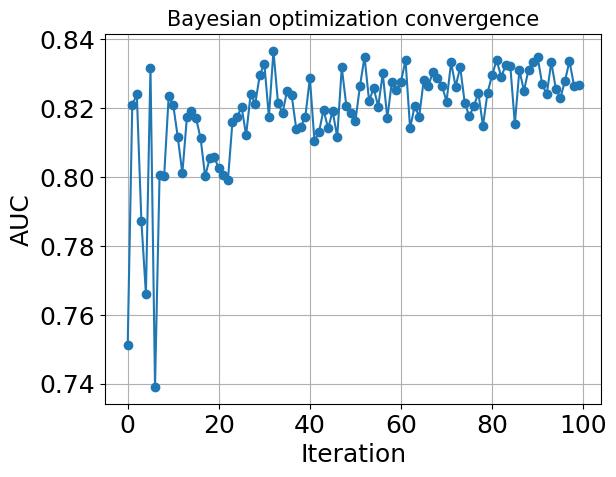}
        \label{subfig: fwow4_plastic}
    }
    \caption{Bayesian Optimization convergence trend of the QML model exploiting the 15-qubits Heisenberg Hamiltonian feature map. In Figure~\ref{subfig: fso4_churn},~\ref{subfig: fwo4_churn}, and~\ref{subfig: fwow4_churn}, we report the convergence trend of the Bayesian Optimization procedure for feature selection ordering, feature weighting, and feature weighting ordering weighting over the Churn dataset. In Figure~\ref{subfig: fso4_virtual},~\ref{subfig: fwo4_virtual}, and~\ref{subfig: fwow4_virtual}, we report the same study for the Virtual Screening dataset and, in Figures~\ref{subfig: fso4_german},~\ref{subfig: fwo4_german}, and~\ref{subfig: fwo4_german}, we present the same analysis but for the German Numeric dataset. Lastly, in Figures~\ref{subfig: fso4_plastic},~\ref{subfig: fwo4_plastic}, and~\ref{subfig: fwow4_plastic}, we report the BO convergence regarding Plastic Astronomy dataset.}
    \label{fig: BO convergence HH}
\end{figure}

In Figure~\ref{fig: BO convergence SE}, we visualize the convergence trend of Bayesian Optimization procedure of the QML model which leverages the 15-qubits Separate Entangled feature map, while, in Figure~\ref{fig: BO convergence HH}, we report the same visualization but for the Heisenberg Hamiltonian feature map. Since we train and test QML models on 10 different batches for each dataset, the $y$-axis shows the average training AUC score computed across these 10 batches at each iteration of Bayesian Optimization (reported on the $x-$axis). We report this type of study for the most complex encoding optimizations, that are feature selection ordering (FSO), feature weighting ordering (FWO), and feature weighting ordering weighting (FWOW), for all three datasets. 
We can notice that, already with 100 iterations of Bayesian Optimization procedure, almost all the optimization procedures show a trend of convergence on average. The only two cases where we do not observe a clear convergence trend are the FWO and FWOW optimizations on the German Numeric dataset for the Separate Entangled feature map. In these scenarios, it seems that the algorithm tries to do more exploration than exploitation. The same optimizations performed employing Heisenberg Hamiltonian feature map shows a more evident convergence pattern demonstrating the effectiveness of this type of feature map and the impact of the feature map choice in the overall analysis of the QML model performance.
This is an interesting analysis since it demonstrates that, in the training phase, feature encoding optimizations are working almost always correctly by making the QML model learning the optimal weights which is then reflected in the optimal encoding of a given set of features. Moreover, the average training AUC obtained by the QML model which uses the Heisenberg Hamiltonian feature map is lower with respect to the Separate Entangled case (except for the German Numeric dataset where the scores are almost comparable). This is probably due to the higher complexity of this type of feature map which would require more iterations to achieve higher scores even if the reported trend is still one of convergence. 

\clearpage
\section{Hyperparameter configurations}
\label{app: hyperparams}
\subsection{Feature maps configurations}
\label{sub: ansatz_hyperparams}
Depending on the feature map, we define different sets of hyperparameter values. For what concern the Separate Entangled feature map, we outline three different configurations that we report in Frame~\ref{lst: SE_config}. The \textit{density} and the number of \textit{blocks} define the number of rotations per qubit and the number of repeated elements of the circuit respectively. We can also define the type of entanglement and the \textit{alpha} value that is the Pauli rotation factor, multiplicative to the Pauli rotation values. The Pauli rotations are defined by the \textit{paulis} hyperparameter, which also define the order of execution for the specific rotations.
\begin{lstlisting}[language=json,firstnumber=1, captionpos=b, caption={Hyperparameter configurations of Separate Entangle feature map}, label={lst: SE_config}]
0: {
    "num_qubits"   : 9-15,
    "blocks"       : 9,
    "density"      : [3, 1],
    "entanglement" : "full",
    "alpha"        : 0.1,
    "paulis"       : ["Y", "X", "Z"],
},
1: {
    "num_qubits"   : 9-15,
    "blocks"       : 9,
    "density"      : [3, 1],
    "entanglement" : "pairwise",
    "alpha"        : 0.1,
    "paulis"       : ["Y", "X", "Z"],
},
2: {
    "num_qubits"   : 9-15,
    "blocks"       : 9,
    "density"      : [2, 1],
    "entanglement" : "pairwise",
    "alpha"        : 0.3,
    "paulis"       : ["Y", "X", "Z"],
}
}
\end{lstlisting}

As already described in Section~\ref{sec:experiments}, we keep almost all the hyperparameter values fixed while varying some values between the three different configurations. On each configuration, we only vary the number of qubits and the value of the density depending on the dataset. In case of the Churn dataset, we set the density to 3 for the first two configurations and to 2 for the last one. In the case of Virtual Screening, we set it to 1 for the first two configurations and to 2 for the last one. As for the German Numeric and Plastic Astronomy datasets, we set the density to 1 for each configuration. The motivation is to have entanglement operations even in the case where we have to encode a low number of features given the number of qubits and blocks. We also choose to test both full and pairwise entanglement with the latter used in 2 configurations out of 3 since this is the most prominent type of entanglement to be implemented and executed on IBM quantum hardware.
In Frames~\ref{lst: HH_config} and~\ref{lst: RepeatedPauli_config}, we report the configuration of the hyperparameter values for the Heisenberg Hamiltonian and Repeated Pauli feature maps respectively.
\begin{lstlisting}[language=json,firstnumber=1, captionpos=b, caption={Hyperparameter configurations of Heisemberg Hamiltonian feature map}, label={lst: HH_config}]
0: {
    "num_qubits"   : 9-15,
    "blocks"       : 1,
    "alpha"        : 0.1
},
1: {
    "num_qubits"   : 9-15,
    "blocks"       : 1,
    "alpha"        : 0.3
}
\end{lstlisting}
Since Heisenberg Hamiltonian feature map is more complex to simulate than the Separate Entangle one, we define and test just 2 configurations instead of 3 as in the Separate Entangled case. The meaning of \textit{blocks} and \textit{alpha} hyperparameters is the same of the Separate Entangled case.
The hyperparameters for the Repeated Pauli feature map are the same of the base Pauli Feature Map ~\cite{qiskit_paulifeaturemap}, where the Pauli operators define the rotation gates that need to be applied before the encoding of the features with phase gates.
\begin{lstlisting}[language=json,firstnumber=1, captionpos=b, caption={Hyperparameter configurations of Repeated Pauli feature map}, label={lst: RepeatedPauli_config}]
0: {
    "num_qubits"   : 9-15,
    "entanglement" : "pairwise",
    "alpha"        : 0.1,
    "paulis"       : ["Y", "XZ"]
}
\end{lstlisting}

In Figure~\ref{fig: entanglement_examples}, we exemplify some of the most common entanglement patterns that may be exploited in feature maps for input data encoding. In our experiments, we apply pairwise and full entanglement techniques but other patterns can be experimented as can be seen from this example. 

\begin{figure}[ht]
\centering
\begin{minipage}{0.45\textwidth}
\centering
\begin{quantikz}[column sep=2mm, row sep=0.3cm]
\lstick{$q_0$}&  & \qw &  \ctrl{1} & \qw      & \qw      & \qw &  \qw\\
\lstick{$q_1$}&     & \qw & \targ{}  & \ctrl{1} & \qw      & \qw & \qw\\
\lstick{$q_2$} &     & \qw & \qw      & \targ{}  & \ctrl{1} & \qw & \qw\\
\lstick{$q_3$} &    & \qw & \qw      & \qw      & \targ{}  & \qw & \qw
\end{quantikz}
\caption*{Linear Entanglement} 
\end{minipage}
\hfill
\begin{minipage}{0.45\textwidth}
\centering
\begin{quantikz}[column sep=2mm, row sep=0.3cm]
\lstick{$q_0$}& & \qw & \ctrl{1} & \qw      & \qw      & \qw & \qw\\
\lstick{$q_1$}&  \qw       & \qw & \targ{}  & \ctrl{1} & \qw      & \qw & \qw\\
\lstick{$q_2$}&  \qw        & \qw & \ctrl{1}      & \targ{}  & \qw & \qw & \qw\\
\lstick{$q_3$}&  \qw       & \qw & \targ{}      & \qw      & \qw & \qw & \qw
\end{quantikz}
\caption*{Pairwise Entanglement}
\end{minipage}

\vspace{0.2cm}

\begin{minipage}{0.45\textwidth}
\centering
\begin{quantikz}[column sep=2mm, row sep=0.3cm]
\lstick{$q_0$}& & \targ{} & \ctrl{1} & \qw      & \qw &  \qw \\
\lstick{$q_1$}& \qw & \qw & \targ{}  & \ctrl{1} & \qw  & \qw\\
\lstick{$q_2$}& \qw & \qw & \qw      & \targ{}  & \ctrl{1}  &\qw\\
\lstick{$q_3$}& \qw & \ctrl{-3} & \qw      & \qw  & \targ{} &\qw
\end{quantikz}
\caption*{Circular Entanglement}
\end{minipage}
\hfill
\begin{minipage}{0.45\textwidth}
\centering
\begin{quantikz}[column sep=2mm, row sep=0.3cm]
\lstick{$q_0$}&  & \qw & \ctrl{1} & \ctrl{2}   & \ctrl{3}     & \qw & \qw & \qw & \qw \\
\lstick{$q_1$}&  \qw       & \qw & \targ{}  & \qw & \qw      & \ctrl{1} & \ctrl{2} & \qw & \qw  \\
\lstick{$q_2$}&  \qw        & \qw & \qw      & \targ{}  & \qw & \targ{} & \qw & \ctrl{1} & \qw  \\
\lstick{$q_3$}&  \qw       & \qw & \qw      & \qw      & \targ{}  & \qw & \targ{} & \targ{} & \qw 
\end{quantikz}
\caption*{Full Entanglement}
\end{minipage}

\caption{Illustrative examples of controlled-$\mathrm{X}$ ($\mathrm{CNOT}$) entanglement patterns on a generic 4-qubits circuit.}
\label{fig: entanglement_examples}
\end{figure}

\subsection{Classifier configurations}
In our experiments, we employ XGBoost and Support Vector Classifier (SVC) as classifier to measure the final performance in terms of AUC. In Frame~\ref{lst: hyperparams}, we report the set of hyperparameter values that we define both for XGBoost and for SVC. For each hyperparameter, we establish a list of possible values to explore optimal model configurations using Grid Search Cross-Validation. In this way, as reported in Section~\ref{subsec: framework implementation}, the execution time of the optimization procedure increases: indeed, using the reported hyperparameter values, we have to test 144 and 195 different XGBoost and SVC models respectively fitted with different hyperparamenter values.
\label{sub: classifier_hyperparams}
\begin{lstlisting}[language=json,firstnumber=1, captionpos=b, caption={Hyperparameter configurations for XGBoost and SVC classifiers.}, label={lst: hyperparams}]
"XGBoost": {
    "max_depth": [2,3,5],
    "n_estimators": [100, 200],
    "learning_rate": [0.1, 0.01, 0.05],
    "subsample": [0.8, 1],
    "colsample_bytree": [0.8, 1],
    "gamma": [0, 0.1]
},
"SVC": {
    "gamma": [0.00390625, 0.0078125, 0.015625, 0.03125, 0.0625, 0.125, 0.25, 0.5, 1.0, 2.0, 4.0, 8.0, 16.0],
    "C": [0.0078125, 0.015625, 0.03125, 0.0625, 0.125, 0.25, 0.5, 1.0, 2.0, 4.0, 8.0, 16.0, 32.0, 64.0,128.0]
}
\end{lstlisting}

\clearpage
\section{Complete set of XGBoost noiseless experiments}
\label{app: XGB experiments appendix}
\subsection{Main numerical results}
\label{sub: main_numerical_results}

In Tables~\ref{tab: XGB_SE0},~\ref{tab: XGB_SE1}, and~\ref{tab: XGB_SE2}, we outline the complete set of results obtained with all the three different configurations of Separate Entangled feature map. In Tables~\ref{tab: XGB_HH0} and~\ref{tab: XGB_HH1}, we report the performance of the QML model with both the configurations defined for the Heisenberg Hamiltonian feature map. In Table \ref{tab: RepeatedPauli_results}, we present the results obtained with the Repeated Pauli feature map. 
These results are useful to explore in a more granular way the behavior of the different optimizations on different feature maps and datasets when the number of qubits varies. Furthermore, the tables reported in Section ~\ref{sec:results} are directly derived from these by calculating the percentage difference with respect to NFO and aggregating the results with the average over all the numbers of qubits (e.g., Table~\ref{tab: pct_diff_results_SE1} is an aggregation of Table~\ref{tab: XGB_SE1}).
The analysis of these results follows the one reported in Section~\ref{subsub: numerical_analyses}. Indeed, we appreciate a substantial performance improvement by using optimization procedures with respect to the NFO baseline, especially for the Churn, Virtual Screening, and Plastic Astronomy datasets. In the case of Virtual Screening encoded with configuration 0 of Separate Entangled (SE$_0$) feature map with 14-qubits, we observe slightly better performance for the NFO baseline. In the case of FS, we also observe this behavior for the 13-qubits case. These particular cases could be points of further investigation. However, these are sporadic cases, where the NFO is only slightly better, almost comparable to the optimization performance cases unlike the scenarios where we get an advantage in using optimizations where the benefit is substantial. Focusing on Virtual Screening results on SE$_1$ feature map, it is interesting to note that we always observe a performance improvement. This highlights once again how not only different feature map, but also differently configured versions of feature maps from the same family, can have distinct impact on the data.

The trend of the results regarding the German Numeric dataset is similar to the one reported in Section~\ref{sec:results}, with feature selection not appearing to be as effective as for the other datasets on the Separate Entangled feature map, showing in this sense the importance of the other types of optimizations. We notice that, for German Numeric dataset, feature selection is more impactful when employing Heisenberg Hamiltonian feature map to encode data. If we look at the overall results, we can better recognize the benefits of our optimizations. Indeed, we always observe a meaningful advantage in optimizing feature encoding with respect to NFO. An interesting result is the FWOW overall score in the SE$_0$ case which appears to be the highest one, further emphasizing the impact of this optimization. We must highlight how the overall FWOW score is similar to the FWO one, underlining how we can achieve comparable results using only one set of weights instead of two for the optimization.

\begin{table*}[h]
\centering
\begin{tabular}{|c|c|c|c|c|c|c|c|}
\hline
\multirow{2}{*}{Feature Map} & \multirow{2}{*}{Dataset} & \multirow{2}{*}{Qubits} & \multicolumn{5}{c|}{Scoring Method} \\ \cline{4-8} 
                      &          &            & NFO        & FS  & FSO & FWO & FWOW   \\ \hline
\multirow{28}{*}{SE$_0$}&\multirow{7}{*}{Churn}
&9&$0.673 \pm 0.007$&$0.712 \pm 0.009$&$0.714 \pm 0.007$&$0.711 \pm 0.009$&$0.715 \pm 0.011$\\ \cline{3-8}
&&10&$0.674 \pm 0.009$&$0.713 \pm 0.009$&$0.710 \pm 0.011$&$0.704 \pm 0.011$&$0.712 \pm 0.011$\\ \cline{3-8}
&&11&$0.653 \pm 0.009$&$0.715 \pm 0.009$&$0.714 \pm 0.011$&$0.714 \pm 0.008$&$0.717 \pm 0.009$\\ \cline{3-8}
&&12&$0.662 \pm 0.009$&$0.713 \pm 0.012$&$0.711 \pm 0.012$&$0.716 \pm 0.011$&$0.718 \pm 0.010$\\ \cline{3-8}
&&13&$0.681 \pm 0.007$&$0.715 \pm 0.010$&$0.710 \pm 0.012$&$0.719 \pm 0.011$&$0.718 \pm 0.011$\\ \cline{3-8}
&&14&$0.674 \pm 0.009$&$0.721 \pm 0.010$&$0.722 \pm 0.008$&$0.716 \pm 0.013$&$0.718 \pm 0.012$\\
\end{tabular}
\caption{Mean ± Average Std Dev AUC Test Scores for QML Method with Different Scoring Methods and Qubits obtained with configuration 0 of the Separate Entangled feature map, reported in Frame~\ref{lst: SE_config} in Appendix~\ref{sub: ansatz_hyperparams}, employing XGBoost as classifier.}
\label{tab: XGB_SE0}
\end{table*}

\begin{table*}[h]
\centering
\begin{tabular}{|c|c|c|c|c|}
\hline
\multirow{2}{*}{Dataset} & \multicolumn{4}{c|}{Scoring Method} \\ \cline{2-5} 
                                           & FS      & FSO & FWO & FWOW  \\ \Xhline{3\arrayrulewidth}
\multirow{1}{*}{Churn} 
&$+6.28\%$ \scriptsize$(1.97\%)$&$+6.15\%$ \scriptsize$(1.97\%)$&$+5.92\%$ \scriptsize$(2.36\%)$& {\boldmath$+6.34\%$} \scriptsize$(2.39\%)$\\ \cline{2-5}
\hline
\multirow{1}{*}{Virtual Screening}
&$+1.77\%$ \scriptsize$(2.05\%)$&$+2.08\%$ \scriptsize$(1.69\%)$&$+2.30\%$ \scriptsize$(1.45\%)$& {\boldmath$+2.43\%$} \scriptsize$(1.55\%)$\\ \cline{2-5}
\hline
\multirow{1}{*}{German Numeric}
&$-1.87\%$ \scriptsize$(2.03\%)$&$ +0.41\%$ \scriptsize$(1.86\%)$&$+1.21\%$ \scriptsize$(2.44\%)$&{\boldmath$+1.75\%$} \scriptsize$(2.53\%)$\\ \cline{2-4}
\hline
\multirow{1}{*}{Plastic Astronomy}
&$+5.69\%$ \scriptsize$(2.57\%)$&$+7.15\%$ \scriptsize$(1.89\%)$&$+6.64\%$ \scriptsize$(1.38\%)$&{\boldmath$+6.66\%$} \scriptsize$(1.29\%)$ \\ \cline{2-4}
\Xhline{3\arrayrulewidth}
\multicolumn{1}{|c|}{\textbf{Overall Average}}  
& $+2.97\%$ \scriptsize$(3.80\%)$& $+3.95\%$ \scriptsize$(3.22\%)$& $+4.02\%$ \scriptsize$(2.67\%)$& {\boldmath$+4.30\%$} \scriptsize$(2.56\%)$ \\ \hline
\end{tabular}
\caption{Percentage difference AUC Test Scores with respect to NFO regarding results reported in Table~\ref{tab: XGB_SE0}.}
\label{tab: pct_diff_results_SE0}
\end{table*}


\begin{table*}[h]
\centering
\begin{tabular}{|c|c|c|c|c|c|c|c|}
\hline
\multirow{2}{*}{Feature Map} & \multirow{2}{*}{Dataset} & \multirow{2}{*}{Qubits} & \multicolumn{5}{c|}{Scoring Method} \\ \cline{4-8} 
                       &  &                         & NFO  & FS      & FSO  & FWO&FWOW      \\ \hline
\multirow{28}{*}{SE$_1$}&\multirow{7}{*}{Churn} 
&9&$0.675 \pm 0.009$&$0.709 \pm 0.009$&$0.710 \pm 0.009$&$0.708 \pm 0.009$&$0.713 \pm 0.009$\\ \cline{3-8}
&&10&$0.676 \pm 0.010$&$0.714 \pm 0.013$&$0.717 \pm 0.011$&$0.707 \pm 0.008$&$0.714 \pm 0.008$\\ \cline{3-8}
&&11&$0.654 \pm 0.007$&$0.716 \pm 0.011$&$0.715 \pm 0.009$&$0.712 \pm 0.009$&$0.715 \pm 0.009$\\ \cline{3-8}
&&12&$0.663 \pm 0.010$&$0.713 \pm 0.012$&$0.708 \pm 0.008$&$0.712 \pm 0.006$&$0.716 \pm 0.010$\\ \cline{3-8}
&&13&$0.684 \pm 0.008$&$0.712 \pm 0.011$&$0.711 \pm 0.011$&$0.713 \pm 0.009$&$0.717 \pm 0.008$\\ \cline{3-8}
&&14&$0.683 \pm 0.008$&$0.721 \pm 0.010$&$0.722 \pm 0.008$&$0.712 \pm 0.011$&$0.716 \pm 0.010$\\
\hline
\end{tabular}
\caption{Mean ± Average Std Dev AUC Test Scores for QML Method with Different Scoring Methods and Qubits obtained with configuration 1 of the Separate Entangled feature map, reported in Frame~\ref{lst: SE_config} in Appendix~\ref{sub: ansatz_hyperparams}, employing XGBoost as classifier.}
\label{tab: XGB_SE1}
\end{table*}

\comment{
\begin{table*}[h]
\centering
\begin{tabular}{|c|c|c|c|c|}
\hline
\multirow{2}{*}{Dataset} & \multicolumn{4}{c|}{Scoring Method} \\ \cline{2-5} 
                                           & FS      & FSO & FWO & FWOW  \\ \Xhline{3\arrayrulewidth}
\multirow{1}{*}{Churn} 
&$+5.73\%$&$+5.70\%$&$+5.03\%$ & $+5.77\%$\\ \cline{2-5}
\hline
\multirow{1}{*}{Virtual Screening}
&$+2.15\%$&$+2.29\%$&$+2.47\%$ & $+2.26\%$ \\ \cline{2-5}
\hline
\multirow{1}{*}{German Numeric} &$-1.43\%$&$ -0.19\%$&$+0.10\%$ &$+0.33\%$ \\ \cline{2-4}
\Xhline{3\arrayrulewidth}
\multicolumn{1}{|c|}{\textbf{Overall Average}}  & $+2.15\%$ & $+2.60\%$ & $+2.53\%$ & $+2.79\%$  \\ \hline
\end{tabular}
\caption{Percentage difference AUC Test Scores with respect to NFO regarding results reported in Table~\ref{tab: XGB_SE1}.}
\label{tab: pct_diff_results_SE1}
\end{table*}
}

\begin{table*}[h]
\centering
\begin{tabular}{|c|c|c|c|c|c|c|c|}
\hline
\multirow{2}{*}{Feature Map} & \multirow{2}{*}{Dataset} & \multirow{2}{*}{Qubits} & \multicolumn{5}{c|}{Scoring Method} \\ \cline{4-8} 
                      &          &            & NFO        & FS  & FSO & FWO & FWOW    \\ \hline
\multirow{28}{*}{SE$_2$}&\multirow{7}{*}{Churn}
&9&$0.646 \pm 0.009$&$0.705 \pm 0.012$&$0.702 \pm 0.009$&$0.693 \pm 0.010$&$0.704 \pm 0.010$\\ \cline{3-8}
&&10&$0.676 \pm 0.009$&$0.712 \pm 0.012$&$0.709 \pm 0.012$&$0.701 \pm 0.013$&$0.705 \pm 0.011$\\ \cline{3-8}
&&11&$0.659 \pm 0.011$&$0.715 \pm 0.010$&$0.709 \pm 0.011$&$0.703 \pm 0.010$&$0.705 \pm 0.007$\\ \cline{3-8}
&&12&$0.659 \pm 0.011$&$0.712 \pm 0.013$&$0.715 \pm 0.010$&$0.706 \pm 0.008$&$0.700 \pm 0.012$\\ \cline{3-8}
&&13&$0.672 \pm 0.007$&$0.713 \pm 0.012$&$0.712 \pm 0.008$&$0.705 \pm 0.009$&$0.711 \pm 0.008$\\ \cline{3-8}
&&14&$0.676 \pm 0.010$&$0.714 \pm 0.011$&$0.712 \pm 0.009$&$0.706 \pm 0.011$&$0.713 \pm 0.011$\\
\end{tabular}
\caption{Mean ± Average Std Dev AUC Test Scores for QML Method with Different Scoring Methods and Qubits obtained with configuration 2 of the Separate Entangled feature map, reported in Frame~\ref{lst: SE_config} in Appendix~\ref{sub: ansatz_hyperparams}, employing XGBoost as classifier.}
\label{tab: XGB_SE2}
\end{table*}

\begin{table*}[h]
\centering
\begin{tabular}{|c|c|c|c|c|}
\hline
\multirow{2}{*}{Dataset} & \multicolumn{4}{c|}{Scoring Method} \\ \cline{2-5} 
                                           & FS      & FSO & FWO & FWOW  \\ \Xhline{3\arrayrulewidth}
\multirow{1}{*}{Churn} 
&{\boldmath$+6.76\%$} \scriptsize$(1.70\%)$&$+6.46\%$ \scriptsize$(1.77\%)$&$+5.30\%$ \scriptsize$(1.64\%)$& $+5.89\%$ \scriptsize$(1.76\%)$\\ \cline{2-5}
\hline
\multirow{1}{*}{Virtual Screening}
&$+1.57\%$ \scriptsize$(1.41\%)$&{\boldmath$+1.75\%$} \scriptsize$(1.07\%)$&$+1.21\%$ \scriptsize$(1.18\%)$& $+1.44\%$ \scriptsize$(1.49\%)$\\ \cline{2-5}
\hline
\multirow{1}{*}{German Numeric}
&$-1.47\%$ \scriptsize$(1.31\%)$&$ -1.10\%$ \scriptsize$(1.03\%)$&$-0.21\%$ \scriptsize$(0.81\%)$ &{\boldmath$+0.24\%$} \scriptsize$(1.15\%)$\\ \cline{2-4}
\hline
\multirow{1}{*}{Plastic Astronomy}
&$+5.68\%$ \scriptsize$(2.16\%)$&{\boldmath$+7.45\%$} \scriptsize$(2.27\%)$&$+5.59\%$ \scriptsize$(1.25\%)$ &$+5.92\%$ \scriptsize$(1.49\%)$\\ \cline{2-4}
\Xhline{3\arrayrulewidth}
\multicolumn{1}{|c|}{\textbf{Overall Average}}  
& $+3.13\%$ \scriptsize$(3.8\%)$& {\boldmath$+3.64\%$} \scriptsize$(4.02\%)$& $+2.97\%$ \scriptsize$(2.91\%)$& $+3.37\%$ \scriptsize$(2.96\%)$ \\ \hline
\end{tabular}
\caption{Percentage difference AUC Test Scores with respect to NFO regarding results reported in Table~\ref{tab: XGB_SE2}.}
\label{tab: pct_diff_results_SE2}
\end{table*}

\begin{table*}[h]
\centering
\begin{tabular}{|c|c|c|c|c|c|c|c|}
\hline
\multirow{2}{*}{Feature Map} & \multirow{2}{*}{Dataset} & \multirow{2}{*}{Qubits} & \multicolumn{5}{c|}{Scoring Method} \\ \cline{4-8} 
                         &  &                       & NFO     & FS   & FSO  & FWO  &FWOW    \\ \hline
\multirow{28}{*}{HH$_0$}&\multirow{7}{*}{Churn} 
&9&$0.654 \pm 0.013$&$0.689 \pm 0.010$&$0.688 \pm 0.012$&$0.682 \pm 0.013$&$0.679 \pm 0.004$\\ \cline{3-8}
&&10&$0.642 \pm 0.009$&$0.698 \pm 0.013$&$0.697 \pm 0.007$&$0.688 \pm 0.011$&$0.679 \pm 0.012$\\ \cline{3-8}
&&11&$0.649 \pm 0.005$&$0.699 \pm 0.007$&$0.699 \pm 0.015$&$0.685 \pm 0.013$&$0.693 \pm 0.010$\\ \cline{3-8}
&&12&$0.657 \pm 0.009$&$0.702 \pm 0.010$&$0.696 \pm 0.016$&$0.692 \pm 0.011$&$0.692 \pm 0.017$\\ \cline{3-8}
&&13&$0.655 \pm 0.008$&$0.701 \pm 0.013$&$0.703 \pm 0.012$&$0.696 \pm 0.013$&$0.688 \pm 0.010$\\ \cline{3-8}
&&14&$0.655 \pm 0.011$&$0.700 \pm 0.011$&$0.704 \pm 0.009$&$0.698 \pm 0.011$&$0.692 \pm 0.009$\\
\end{tabular}
\caption{Mean ± Average Std Dev AUC Test Scores for QML Method with Different Scoring Methods and Qubits obtained with configuration 0 of Heisenberg Hamiltonian feature map, defined in Frame~\ref{lst: HH_config}, employing XGBoost as classifier.}
\label{tab: XGB_HH0}
\end{table*}
\begin{table*}[h]
\centering
\begin{tabular}{|c|c|c|c|c|}
\hline
\multirow{2}{*}{Dataset} & \multicolumn{4}{c|}{Scoring Method} \\ \cline{2-5} 
                                           & FS      & FSO & FWO & FWOW  \\ 
                                           \Xhline{3\arrayrulewidth}
\multirow{1}{*}{Churn} 
&{\boldmath$+6.86\%$} \scriptsize$(1.16\%)$&$+6.79\%$ \scriptsize$(1.27\%)$&$+5.59\%$ \scriptsize$(1.09\%)$& $+5.25\%$ \scriptsize$(0.92\%)$\\ \cline{2-5}
\hline
\multirow{1}{*}{Virtual Screening}
&{\boldmath$+1.95\%$} \scriptsize$(1.21\%)$&$+1.92\%$ \scriptsize$(1.07\%)$&$+1.70\%$ \scriptsize$(0.94\%)$& $+1.53\%$ \scriptsize$(1.01\%)$\\ \cline{2-5}
\hline
\multirow{1}{*}{German Numeric} &$+1.22\%$ \scriptsize$(2.44\%)$&$+1.61\%$ \scriptsize$(2.00\%)$&{\boldmath$+2.23\%$} \scriptsize$(1.94\%)$& $+1.70\%$ \scriptsize$(1.99\%)$\\ \cline{2-5}
\hline
\multirow{1}{*}{Plastic Astronomy} &$+11.08\%$ \scriptsize$(3.96\%)$&{\boldmath$+11.97\%$} \scriptsize$(4.06\%)$&$+10.46\%$ \scriptsize$(3.97\%)$& $+9.62\%$ \scriptsize$(3.57\%)$\\ \cline{2-5}
\Xhline{3\arrayrulewidth}
\multicolumn{1}{|c|}{\textbf{Overall Average}}  &$+5.27\%$ \scriptsize$(4.61\%)$& {\boldmath$+5.57\%$} \scriptsize$(4.88\%)$& $+4.99\%$ \scriptsize$(4.03\%)$&  $+4.52\%$ \scriptsize$(3.80\%)$\\ \hline
\end{tabular}
\caption{AUC Percentage Change Test Scores with respect to NFO for QML model using configuration 0 of Heisenberg Hamilton feature map reported in Frame~\ref{lst: HH_config} in Appendix~\ref{sub: ansatz_hyperparams}. Each score is computed as the average percentage variation of the specific optimization exploiting Heisenberg Hamiltonian feature map, with qubits ranging from 9 to 15, relative to the baseline.}
\label{tab: pct_diff_results_HH0}
\end{table*}
\comment{
\begin{table*}[h]
\centering
\begin{tabular}{|c|c|c|c|c|}
\hline
\multirow{2}{*}{Dataset} & \multicolumn{4}{c|}{Scoring Method} \\ \cline{2-5} 
                                           & FS      & FSO & FWO & FWOW  \\ 
                                           \Xhline{3\arrayrulewidth}
\multirow{1}{*}{Churn} 
&$+6.86\%$&$+6.79\%$&$+5.59\%$ & $+5.25\%$ \\ \cline{2-5}
\hline
\multirow{1}{*}{Virtual Screening}
&$+1.95\%$&$+1.92\%$&$+1.70\%$ & $+1.53\%$\\ \cline{2-5}
\hline
\multirow{1}{*}{German Numeric} &$+1.22\%$&$+1.61\%$&$+2.23\%$ & $+1.70\%$\\ \cline{2-5}
\Xhline{3\arrayrulewidth}
\multicolumn{1}{|c|}{\textbf{Overall Average}}  &$+3.34\%$ & $+3.44\%$ & $+3.18\%$ &  $+2.83\%$\\ \hline
\end{tabular}
\caption{Percentage difference AUC Test Scores with respect to NFO regarding results reported in Table~\ref{tab: XGB_HH0}.}
\label{tab: pct_diff_results_HH0}
\end{table*}
}

\begin{table*}[h]
\centering
\begin{tabular}{|c|c|c|c|c|c|c|c|}
\hline
\multirow{2}{*}{Feature Map} & \multirow{2}{*}{Dataset} & \multirow{2}{*}{Qubits} & \multicolumn{5}{c|}{Scoring Method} \\ \cline{4-8} 
                      &          &            & NFO        & FS  & FSO & FWO  &FWOW   \\ \hline
\multirow{28}{*}{HH$_1$}&\multirow{7}{*}{Churn}
&9&$0.613 \pm 0.013$&$0.687 \pm 0.012$&$0.685 \pm 0.009$&$0.681 \pm 0.009$&$0.680 \pm 0.012$\\ \cline{3-8}
&&10&$0.619 \pm 0.009$&$0.690 \pm 0.011$&$0.692 \pm 0.009$&$0.678 \pm 0.011$&$0.684 \pm 0.011$\\ \cline{3-8}
&&11&$0.645 \pm 0.010$&$0.694 \pm 0.008$&$0.696 \pm 0.013$&$0.688 \pm 0.008$&$0.684 \pm 0.012$\\ \cline{3-8}
&&12&$0.661 \pm 0.007$&$0.693 \pm 0.012$&$0.696 \pm 0.014$&$0.690 \pm 0.007$&$0.689 \pm 0.008$\\ \cline{3-8}
&&13&$0.637 \pm 0.011$&$0.702 \pm 0.011$&$0.697 \pm 0.009$&$0.691 \pm 0.009$&$0.691 \pm 0.009$\\ \cline{3-8}
&&14&$0.652 \pm 0.012$&$0.698 \pm 0.010$&$0.701 \pm 0.014$&$0.694 \pm 0.010$&$0.692 \pm 0.010$\\
\end{tabular}
\caption{Mean ± Average Std Dev AUC Test Scores for QML Method with Different Scoring Methods and Qubits obtained with configuration 1 of the Heisenberg Hamiltonian feature map, reported in Frame~\ref{lst: HH_config} in Appendix~\ref{sub: ansatz_hyperparams}, employing XGBoost as classifier.}
\label{tab: XGB_HH1}
\end{table*}

\comment{
\begin{table*}[h]
\centering
\begin{tabular}{|c|c|c|c|c|}
\hline
\multirow{2}{*}{Dataset} & \multicolumn{4}{c|}{Scoring Method} \\ \cline{2-5} 
                                           & FS      & FSO & FWO & FWOW  \\ 
                                           \Xhline{3\arrayrulewidth}
\multirow{1}{*}{Churn} 
&$+8.68\%$&$+8.71\%$&$+7.69\%$ & $+7.53\%$ \\ \cline{2-5}
\hline
\multirow{1}{*}{Virtual Screening}
&$+1.92\%$&$+1.91\%$&$+1.81\%$ & $+1.54\%$\\ \cline{2-5}
\hline
\multirow{1}{*}{German Numeric} 
&$+1.68\%$&$+1.97\%$&$+1.99\%$ & $+1.65\%$\\ \cline{2-5}
\Xhline{3\arrayrulewidth}
\multicolumn{1}{|c|}{\textbf{Overall Average}}
&$+4.09\%$ & $+4.20\%$ & $+3.82\%$ &  $+3.58\%$\\ \hline
\end{tabular}
\caption{Percentage difference AUC Test Scores with respect to NFO regarding results reported in Table~\ref{tab: XGB_HH1}.}
\label{tab: pct_diff_results_HH1}
\end{table*}
}

\begin{table*}[h]
\centering
\begin{tabular}{|c|c|c|c|c|c|c|c|}
\hline
\multirow{2}{*}{Feature Map} & \multirow{2}{*}{Dataset} & \multirow{2}{*}{Qubits} & \multicolumn{5}{c|}{Scoring Method} \\ \cline{4-8} 
                      &          &            & NFO        & FS  & FSO & FWO  &FWOW   \\ \hline
\multirow{28}{*}{RP$_0$}&\multirow{7}{*}{Churn} 
&9&$0.661 \pm 0.007$&$0.702 \pm 0.010$&$0.702 \pm 0.012$&$0.700 \pm 0.009$&$0.698 \pm 0.012$\\ \cline{3-8}
&&10&$0.667 \pm 0.008$&$0.704 \pm 0.009$&$0.705 \pm 0.009$&$0.700 \pm 0.014$&$0.702 \pm 0.012$\\ \cline{3-8}
&&11&$0.672 \pm 0.010$&$0.709 \pm 0.010$&$0.707 \pm 0.012$&$0.704 \pm 0.012$&$0.703 \pm 0.011$\\ \cline{3-8}
&&12&$0.684 \pm 0.011$&$0.709 \pm 0.010$&$0.713 \pm 0.009$&$0.703 \pm 0.013$&$0.705 \pm 0.012$\\ \cline{3-8}
&&13&$0.670 \pm 0.008$&$0.709 \pm 0.009$&$0.710 \pm 0.007$&$0.704 \pm 0.013$&$0.702 \pm 0.012$\\ \cline{3-8}
&&14&$0.692 \pm 0.008$&$0.717 \pm 0.013$&$0.711 \pm 0.008$&$0.707 \pm 0.009$&$0.709 \pm 0.010$\\
\hline
\end{tabular}
\caption{Mean ± Average Std Dev AUC Test Scores for QML Method with Different Scoring Methods and Qubits obtained with configuration 0 of Repeated Pauli feature map, defined in Frame~\ref{lst: RepeatedPauli_config}, employing XGBoost as classifier.}
\label{tab: RepeatedPauli_results}
\end{table*}

\clearpage
\subsection{Ablation study numerical results}
\label{sub: ablation_numerical_results}

\begin{table*}[h]
\centering
\begin{tabular}{|c|c|c|c|c|c|c|}
\hline
\multirow{2}{*}{Feature Map} & \multirow{2}{*}{Dataset} & \multirow{2}{*}{Qubits} & \multicolumn{4}{c|}{Scoring Method} \\ \cline{4-7} 
                         &  &                       & NFO        & FO  & FW & FWOW     \\ \hline
\multirow{14}{*}{SE$_0$}&\multirow{7}{*}{Churn}
&9&$0.673 \pm 0.007$&$0.711 \pm 0.011$&$0.705 \pm 0.010$&$0.715 \pm 0.011$\\ \cline{3-7}
&&10&$0.674 \pm 0.009$&$0.709 \pm 0.010$&$0.683 \pm 0.010$&$0.712 \pm 0.011$\\ \cline{3-7}
&&11&$0.653 \pm 0.009$&$0.713 \pm 0.009$&$0.673 \pm 0.010$&$0.717 \pm 0.009$\\ \cline{3-7}
&&12&$0.662 \pm 0.009$&$0.718 \pm 0.007$&$0.667 \pm 0.011$&$0.718 \pm 0.010$\\ \cline{3-7}
&&13&$0.681 \pm 0.007$&$0.720 \pm 0.011$&$0.695 \pm 0.012$&$0.718 \pm 0.011$\\ \cline{3-7}
&&14&$0.674 \pm 0.009$&$0.717 \pm 0.009$&$0.688 \pm 0.011$&$0.718 \pm 0.012$\\ \cline{3-7}
&&15&$0.698 \pm 0.010$&$0.713 \pm 0.009$&$0.709 \pm 0.010$&$0.714 \pm 0.012$\\ \cline{3-7}
\Xcline{2-7}{1pt}
&\multirow{7}{*}{German Numeric}
&9&$0.772 \pm 0.032$&$0.759 \pm 0.024$&$0.789 \pm 0.032$&$0.766 \pm 0.028$\\ \cline{3-7}
&&10&$0.779 \pm 0.023$&$0.768 \pm 0.024$&$0.775 \pm 0.018$&$0.781 \pm 0.027$\\ \cline{3-7}
&&11&$0.770 \pm 0.023$&$0.761 \pm 0.033$&$0.772 \pm 0.031$&$0.765 \pm 0.026$\\ \cline{3-7}
&&12&$0.758 \pm 0.021$&$0.777 \pm 0.028$&$0.759 \pm 0.032$&$0.778 \pm 0.024$\\ \cline{3-7}
&&13&$0.738 \pm 0.024$&$0.773 \pm 0.024$&$0.759 \pm 0.022$&$0.784 \pm 0.013$\\ \cline{3-7}
&&14&$0.767 \pm 0.023$&$0.773 \pm 0.035$&$0.759 \pm 0.026$&$0.777 \pm 0.025$\\ \cline{3-7}
&&15&$0.746 \pm 0.019$&$0.766 \pm 0.027$&$0.763 \pm 0.024$&$0.771 \pm 0.022$\\ \cline{3-7}
\Xhline{3\arrayrulewidth}
\multicolumn{3}{|c|}{\textbf{Overall Average}} & $0.717 \pm 0.016$ &$0.741 \pm 0.018$  &$0.728 \pm 0.018$  & $0.745 \pm 0.017$ \\ \hline
\end{tabular}
\caption{Mean ± Average Std Dev AUC Test Scores for QML Method with Different Scoring Methods and Qubits obtained with configuration 0 of the Separate Entangled feature map, defined in Frame~\ref{lst: SE_config}, employing XGBoost as classifier.}
\label{tab: XGB_SE0_ablation_numerical}
\end{table*}

\begin{table*}[h]
\centering
\begin{tabular}{|c|c|c|c|}
\hline
\multirow{2}{*}{Dataset} & \multicolumn{3}{c|}{Scoring Method} \\ \cline{2-4} 
                                           & FO      & FW & FWOW  \\ 
                                           \Xhline{3\arrayrulewidth}
\multirow{1}{*}{Churn} 
&$+6.07\%$ \scriptsize$(2.29\%)$&$+2.21\%$ \scriptsize$(1.36\%)$&{\boldmath$+6.34\%$} \scriptsize$(2.39\%)$\\ \cline{2-4}
\hline
\multirow{1}{*}{German Numeric} &$+0.94\%$ \scriptsize$(2.50\%)$&$+0.90\%$ \scriptsize$(1.54\%)$&{\boldmath$+1.75\%$} \scriptsize$(2.53\%)$\\ \cline{2-4}
\Xhline{3\arrayrulewidth}
\multicolumn{1}{|c|}{\textbf{Overall Average}}  &$+3.51\%$ \scriptsize$(3.62\%)$& $+1.55\%$ \scriptsize$(0.93\%)$& {\boldmath$+4.05\%$} \scriptsize$(3.25\%)$\\ \hline
\end{tabular}
\caption{AUC Percentage Change Test Scores with respect to NFO for QML model using configuration 0 of Separate Entangled feature map reported in Frame~\ref{lst: SE_config} in Appendix~\ref{sub: ansatz_hyperparams}. Each score is computed as the average percentage variation of the specific optimization exploiting Separate Entangled feature map, with qubits ranging from 9 to 15, relative to the baseline.}
\label{tab: pct_diff_results_SE0_ablation}
\end{table*}


\begin{table*}[h]
\centering
\begin{tabular}{|c|c|c|c|c|c|c|}
\hline
\multirow{2}{*}{Feature Map} & \multirow{2}{*}{Dataset} & \multirow{2}{*}{Qubits} & \multicolumn{4}{c|}{Scoring Method} \\ \cline{4-7} 
                         &  &                       & NFO        & FO  & FW & FWOW     \\ \hline
\multirow{14}{*}{SE$_1$}&\multirow{7}{*}{Churn}
&9&$0.675 \pm 0.009$&$0.712 \pm 0.009$&$0.707 \pm 0.009$&$0.713 \pm 0.009$\\ \cline{3-7}
&&10&$0.676 \pm 0.010$&$0.711 \pm 0.007$&$0.685 \pm 0.009$&$0.714 \pm 0.008$\\ \cline{3-7}
&&11&$0.654 \pm 0.007$&$0.713 \pm 0.009$&$0.687 \pm 0.014$&$0.715 \pm 0.009$\\ \cline{3-7}
&&12&$0.663 \pm 0.010$&$0.714 \pm 0.009$&$0.680 \pm 0.010$&$0.716 \pm 0.010$\\ \cline{3-7}
&&13&$0.684 \pm 0.008$&$0.715 \pm 0.007$&$0.695 \pm 0.010$&$0.717 \pm 0.008$\\ \cline{3-7}
&&14&$0.683 \pm 0.008$&$0.714 \pm 0.011$&$0.685 \pm 0.010$&$0.716 \pm 0.010$\\ \cline{3-7}
&&15&$0.703 \pm 0.010$&$0.716 \pm 0.010$&$0.711 \pm 0.010$&$0.717 \pm 0.010$\\ \cline{3-7}
\Xcline{2-7}{1pt}
&\multirow{7}{*}{German Numeric}
&9&$0.770 \pm 0.018$&$0.758 \pm 0.031$&$0.775 \pm 0.022$&$0.772 \pm 0.022$\\ \cline{3-7}
&&10&$0.767 \pm 0.020$&$0.754 \pm 0.031$&$0.760 \pm 0.029$&$0.765 \pm 0.022$\\ \cline{3-7}
&&11&$0.776 \pm 0.024$&$0.764 \pm 0.021$&$0.769 \pm 0.021$&$0.769 \pm 0.024$\\ \cline{3-7}
&&12&$0.773 \pm 0.023$&$0.769 \pm 0.019$&$0.763 \pm 0.024$&$0.781 \pm 0.024$\\ \cline{3-7}
&&13&$0.765 \pm 0.028$&$0.774 \pm 0.019$&$0.761 \pm 0.020$&$0.775 \pm 0.028$\\ \cline{3-7}
&&14&$0.770 \pm 0.030$&$0.764 \pm 0.023$&$0.764 \pm 0.025$&$0.777 \pm 0.025$\\ \cline{3-7}
&&15&$0.764 \pm 0.031$&$0.765 \pm 0.028$&$0.773 \pm 0.023$&$0.764 \pm 0.034$\\ \cline{3-7}
\Xhline{3\arrayrulewidth}
\multicolumn{3}{|c|}{\textbf{Overall Average}} & $0.723 \pm 0.017$ &$0.739 \pm 0.017$  &$0.729 \pm 0.017$  & $0.744 \pm 0.017$ \\ \hline
\end{tabular}
\caption{Mean ± Average Std Dev AUC Test Scores for QML Method with Different Scoring Methods and Qubits obtained with configuration 1 of the Separate Entangled feature map, defined in Frame~\ref{lst: SE_config}, employing XGBoost as classifier.}
\label{tab: XGB_SE1_ablation_numerical}
\end{table*}

\comment{
\begin{table*}[h]
\centering
\begin{tabular}{|c|c|c|c|}
\hline
\multirow{2}{*}{Dataset} & \multicolumn{3}{c|}{Scoring Method} \\ \cline{2-4} 
                                           & FO      & FW & FWOW  \\ 
                                           \Xhline{3\arrayrulewidth}
\multirow{1}{*}{Churn} 
&$+5.47\%$&$+2.39\%$&\boldmath$+5.77\%$ \\ \cline{2-4}
\hline
\multirow{1}{*}{German Numeric} &$-0.69\%$&$-0.37\%$&\boldmath$+0.33\%$\\ \cline{2-4}
\Xhline{3\arrayrulewidth}
\multicolumn{1}{|c|}{\textbf{Overall Average}}  &$+2.39\%$ & $+1.01\%$ & \boldmath$+3.05\%$\\ \hline
\end{tabular}
\caption{AUC Percentage Change Test Scores with respect to NFO for QML model using configuration 1 of Separate Entangled feature map reported in Frame~\ref{lst: SE_config} in Appendix~\ref{sub: ansatz_hyperparams}. Each score is computed as the average percentage variation of the specific optimization exploiting Separate Entangled feature map, with qubits ranging from 9 to 15, relative to the baseline.}
\label{tab: pct_diff_results_SE1_ablation}
\end{table*}}


\begin{table*}[h]
\centering
\begin{tabular}{|c|c|c|c|c|c|c|}
\hline
\multirow{2}{*}{Feature Map} & \multirow{2}{*}{Dataset} & \multirow{2}{*}{Qubits} & \multicolumn{4}{c|}{Scoring Method} \\ \cline{4-7} 
                         &  &                       & NFO        & FO  & FW & FWOW     \\ \hline
\multirow{14}{*}{SE$_2$}&\multirow{7}{*}{Churn}
&9&$0.646 \pm 0.009$&$0.690 \pm 0.011$&$0.694 \pm 0.009$&$0.704 \pm 0.010$\\ \cline{3-7}
&&10&$0.676 \pm 0.009$&$0.692 \pm 0.010$&$0.693 \pm 0.011$&$0.705 \pm 0.011$\\ \cline{3-7}
&&11&$0.659 \pm 0.011$&$0.689 \pm 0.008$&$0.686 \pm 0.006$&$0.705 \pm 0.007$\\ \cline{3-7}
&&12&$0.659 \pm 0.011$&$0.688 \pm 0.009$&$0.687 \pm 0.010$&$0.700 \pm 0.012$\\ \cline{3-7}
&&13&$0.672 \pm 0.007$&$0.705 \pm 0.011$&$0.690 \pm 0.009$&$0.711 \pm 0.008$\\ \cline{3-7}
&&14&$0.676 \pm 0.010$&$0.697 \pm 0.008$&$0.684 \pm 0.008$&$0.713 \pm 0.011$\\ \cline{3-7}
&&15&$0.683 \pm 0.007$&$0.698 \pm 0.012$&$0.703 \pm 0.013$&$0.708 \pm 0.010$\\ \cline{3-7}
\Xcline{2-7}{1pt}
&\multirow{7}{*}{German Numeric}
&9&$0.771 \pm 0.018$&$0.764 \pm 0.020$&$0.776 \pm 0.021$&$0.767 \pm 0.023$\\ \cline{3-7}
&&10&$0.766 \pm 0.018$&$0.763 \pm 0.031$&$0.750 \pm 0.016$&$0.775 \pm 0.025$\\ \cline{3-7}
&&11&$0.774 \pm 0.020$&$0.766 \pm 0.024$&$0.767 \pm 0.024$&$0.765 \pm 0.021$\\ \cline{3-7}
&&12&$0.769 \pm 0.025$&$0.773 \pm 0.020$&$0.765 \pm 0.028$&$0.782 \pm 0.028$\\ \cline{3-7}
&&13&$0.762 \pm 0.031$&$0.772 \pm 0.025$&$0.768 \pm 0.030$&$0.768 \pm 0.024$\\ \cline{3-7}
&&14&$0.772 \pm 0.024$&$0.769 \pm 0.018$&$0.763 \pm 0.023$&$0.764 \pm 0.024$\\ \cline{3-7}
&&15&$0.771 \pm 0.026$&$0.767 \pm 0.031$&$0.761 \pm 0.023$&$0.777 \pm 0.020$\\ \cline{3-7}
\Xhline{3\arrayrulewidth}
\multicolumn{3}{|c|}{\textbf{Overall Average}} & $0.718 \pm 0.016$ &$0.731 \pm 0.017$  &$0.728 \pm 0.016$  & $0.739 \pm 0.017$  \\ \hline
\end{tabular}
\caption{Mean ± Average Std Dev AUC Test Scores for QML Method with Different Scoring Methods and Qubits obtained with configuration 2 of the Separate Entangled feature map, defined in Frame~\ref{lst: SE_config}, employing XGBoost as classifier.}
\label{tab: XGB_SE2_ablation_numerical}
\end{table*}

\begin{table*}[h]
\centering
\begin{tabular}{|c|c|c|c|}
\hline
\multirow{2}{*}{Dataset} & \multicolumn{3}{c|}{Scoring Method} \\ \cline{2-4} 
                                           & FO      & FW & FWOW  \\ 
                                           \Xhline{3\arrayrulewidth}
\multirow{1}{*}{Churn} 
&$+4.00\%$ \scriptsize$(1.59\%)$&$+3.55\%$ \scriptsize$(1.95\%)$&{\boldmath$+5.88\%$} \scriptsize$(1.76\%)$\\ \cline{2-4}
\hline
\multirow{1}{*}{German Numeric} &$-0.21\%$ \scriptsize$(0.87\%)$&$-0.64\%$ \scriptsize$(1.02\%)$&{\boldmath$+0.23\%$} \scriptsize$(1.15\%)$\\ \cline{2-4}
\Xhline{3\arrayrulewidth}
\multicolumn{1}{|c|}{\textbf{Overall Average}}  &$+1.90\%$ \scriptsize$(2.98\%)$& $+1.46\%$\scriptsize$(2.96\%)$& {\boldmath$+3.06\%$} \scriptsize$(3.99\%)$\\ \hline
\end{tabular}
\caption{AUC Percentage Change Test Scores with respect to NFO for QML model using configuration 2 of Separate Entangled feature map reported in Frame~\ref{lst: SE_config} in Appendix~\ref{sub: ansatz_hyperparams}. Each score is computed as the average percentage variation of the specific optimization exploiting Separate Entangled feature map, with qubits ranging from 9 to 15, relative to the baseline.}
\label{tab: pct_diff_results_SE2_ablation}
\end{table*}


\begin{table*}[h]
\centering
\begin{tabular}{|c|c|c|c|c|c|c|}
\hline
\multirow{2}{*}{Feature Map} & \multirow{2}{*}{Dataset} & \multirow{2}{*}{Qubits} & \multicolumn{4}{c|}{Scoring Method} \\ \cline{4-7} 
                         &     &                    & NFO        & FO  & FW &FWOW     \\ \hline
\multirow{14}{*}{HH$_0$}&\multirow{7}{*}{Churn} &9&$0.654 \pm 0.013$&$0.664 \pm 0.011$&$0.685 \pm 0.011$&$0.679 \pm 0.004$\\ \cline{3-7}
&&10&$0.642 \pm 0.009$&$0.667 \pm 0.009$&$0.691 \pm 0.010$&$0.679 \pm 0.012$\\ \cline{3-7}
&&11&$0.649 \pm 0.005$&$0.672 \pm 0.012$&$0.692 \pm 0.010$&$0.693 \pm 0.010$\\ \cline{3-7}
&&12&$0.657 \pm 0.009$&$0.680 \pm 0.010$&$0.692 \pm 0.009$&$0.692 \pm 0.017$\\ \cline{3-7}
&&13&$0.655 \pm 0.008$&$0.683 \pm 0.009$&$0.698 \pm 0.011$&$0.688 \pm 0.010$\\ \cline{3-7}
&&14&$0.655 \pm 0.011$&$0.682 \pm 0.007$&$0.696 \pm 0.011$&$0.692 \pm 0.009$\\ \cline{3-7}
&&15&$0.669 \pm 0.010$&$0.686 \pm 0.009$&$0.704 \pm 0.010$&$0.699 \pm 0.013$\\ \cline{3-7}
\Xcline{2-7}{1pt}
&\multirow{7}{*}{German Numeric} &9&$0.730 \pm 0.022$&$0.746 \pm 0.029$&$0.761 \pm 0.023$&$0.762 \pm 0.024$\\ \cline{3-7}
&&10&$0.742 \pm 0.036$&$0.758 \pm 0.022$&$0.759 \pm 0.021$&$0.760 \pm 0.027$\\ \cline{3-7}
&&11&$0.747 \pm 0.028$&$0.750 \pm 0.026$&$0.761 \pm 0.023$&$0.757 \pm 0.032$\\ \cline{3-7}
&&12&$0.747 \pm 0.018$&$0.757 \pm 0.028$&$0.757 \pm 0.029$&$0.757 \pm 0.022$\\ \cline{3-7}
&&13&$0.750 \pm 0.020$&$0.759 \pm 0.025$&$0.753 \pm 0.024$&$0.763 \pm 0.023$\\ \cline{3-7}
&&14&$0.742 \pm 0.031$&$0.757 \pm 0.028$&$0.747 \pm 0.027$&$0.762 \pm 0.029$\\ \cline{3-7}
&&15&$0.784 \pm 0.025$&$0.766 \pm 0.024$&$0.766 \pm 0.025$&$0.767 \pm 0.024$\\ \cline{3-7}
\Xhline{3\arrayrulewidth}
\multicolumn{3}{|c|}{\textbf{Overall Average}} & $0.701 \pm 0.017$ &$0.716 \pm 0.018$  &$0.726 \pm 0.017$  & $0.725 \pm 0.018$ \\ \hline
\end{tabular}
\caption{Mean ± Average Std Dev AUC Test Scores for QML Method with Different Scoring Methods and Qubits obtained with configuration 0 of Heisemberg Hamiltonian feature map, defined in Frame~\ref{lst: HH_config}, employing XGBoost as classifier.}
\label{tab: XGB_HH0_numerical_ablation}
\end{table*}

\begin{table*}[h]
\centering
\begin{tabular}{|c|c|c|c|}
\hline
\multirow{2}{*}{Dataset} & \multicolumn{3}{c|}{Scoring Method} \\ \cline{2-4} 
                                           & FO      & FW & FWOW  \\ 
                                           \Xhline{3\arrayrulewidth}
\multirow{1}{*}{Churn} 
&$+3.33\%$ \scriptsize$(0.94\%)$&{\boldmath$+6.07\%$} \scriptsize$(0.98\%)$&$+5.25\%$ \scriptsize$(0.92\%)$\\ \cline{2-4}
\hline
\multirow{1}{*}{German Numeric} &$+1.07\%$ \scriptsize$(1.57\%)$&$+1.26\%$ \scriptsize$(2.01\%)$&{\boldmath$+1.70\%$} \scriptsize$(1.99\%)$\\ \cline{2-4}
\Xhline{3\arrayrulewidth}
\multicolumn{1}{|c|}{\textbf{Overall Average}}  &$+2.20\%$ \scriptsize$(1.60\%)$& {\boldmath$+3.66\%$} \scriptsize$(3.40\%)$&$+3.47\%$ \scriptsize$(2.51\%)$\\ \hline
\end{tabular}
\caption{AUC Percentage Change Test Scores with respect to NFO for QML model using configuration 0 of Heisenberg Hamiltonian feature map reported in Frame~\ref{lst: HH_config} in Appendix~\ref{sub: ansatz_hyperparams}. Each score is computed as the average percentage variation of the specific optimization exploiting Heisenberg Hamiltonian feature map, with qubits ranging from 9 to 15, relative to the baseline.}
\label{tab: pct_diff_results_HH0_ablation}
\end{table*}


\begin{table*}[h]
\centering
\begin{tabular}{|c|c|c|c|c|c|c|}
\hline
\multirow{2}{*}{Feature Map} & \multirow{2}{*}{Dataset} & \multirow{2}{*}{Qubits} & \multicolumn{4}{c|}{Scoring Method} \\ \cline{4-7} 
                         &     &                    & NFO        & FO  & FW &FWOW     \\ \hline
\multirow{14}{*}{HH$_1$}&\multirow{7}{*}{Churn} 
&9&$0.613 \pm 0.013$&$0.657 \pm 0.012$&$0.683 \pm 0.009$&$0.680 \pm 0.012$\\ \cline{3-7}
&&10&$0.619 \pm 0.009$&$0.661 \pm 0.012$&$0.693 \pm 0.009$&$0.684 \pm 0.011$\\ \cline{3-7}
&&11&$0.645 \pm 0.010$&$0.664 \pm 0.015$&$0.684 \pm 0.005$&$0.684 \pm 0.012$\\ \cline{3-7}
&&12&$0.661 \pm 0.007$&$0.666 \pm 0.011$&$0.689 \pm 0.014$&$0.689 \pm 0.008$\\ \cline{3-7}
&&13&$0.637 \pm 0.011$&$0.673 \pm 0.011$&$0.703 \pm 0.009$&$0.691 \pm 0.009$\\ \cline{3-7}
&&14&$0.652 \pm 0.012$&$0.673 \pm 0.010$&$0.692 \pm 0.010$&$0.692 \pm 0.010$\\ \cline{3-7}
&&15&$0.654 \pm 0.009$&$0.677 \pm 0.012$&$0.707 \pm 0.009$&$0.697 \pm 0.013$\\ \cline{3-7}
\Xcline{2-7}{1pt}
&\multirow{7}{*}{German Numeric} 
&9&$0.709 \pm 0.029$&$0.744 \pm 0.035$&$0.759 \pm 0.019$&$0.748 \pm 0.037$\\ \cline{3-7}
&&10&$0.730 \pm 0.036$&$0.743 \pm 0.029$&$0.758 \pm 0.031$&$0.758 \pm 0.021$\\ \cline{3-7}
&&11&$0.757 \pm 0.023$&$0.748 \pm 0.019$&$0.766 \pm 0.026$&$0.752 \pm 0.013$\\ \cline{3-7}
&&12&$0.765 \pm 0.026$&$0.754 \pm 0.030$&$0.761 \pm 0.030$&$0.764 \pm 0.033$\\ \cline{3-7}
&&13&$0.756 \pm 0.031$&$0.763 \pm 0.032$&$0.748 \pm 0.029$&$0.759 \pm 0.025$\\ \cline{3-7}
&&14&$0.741 \pm 0.030$&$0.763 \pm 0.021$&$0.755 \pm 0.022$&$0.768 \pm 0.020$\\ \cline{3-7}
&&15&$0.769 \pm 0.026$&$0.759 \pm 0.021$&$0.770 \pm 0.024$&$0.759 \pm 0.028$\\ \cline{3-7}
\Xhline{3\arrayrulewidth}
\multicolumn{3}{|c|}{\textbf{Overall Average}} & $0.693 \pm 0.019$ &$0.710 \pm 0.019$  &$0.726 \pm 0.018$  & $0.723 \pm 0.018$\\ \hline
\end{tabular}
\caption{Mean ± Average Std Dev AUC Test Scores for QML Method with Different Scoring Methods and Qubits obtained with configuration 1 of Heisemberg Hamiltonian feature map, defined in Frame~\ref{lst: HH_config}, employing XGBoost as classifier.}
\label{tab: XGB_HH1_numerical_ablation}
\end{table*}

\comment{
\begin{table*}[h]
\centering
\begin{tabular}{|c|c|c|c|}
\hline
\multirow{2}{*}{Dataset} & \multicolumn{3}{c|}{Scoring Method} \\ \cline{2-4} 
                                           & FO      & FW & FWOW  \\ 
                                           \Xhline{3\arrayrulewidth}
\multirow{1}{*}{Churn} 
&$+4.30\%$&\boldmath$+8.31\%$&$+7.53\%$ \\ \cline{2-4}
\hline
\multirow{1}{*}{German Numeric} &$+0.96\%$&\boldmath$+1.81\%$&$+1.65\%$\\ \cline{2-4}
\Xhline{3\arrayrulewidth}
\multicolumn{1}{|c|}{\textbf{Overall Average}}  &$+2.63\%$ & \boldmath$+5.06\%$ &$+4.59\%$\\ \hline
\end{tabular}
\caption{AUC Percentage Change Test Scores with respect to NFO for QML model using configuration 1 of Heisenberg Hamiltonian feature map reported in Frame~\ref{lst: HH_config} in Appendix~\ref{sub: ansatz_hyperparams}. Each score is computed as the average percentage variation of the specific optimization exploiting Heisenberg Hamiltonian feature map, with qubits ranging from 9 to 15, relative to the baseline.}
\label{tab: pct_diff_results_HH1_ablation}
\end{table*}}

\clearpage
\section{SVC experiments}
\label{app: svc}
As reported in Section~\ref{subsub:classifier}, we use mainly use XGBoost to measure the performance of our QML model but we also test the performance using SVC. In this section, in Table~\ref{tab: SE0_results_SVC}, we report all the results obtained with the SVC experiments. We perform a subset of experiments performed for XGBoost since the goal was not to compare SVC and XGboost but to verify that, even using different models, we obtained benefits in using quantum feature encoding optimization. We employ SVC just on the Churn dataset using configurations 0 and 1 of the Separate Entangled feature map reported in Frame~\ref{lst: SE_config} to encode input data and performing feature selection, feature ordering, and feature selection ordering as data manipulation techniques. In terms of comparison with the XGBoost results, we obtain better performance with XGBoost classifier which means that XGBoost is more suited to deal with this dataset considering also the set of hyperparameter values defined in Frame~\ref{lst: hyperparams} both for XGBoost and SVC. However, as for the XGBoost experiments, we always observe better AUC scores when employing QFEO framework with respect to NFO confirming the benefits of these approaches. 
\subsection{Churn Results}
\label{subsub: churn_result_SVC}
\begin{table*}[h]
\centering
\begin{tabular}{|c|c|c|c|c|c|c|}
\hline
\multirow{2}{*}{Feature Map} & \multirow{2}{*}{Dataset} & \multirow{2}{*}{Qubits} & \multicolumn{4}{c|}{Scoring Method} \\ \cline{4-7} 
                         &          &            & NFO        & FS  & FO & FSO     \\ \hline
\multirow{7}{*}{SE$_0$}&\multirow{7}{*}{Churn}
&9&$0.668 \pm 0.009$&$0.685 \pm 0.007$&$0.700 \pm 0.009$&$0.694 \pm 0.005$\\ \cline{3-7}
&&10&$0.672 \pm 0.007$&$0.683 \pm 0.013$&$0.698 \pm 0.008$&$0.690 \pm 0.013$\\ \cline{3-7}
&&11&$0.656 \pm 0.010$&$0.687 \pm 0.010$&$0.700 \pm 0.010$&$0.694 \pm 0.005$\\ \cline{3-7}
&&12&$0.654 \pm 0.012$&$0.688 \pm 0.006$&$0.704 \pm 0.006$&$0.688 \pm 0.011$\\ \cline{3-7}
&&13&$0.689 \pm 0.006$&$0.692 \pm 0.008$&$0.703 \pm 0.009$&$0.689 \pm 0.008$\\ \cline{3-7}
&&14&$0.672 \pm 0.010$&$0.703 \pm 0.008$&$0.704 \pm 0.008$&$0.709 \pm 0.008$\\ \cline{3-7}
&&15&$0.686 \pm 0.008$&$0.700 \pm 0.007$&$0.698 \pm 0.007$&$0.707 \pm 0.009$\\ \cline{3-7}
\Xhline{3\arrayrulewidth}
\multicolumn{3}{|c|}{\textbf{Overall Average}} & $0.671 \pm 0.009$  & $0.691 \pm 0.008$ & $0.701 \pm 0.008$&$0.696 \pm 0.008$  \\ \hline
\end{tabular}
\caption{Mean ± Average Std Dev AUC Test Scores for QML Method with Different Scoring Methods and Qubits obtained with configuration 0 of Separate Entangled feature map reported in Frame~\ref{lst: SE_config} and with SVC classifier.}
\label{tab: SE0_results_SVC}
\end{table*}

\begin{table*}[h]
\centering
\begin{tabular}{|c|c|c|c|c|c|c|}
\hline
\multirow{2}{*}{Feature Map} & \multirow{2}{*}{Dataset} & \multirow{2}{*}{Qubits} & \multicolumn{4}{c|}{Scoring Method} \\ \cline{4-7} 
                         &          &            & NFO        & FS  & FO & FSO     \\ \hline
\multirow{7}{*}{SE$_1$}&\multirow{7}{*}{Churn}
&9&$0.673 \pm 0.007$&$0.688 \pm 0.007$&$0.701 \pm 0.012$&$0.698 \pm 0.011$\\ \cline{3-7}
&&10&$0.676 \pm 0.009$&$0.690 \pm 0.007$&$0.694 \pm 0.007$&$0.697 \pm 0.008$\\ \cline{3-7}
&&11&$0.657 \pm 0.011$&$0.696 \pm 0.007$&$0.701 \pm 0.011$&$0.696 \pm 0.009$\\ \cline{3-7}
&&12&$0.656 \pm 0.011$&$0.697 \pm 0.009$&$0.705 \pm 0.011$&$0.699 \pm 0.008$\\ \cline{3-7}
&&13&$0.685 \pm 0.008$&$0.695 \pm 0.009$&$0.707 \pm 0.007$&$0.696 \pm 0.009$\\ \cline{3-7}
&&14&$0.670 \pm 0.009$&$0.703 \pm 0.008$&$0.702 \pm 0.009$&$0.709 \pm 0.008$\\ \cline{3-7}
&&15&$0.692 \pm 0.007$&$0.700 \pm 0.007$&$0.698 \pm 0.009$&$0.707 \pm 0.009$\\ \cline{3-7}
\Xhline{3\arrayrulewidth}
\multicolumn{3}{|c|}{\textbf{Overall Average}} & $0.673 \pm 0.009$ & $0.695 \pm 0.008$ & $0.701 \pm 0.009$ & $0.700 \pm 0.009$ \\\hline
\end{tabular}
\caption{Mean ± Average Std Dev AUC Test Scores for QML Method with Different Scoring Methods and Qubits obtained with configuration 1 of Separate Entangled feature map reported in Frame~\ref{lst: SE_config} and with SVC classifier.}
\label{tab: SE1_results_SVC}
\end{table*}

\begin{table*}[h]
\centering
\begin{tabular}{|c|c|c|c|c|}
\hline
\multirow{2}{*}{Dataset} &\multirow{2}{*}{Feature Map} & \multicolumn{3}{c|}{Scoring Method} \\ \cline{3-5} 
                                          & & FS      & FO & FSO  \\ 
                                           \Xhline{3\arrayrulewidth}
\multirow{2}{*}{Churn} & \multirow{1}{*}{SE$_0$} 
&$+3.06\%$ \scriptsize{($1.82\%$)}&{\boldmath$+4.49\%$} \scriptsize{($2.18\%$)}&$+3.72\%$ \scriptsize{($2.02\%$)}\\ \cline{2-5}
 & \multirow{1}{*}{SE$_1$} 
&$+3.42\%$ \scriptsize{($2.22\%$)}&{\boldmath$+4.25\%$} \scriptsize{($2.32\%$)}&$+4.09\%$ \scriptsize{($1.99\%$)}\\ \cline{2-5}\hline
\end{tabular}
\caption{AUC Percentage Change Test Scores with respect to NFO.}
\label{tab: pct_diff_results_churn_svc}
\end{table*}

\subsection{German Numeric Results}
\label{subsub: german_result_SVC}

\subsubsection{Main results}
\begin{table*}[h]
\centering
\begin{tabular}{|c|c|c|c|c|c|c|c|}
\hline
\multirow{2}{*}{Feature Map} & \multirow{2}{*}{Dataset} & \multirow{2}{*}{Qubits} & \multicolumn{5}{c|}{Scoring Method} \\ \cline{4-8} 
                         &          &            & NFO        & FS & FSO&FWO&FWOW     \\ \hline
\multirow{7}{*}{SE$_0$}&\multirow{7}{*}{German Numeric}
&9&$0.769 \pm 0.025$&$0.763 \pm 0.027$&$0.779 \pm 0.025$&$0.757 \pm 0.017$&$0.768 \pm 0.026$\\ \cline{3-8}
&&10&$0.761 \pm 0.016$&$0.758 \pm 0.031$&$0.767 \pm 0.029$&$0.764 \pm 0.020$&$0.764 \pm 0.031$\\ \cline{3-8}
&&11&$0.762 \pm 0.024$&$0.752 \pm 0.028$&$0.767 \pm 0.031$&$0.765 \pm 0.025$&$0.755 \pm 0.036$\\ \cline{3-8}
&&12&$0.750 \pm 0.022$&$0.739 \pm 0.027$&$0.768 \pm 0.022$&$0.779 \pm 0.022$&$0.767 \pm 0.029$\\ \cline{3-8}
&&13&$0.750 \pm 0.019$&$0.747 \pm 0.025$&$0.766 \pm 0.027$&$0.772 \pm 0.018$&$0.765 \pm 0.028$\\ \cline{3-8}
&&14&$0.755 \pm 0.022$&$0.741 \pm 0.028$&$0.752 \pm 0.031$&$0.769 \pm 0.022$&$0.778 \pm 0.029$\\ \cline{3-8}
&&15&$0.744 \pm 0.021$&$0.750 \pm 0.036$&$0.759 \pm 0.030$&$0.776 \pm 0.021$&$0.774 \pm 0.029$\\ \cline{3-8}
\Xhline{3\arrayrulewidth}
\multicolumn{3}{|c|}{\textbf{Overall Average}} & $0.756 \pm 0.021$ & $0.749 \pm 0.029$ & $0.765 \pm 0.028$ & $0.769 \pm 0.021$ & $0.767 \pm 0.029$ \\\hline
\end{tabular}
\caption{Mean ± Average Std Dev AUC Test Scores for QML Method with Different Scoring Methods and Qubits obtained with configuration 0 of Separate Entangled feature map reported in Frame~\ref{lst: SE_config} and with SVC classifier.}
\label{tab: SE0_results_SVC_german}
\end{table*}

\begin{table*}[h]
\centering
\begin{tabular}{|c|c|c|c|c|c|c|c|}
\hline
\multirow{2}{*}{Feature Map} & \multirow{2}{*}{Dataset} & \multirow{2}{*}{Qubits} & \multicolumn{5}{c|}{Scoring Method} \\ \cline{4-8} 
                         &          &            & NFO        & FS & FSO&FWO&FWOW     \\ \hline
\multirow{7}{*}{SE$_1$}&\multirow{7}{*}{German Numeric}
&9&$0.756 \pm 0.025$&$0.758 \pm 0.024$&$0.771 \pm 0.029$&$0.763 \pm 0.033$&$0.762 \pm 0.017$\\ \cline{3-8}
&&10&$0.757 \pm 0.018$&$0.751 \pm 0.024$&$0.767 \pm 0.025$&$0.766 \pm 0.026$&$0.772 \pm 0.027$\\ \cline{3-8}
&&11&$0.754 \pm 0.021$&$0.750 \pm 0.029$&$0.771 \pm 0.025$&$0.751 \pm 0.029$&$0.758 \pm 0.029$\\ \cline{3-8}
&&12&$0.762 \pm 0.021$&$0.758 \pm 0.026$&$0.761 \pm 0.027$&$0.780 \pm 0.028$&$0.776 \pm 0.030$\\ \cline{3-8}
&&13&$0.759 \pm 0.020$&$0.751 \pm 0.035$&$0.763 \pm 0.012$&$0.771 \pm 0.024$&$0.768 \pm 0.023$\\ \cline{3-8}
&&14&$0.766 \pm 0.029$&$0.743 \pm 0.027$&$0.753 \pm 0.020$&$0.772 \pm 0.032$&$0.762 \pm 0.029$\\ \cline{3-8}
&&15&$0.759 \pm 0.021$&$0.749 \pm 0.029$&$0.756 \pm 0.030$&$0.771 \pm 0.025$&$0.767 \pm 0.027$\\ \cline{3-8}
\Xhline{3\arrayrulewidth}
\multicolumn{3}{|c|}{\textbf{Overall Average}} & $0.759 \pm 0.022$ & $0.752 \pm 0.028$ & $0.763 \pm 0.024$ & $0.768 \pm 0.028$ & $0.766 \pm 0.026$ \\\hline
\end{tabular}
\caption{Mean ± Average Std Dev AUC Test Scores for QML Method with Different Scoring Methods and Qubits obtained with configuration 1 of Separate Entangled feature map reported in Frame~\ref{lst: SE_config} and with SVC classifier.}
\label{tab: SE1_results_SVC_german}
\end{table*}

\begin{table*}[h]
\centering
\begin{tabular}{|c|c|c|c|c|c|c|c|}
\hline
\multirow{2}{*}{Feature Map} & \multirow{2}{*}{Dataset} & \multirow{2}{*}{Qubits} & \multicolumn{5}{c|}{Scoring Method} \\ \cline{4-8} 
                         &          &            & NFO        & FS & FSO&FWO&FWOW     \\ \hline
\multirow{7}{*}{SE$_2$}&\multirow{7}{*}{German Numeric}
&9&$0.749 \pm 0.031$&$0.761 \pm 0.031$&$0.770 \pm 0.019$&$0.767 \pm 0.026$&$0.754 \pm 0.024$\\ \cline{3-8}
&&10&$0.757 \pm 0.023$&$0.758 \pm 0.032$&$0.766 \pm 0.018$&$0.765 \pm 0.026$&$0.750 \pm 0.031$\\ \cline{3-8}
&&11&$0.755 \pm 0.020$&$0.750 \pm 0.040$&$0.769 \pm 0.029$&$0.753 \pm 0.034$&$0.757 \pm 0.028$\\ \cline{3-8}
&&12&$0.766 \pm 0.022$&$0.752 \pm 0.022$&$0.762 \pm 0.036$&$0.769 \pm 0.021$&$0.774 \pm 0.024$\\ \cline{3-8}
&&13&$0.767 \pm 0.022$&$0.756 \pm 0.028$&$0.765 \pm 0.031$&$0.763 \pm 0.023$&$0.769 \pm 0.015$\\ \cline{3-8}
&&14&$0.767 \pm 0.025$&$0.757 \pm 0.026$&$0.755 \pm 0.029$&$0.768 \pm 0.030$&$0.767 \pm 0.026$\\ \cline{3-8}
&&15&$0.774 \pm 0.021$&$0.742 \pm 0.031$&$0.759 \pm 0.026$&$0.766 \pm 0.026$&$0.771 \pm 0.031$\\ \cline{3-8}
\Xhline{3\arrayrulewidth}
\multicolumn{3}{|c|}{\textbf{Overall Average}} & $0.762 \pm 0.023$ & $0.754 \pm 0.029$ & $0.764 \pm 0.027$ & $0.765 \pm 0.027$ & $0.763 \pm 0.026$ \\\hline
\end{tabular}
\caption{Mean ± Average Std Dev AUC Test Scores for QML Method with Different Scoring Methods and Qubits obtained with configuration 2 of Separate Entangled feature map reported in Frame~\ref{lst: SE_config} and with SVC classifier.}
\label{tab: SE2_results_SVC_german}
\end{table*}

\begin{table*}[h]
\centering
\begin{tabular}{|c|c|c|c|c|c|c|c|}
\hline
\multirow{2}{*}{Feature Map} & \multirow{2}{*}{Dataset} & \multirow{2}{*}{Qubits} & \multicolumn{5}{c|}{Scoring Method} \\ \cline{4-8} 
                         &          &            & NFO        & FS & FSO&FWO&FWOW     \\ \hline
\multirow{7}{*}{HH$_0$}&\multirow{7}{*}{German Numeric}
&9&$0.697 \pm 0.025$&$0.744 \pm 0.028$&$0.761 \pm 0.029$&$0.764 \pm 0.030$&$0.763 \pm 0.029$\\ \cline{3-8}
&&10&$0.726 \pm 0.032$&$0.758 \pm 0.029$&$0.758 \pm 0.036$&$0.755 \pm 0.033$&$0.762 \pm 0.025$\\ \cline{3-8}
&&11&$0.742 \pm 0.021$&$0.767 \pm 0.029$&$0.768 \pm 0.028$&$0.764 \pm 0.027$&$0.759 \pm 0.035$\\ \cline{3-8}
&&12&$0.737 \pm 0.017$&$0.772 \pm 0.032$&$0.761 \pm 0.034$&$0.759 \pm 0.041$&$0.772 \pm 0.022$\\ \cline{3-8}
&&13&$0.740 \pm 0.024$&$0.769 \pm 0.023$&$0.769 \pm 0.020$&$0.775 \pm 0.029$&$0.769 \pm 0.026$\\ \cline{3-8}
&&14&$0.715 \pm 0.038$&$0.752 \pm 0.028$&$0.768 \pm 0.038$&$0.767 \pm 0.034$&$0.772 \pm 0.029$\\ \cline{3-8}
&&15&$0.764 \pm 0.026$&$0.764 \pm 0.020$&$0.767 \pm 0.032$&$0.776 \pm 0.028$&$0.772 \pm 0.026$\\ \cline{3-8}
\Xhline{3\arrayrulewidth}
\multicolumn{3}{|c|}{\textbf{Overall Average}} & $0.732 \pm 0.026$ & $0.761 \pm 0.027$ & $0.764 \pm 0.031$ & $0.766 \pm 0.032$ & $0.767 \pm 0.027$ \\\hline
\end{tabular}
\caption{Mean ± Average Std Dev AUC Test Scores for QML Method with Different Scoring Methods and Qubits obtained with configuration 0 of Heisenberg Hamiltonian feature map reported in Frame~\ref{lst: HH_config} and with SVC classifier.}
\label{tab: HH0_results_SVC_german}
\end{table*}

\begin{table*}[h]
\centering
\begin{tabular}{|c|c|c|c|c|c|c|c|}
\hline
\multirow{2}{*}{Feature Map} & \multirow{2}{*}{Dataset} & \multirow{2}{*}{Qubits} & \multicolumn{5}{c|}{Scoring Method} \\ \cline{4-8} 
                         &          &            & NFO        & FS & FSO&FWO&FWOW     \\ \hline
\multirow{7}{*}{HH$_1$}&\multirow{7}{*}{German Numeric}
&9&$0.694 \pm 0.023$&$0.722 \pm 0.033$&$0.744 \pm 0.028$&$0.750 \pm 0.025$&$0.752 \pm 0.027$\\ \cline{3-8}
&&10&$0.722 \pm 0.044$&$0.747 \pm 0.027$&$0.758 \pm 0.033$&$0.758 \pm 0.023$&$0.747 \pm 0.039$\\ \cline{3-8}
&&11&$0.734 \pm 0.028$&$0.762 \pm 0.029$&$0.752 \pm 0.031$&$0.748 \pm 0.028$&$0.752 \pm 0.031$\\ \cline{3-8}
&&12&$0.755 \pm 0.036$&$0.754 \pm 0.034$&$0.752 \pm 0.031$&$0.763 \pm 0.022$&$0.766 \pm 0.027$\\ \cline{3-8}
&&13&$0.754 \pm 0.029$&$0.759 \pm 0.025$&$0.768 \pm 0.027$&$0.760 \pm 0.026$&$0.770 \pm 0.034$\\ \cline{3-8}
&&14&$0.729 \pm 0.039$&$0.750 \pm 0.024$&$0.765 \pm 0.033$&$0.762 \pm 0.030$&$0.773 \pm 0.023$\\ \cline{3-8}
&&15&$0.765 \pm 0.036$&$0.764 \pm 0.029$&$0.768 \pm 0.029$&$0.773 \pm 0.030$&$0.768 \pm 0.031$\\ \cline{3-8}
\Xhline{3\arrayrulewidth}
\multicolumn{3}{|c|}{\textbf{Overall Average}} & $0.736 \pm 0.034$ & $0.751 \pm 0.029$ & $0.758 \pm 0.030$ & $0.759 \pm 0.026$ & $0.761 \pm 0.030$ \\\hline
\end{tabular}
\caption{Mean ± Average Std Dev AUC Test Scores for QML Method with Different Scoring Methods and Qubits obtained with configuration 1 of Heisenberg Hamiltonian feature map reported in Frame~\ref{lst: HH_config} and with SVC classifier.}
\label{tab: HH1_results_SVC_german}
\end{table*}

\begin{table*}[h]
\centering
\begin{tabular}{|c|c|c|c|c|c|}
\hline
\multirow{2}{*}{Dataset} &\multirow{2}{*}{Feature Map} & \multicolumn{4}{c|}{Scoring Method} \\ \cline{3-6} 
                                          & & FS      & FSO & FWO&FWOW  \\ 
                                           \Xhline{3\arrayrulewidth}
\multirow{5}{*}{German Numeric} & \multirow{1}{*}{SE$_0$} 
&$-0.78\%$ \scriptsize{($0.90\%$)}&$+1.28\%$ \scriptsize{($1.01\%$)}&{\boldmath$+1.75\%$} \scriptsize{($2.11\%$)}& $+1.53\%$ \scriptsize{($1.78\%$)}\\ \cline{2-6}
 & \multirow{1}{*}{SE$_1$} 
&$-0.96\%$ \scriptsize{($0.99\%$)}&$+0.54\%$ \scriptsize{($1.39\%$)}&{\boldmath$+1.16\%$} \scriptsize{($0.90\%$)}& $+1.00\%$ \scriptsize{($0.83\%$)}\\ \cline{2-6}
 & \multirow{1}{*}{SE$_2$} 
&$-1.11\%$ \scriptsize{($1.81\%$)}&$+0.19\%$ \scriptsize{($1.80\%$)}&{\boldmath$+0.30\%$} \scriptsize{($1.15\%$)}& $+0.11\%$ \scriptsize{($0.67\%$)}\\ \cline{2-6}
 & \multirow{1}{*}{HH$_0$} 
&$+4.06\%$ \scriptsize{($2.07\%$)}&$+4.56\%$ \scriptsize{($2.89\%$)}&$+4.73\%$ \scriptsize{($2.79\%$)} &{\boldmath$+4.92\%$} \scriptsize{($2.98\%$)}\\ \cline{2-6}
 & \multirow{1}{*}{HH$_1$} 
&$+2.07\%$ \scriptsize{($1.90\%$)}&$+3.06\%$ \scriptsize{($2.77\%$)}&$+3.18\%$ \scriptsize{($2.76\%$)}&{\boldmath$+3.46\%$} \scriptsize{($2.81\%$)}\\ \cline{2-6} \hline
\end{tabular}
\caption{AUC Percentage Change Test Scores with respect to NFO.}
\label{tab: pct_diff_results_german_svc}
\end{table*}

\clearpage
\subsubsection{Ablation results}

\begin{table*}[h]
\centering
\begin{tabular}{|c|c|c|c|c|c|c|}
\hline
\multirow{2}{*}{Feature Map} & \multirow{2}{*}{Dataset} & \multirow{2}{*}{Qubits} & \multicolumn{4}{c|}{Scoring Method} \\ \cline{4-7} 
                         &  &                       & NFO        & FO  & FW & FWOW     \\ \hline
\multirow{7}{*}{SE$_0$}&\multirow{7}{*}{German Numeric}
&9&$0.769 \pm 0.025$&$0.765 \pm 0.030$&$0.769 \pm 0.028$&$0.768 \pm 0.026$\\ \cline{3-7}
&&10&$0.761 \pm 0.016$&$0.772 \pm 0.030$&$0.752 \pm 0.032$&$0.764 \pm 0.031$\\ \cline{3-7}
&&11&$0.762 \pm 0.024$&$0.772 \pm 0.036$&$0.753 \pm 0.027$&$0.755 \pm 0.036$\\ \cline{3-7}
&&12&$0.750 \pm 0.022$&$0.776 \pm 0.021$&$0.757 \pm 0.025$&$0.767 \pm 0.029$\\ \cline{3-7}
&&13&$0.750 \pm 0.019$&$0.776 \pm 0.025$&$0.757 \pm 0.027$&$0.765 \pm 0.028$\\ \cline{3-7}
&&14&$0.755 \pm 0.022$&$0.781 \pm 0.020$&$0.759 \pm 0.028$&$0.778 \pm 0.029$\\ \cline{3-7}
&&15&$0.744 \pm 0.021$&$0.771 \pm 0.021$&$0.762 \pm 0.019$&$0.774 \pm 0.029$\\ \cline{3-7}
\Xhline{3\arrayrulewidth}
\multicolumn{3}{|c|}{\textbf{Overall Average}} & $0.756 \pm 0.021$ &$0.773 \pm 0.026$  &$0.758 \pm 0.027$  & $0.767 \pm 0.029$ \\\hline
\end{tabular}
\caption{Mean ± Average Std Dev AUC Test Scores for QML Method with Different Scoring Methods and Qubits obtained with configuration 0 of Separate Entangled feature map reported in Frame~\ref{lst: SE_config} and with SVC classifier.}
\label{tab: SE0_results_SVC_german_ablation}
\end{table*}

\begin{table*}[h]
\centering
\begin{tabular}{|c|c|c|c|c|c|c|}
\hline
\multirow{2}{*}{Feature Map} & \multirow{2}{*}{Dataset} & \multirow{2}{*}{Qubits} & \multicolumn{4}{c|}{Scoring Method} \\ \cline{4-7} 
                         &  &                       & NFO        & FO  & FW & FWOW     \\ \hline
\multirow{7}{*}{SE$_1$}&\multirow{7}{*}{German Numeric}
&9&$0.756 \pm 0.025$&$0.747 \pm 0.036$&$0.757 \pm 0.031$&$0.762 \pm 0.017$\\ \cline{3-7}
&&10&$0.757 \pm 0.018$&$0.779 \pm 0.021$&$0.758 \pm 0.018$&$0.772 \pm 0.027$\\ \cline{3-7}
&&11&$0.754 \pm 0.021$&$0.768 \pm 0.023$&$0.757 \pm 0.026$&$0.758 \pm 0.029$\\ \cline{3-7}
&&12&$0.762 \pm 0.021$&$0.778 \pm 0.027$&$0.765 \pm 0.022$&$0.776 \pm 0.030$\\ \cline{3-7}
&&13&$0.759 \pm 0.020$&$0.773 \pm 0.019$&$0.763 \pm 0.025$&$0.768 \pm 0.023$\\ \cline{3-7}
&&14&$0.766 \pm 0.029$&$0.770 \pm 0.021$&$0.764 \pm 0.023$&$0.762 \pm 0.029$\\ \cline{3-7}
&&15&$0.759 \pm 0.021$&$0.770 \pm 0.023$&$0.764 \pm 0.025$&$0.767 \pm 0.027$\\ \cline{3-7}
\Xhline{3\arrayrulewidth}
\multicolumn{3}{|c|}{\textbf{Overall Average}} & $0.759 \pm 0.022$ &$0.769 \pm 0.024$  &$0.761 \pm 0.024$  & $0.766 \pm 0.026$ \\\hline
\end{tabular}
\caption{Mean ± Average Std Dev AUC Test Scores for QML Method with Different Scoring Methods and Qubits obtained with configuration 1 of Separate Entangled feature map reported in Frame~\ref{lst: SE_config} and with SVC classifier.}
\label{tab: SE1_results_SVC_german_ablation}
\end{table*}

\begin{table*}[h]
\centering
\begin{tabular}{|c|c|c|c|c|c|c|}
\hline
\multirow{2}{*}{Feature Map} & \multirow{2}{*}{Dataset} & \multirow{2}{*}{Qubits} & \multicolumn{4}{c|}{Scoring Method} \\ \cline{4-7} 
                         &  &                       & NFO        & FO  & FW & FWOW     \\ \hline
\multirow{7}{*}{SE$_2$}&\multirow{7}{*}{German Numeric}
&9&$0.749 \pm 0.031$&$0.768 \pm 0.026$&$0.756 \pm 0.027$&$0.754 \pm 0.024$\\ \cline{3-7}
&&10&$0.757 \pm 0.023$&$0.754 \pm 0.028$&$0.753 \pm 0.025$&$0.750 \pm 0.031$\\ \cline{3-7}
&&11&$0.755 \pm 0.020$&$0.760 \pm 0.028$&$0.763 \pm 0.024$&$0.757 \pm 0.028$\\ \cline{3-7}
&&12&$0.766 \pm 0.022$&$0.772 \pm 0.025$&$0.752 \pm 0.028$&$0.774 \pm 0.024$\\ \cline{3-7}
&&13&$0.767 \pm 0.022$&$0.766 \pm 0.024$&$0.760 \pm 0.024$&$0.769 \pm 0.015$\\ \cline{3-7}
&&14&$0.767 \pm 0.025$&$0.765 \pm 0.026$&$0.759 \pm 0.029$&$0.767 \pm 0.026$\\ \cline{3-7}
&&15&$0.774 \pm 0.021$&$0.766 \pm 0.027$&$0.768 \pm 0.024$&$0.771 \pm 0.031$\\ \cline{3-7}
\Xhline{3\arrayrulewidth}
\multicolumn{3}{|c|}{\textbf{Overall Average}} & $0.762 \pm 0.023$ &$0.764 \pm 0.026$  &$0.759 \pm 0.026$  & $0.763 \pm 0.026$ \\\hline
\end{tabular}
\caption{Mean ± Average Std Dev AUC Test Scores for QML Method with Different Scoring Methods and Qubits obtained with configuration 2 of Separate Entangled feature map reported in Frame~\ref{lst: SE_config} and with SVC classifier.}
\label{tab: SE2_results_SVC_german_ablation}
\end{table*}

\begin{table*}[h]
\centering
\begin{tabular}{|c|c|c|c|c|c|c|}
\hline
\multirow{2}{*}{Feature Map} & \multirow{2}{*}{Dataset} & \multirow{2}{*}{Qubits} & \multicolumn{4}{c|}{Scoring Method} \\ \cline{4-7} 
                         &  &                       & NFO        & FO  & FW & FWOW     \\ \hline
\multirow{7}{*}{HH$_0$}&\multirow{7}{*}{German Numeric}
&9&$0.697 \pm 0.025$&$0.754 \pm 0.025$&$0.764 \pm 0.022$&$0.763 \pm 0.029$\\ \cline{3-7}
&&10&$0.726 \pm 0.032$&$0.747 \pm 0.031$&$0.754 \pm 0.033$&$0.762 \pm 0.025$\\ \cline{3-7}
&&11&$0.742 \pm 0.021$&$0.759 \pm 0.024$&$0.763 \pm 0.022$&$0.759 \pm 0.035$\\ \cline{3-7}
&&12&$0.737 \pm 0.017$&$0.765 \pm 0.027$&$0.747 \pm 0.036$&$0.772 \pm 0.022$\\ \cline{3-7}
&&13&$0.740 \pm 0.024$&$0.763 \pm 0.030$&$0.785 \pm 0.023$&$0.769 \pm 0.026$\\ \cline{3-7}
&&14&$0.715 \pm 0.038$&$0.757 \pm 0.026$&$0.742 \pm 0.037$&$0.772 \pm 0.029$\\ \cline{3-7}
&&15&$0.764 \pm 0.026$&$0.768 \pm 0.030$&$0.773 \pm 0.030$&$0.772 \pm 0.026$\\ \cline{3-7}
\Xhline{3\arrayrulewidth}
\multicolumn{3}{|c|}{\textbf{Overall Average}} & $0.732 \pm 0.026$ &$0.759 \pm 0.028$  &$0.761 \pm 0.029$  & $0.767 \pm 0.027$ \\\hline
\end{tabular}
\caption{Mean ± Average Std Dev AUC Test Scores for QML Method with Different Scoring Methods and Qubits obtained with configuration 0 of Heisenberg Hamiltonian feature map reported in Frame~\ref{lst: SE_config} and with SVC classifier.}
\label{tab: HH0_results_SVC_german_ablation}
\end{table*}

\begin{table*}[h]
\centering
\begin{tabular}{|c|c|c|c|c|c|c|}
\hline
\multirow{2}{*}{Feature Map} & \multirow{2}{*}{Dataset} & \multirow{2}{*}{Qubits} & \multicolumn{4}{c|}{Scoring Method} \\ \cline{4-7} 
                         &  &                       & NFO        & FO  & FW & FWOW     \\ \hline
\multirow{7}{*}{HH$_1$}&\multirow{7}{*}{German Numeric}
&9&$0.694 \pm 0.023$&$0.742 \pm 0.027$&$0.761 \pm 0.026$&$0.752 \pm 0.027$\\ \cline{3-7}
&&10&$0.722 \pm 0.044$&$0.745 \pm 0.026$&$0.752 \pm 0.028$&$0.747 \pm 0.039$\\ \cline{3-7}
&&11&$0.734 \pm 0.028$&$0.752 \pm 0.026$&$0.753 \pm 0.031$&$0.752 \pm 0.031$\\ \cline{3-7}
&&12&$0.755 \pm 0.036$&$0.749 \pm 0.030$&$0.758 \pm 0.036$&$0.766 \pm 0.027$\\ \cline{3-7}
&&13&$0.754 \pm 0.029$&$0.765 \pm 0.027$&$0.776 \pm 0.027$&$0.770 \pm 0.034$\\ \cline{3-7}
&&14&$0.729 \pm 0.039$&$0.757 \pm 0.026$&$0.739 \pm 0.033$&$0.773 \pm 0.023$\\ \cline{3-7}
&&15&$0.765 \pm 0.036$&$0.754 \pm 0.027$&$0.761 \pm 0.035$&$0.768 \pm 0.031$\\ \cline{3-7}
\Xhline{3\arrayrulewidth}
\multicolumn{3}{|c|}{\textbf{Overall Average}} & $0.736 \pm 0.034$ &$0.752 \pm 0.027$  &$0.757 \pm 0.031$  & $0.761 \pm 0.030$ \\\hline
\end{tabular}
\caption{Mean ± Average Std Dev AUC Test Scores for QML Method with Different Scoring Methods and Qubits obtained with configuration 1 of Heisenberg Hamiltonian feature map reported in Frame~\ref{lst: SE_config} and with SVC classifier.}
\label{tab: HH1_results_SVC_german_ablation}
\end{table*}

\begin{table*}[h]
\centering
\begin{tabular}{|c|c|c|c|c|}
\hline
\multirow{2}{*}{Dataset} &\multirow{2}{*}{Feature Map} & \multicolumn{3}{c|}{Scoring Method} \\ \cline{3-5} 
                                          & & FO      & FW & FWOW  \\ 
                                           \Xhline{3\arrayrulewidth}
\multirow{5}{*}{German Numeric} & \multirow{1}{*}{SE$_0$} 
&{\boldmath$+2.32\%$} \scriptsize{($1.61\%$)}&$+0.35\%$ \scriptsize{($1.26\%$)}&$+1.53\%$ \scriptsize{($1.78\%$)}\\ \cline{2-5}
 & \multirow{1}{*}{SE$_1$} 
&{\boldmath$+1.35\%$} \scriptsize{($1.31\%$)}&$+0.31\%$ \scriptsize{($0.31\%$)}&$+1.00\%$ \scriptsize{($0.83\%$)}\\ \cline{2-5}
 & \multirow{1}{*}{SE$_2$} 
&{\boldmath$+0.29\%$} \scriptsize{($1.20\%$)}&$-0.46\%$ \scriptsize{($1.07\%$)}&$+0.11\%$ \scriptsize{($0.67\%$)}\\ \cline{2-5}
 & \multirow{1}{*}{HH$_0$} 
&$+3.84\%$ \scriptsize{($2.49\%$)}&$+4.10\%$ \scriptsize{($2.94\%$)}&{\boldmath$+4.92\%$} \scriptsize{($2.98\%$)}\\ \cline{2-5}
 & \multirow{1}{*}{HH$_1$} 
&$+2.25\%$ \scriptsize{($2.86\%$)}&$+2.94\%$ \scriptsize{($3.36\%$)}&{\boldmath$+3.46\%$} \scriptsize{($2.81\%$)}\\ \cline{2-5} \hline
\end{tabular}
\caption{AUC Percentage Change Test Scores with respect to NFO.}
\label{tab: pct_diff_results_german_svc_ablation}
\end{table*}

\clearpage
\section{Real Hardware experiments}
\label{app: real_hardware_experiments}
\subsection{Data reloading setup}
\label{app: data_reloading_setup}
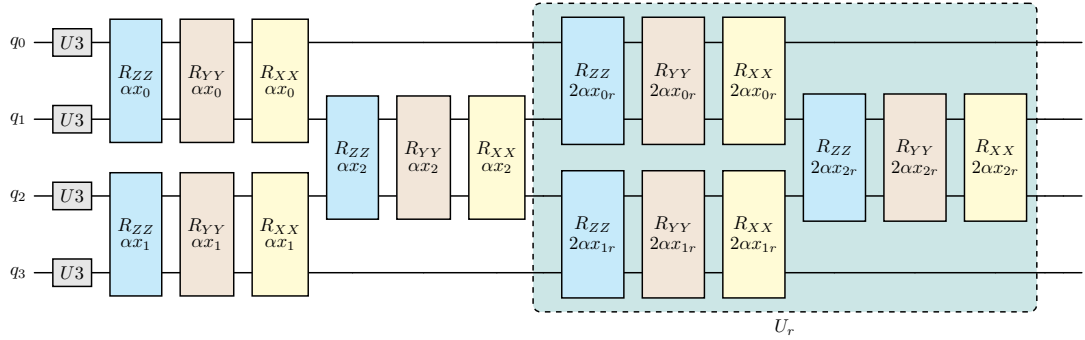
\begin{figure}[h]
    \centering
    \scalebox{0.7}{
    \begin{quantikz}[column sep=3.5mm]
            \lstick{$q_0$}& 
            \gate[style={fill=gray!20}]{U3} & 
            \gate[2, style={fill=cyan!20}]{\shortstack{$R_{ZZ}$ \\ $\alpha x_0$}} & 
            \gate[2, style={fill=brown!20}]{\shortstack{$R_{YY}$ \\ $ \alpha x_0$}} & 
            \gate[2, style={fill=yellow!20}]{\shortstack{$R_{XX}$ \\ $ \alpha x_0$}} & 
            {} & 
            {} & 
            {} & 
            \gategroup[4,steps=7,style={dashed,rounded
                        corners,fill=teal!20, inner
                        xsep=2pt},background,label style={label position=below,anchor=north,yshift=-0.2cm}]{{\sc
                        $U_{r}$}}
            &
            \gate[2, style={fill=cyan!20}]{\shortstack{$R_{ZZ}$ \\ $2\alpha x_{0r}$}} & 
            \gate[2, style={fill=brown!20}]{\shortstack{$R_{YY}$ \\ $ 2\alpha x_{0r}$}} & 
            \gate[2, style={fill=yellow!20}]{\shortstack{$R_{XX}$ \\ $ 2\alpha x_{0r}$}}
            & {} & 
            {} & 
            {} & 
            {} & 
            {} & \\
            \lstick{$q_1$} & 
            \gate[style={fill=gray!20}]{U3} &
            {} & 
            {}& 
            {} & 
            \gate[2, style={fill=cyan!20}]{\shortstack{$R_{ZZ}$ \\ $\alpha x_2$}} & 
            \gate[2, style={fill=brown!20}]{\shortstack{$R_{YY}$ \\ $ \alpha x_2$}} & 
            \gate[2, style={fill=yellow!20}]{\shortstack{$R_{XX}$ \\ $ \alpha x_2$}} &
            {} & 
            {}& 
            {} &
            {}
            &
            \gate[2, style={fill=cyan!20}]{\shortstack{$R_{ZZ}$ \\ $2\alpha x_{2r}$}} & 
            \gate[2, style={fill=brown!20}]{\shortstack{$R_{YY}$ \\ $ 2\alpha x_{2r}$}} & 
            \gate[2, style={fill=yellow!20}]{\shortstack{$R_{XX}$ \\ $ 2\alpha x_{2r}$}} && &
            \\
            \lstick{$q_2$}& 
             \gate[style={fill=gray!20}]{U3} & 
             \gate[2, style={fill=cyan!20}]{\shortstack{$R_{ZZ}$ \\ $\alpha x_1$}} & 
            \gate[2, style={fill=brown!20}]{\shortstack{$R_{YY}$ \\ $ \alpha x_1$}} & 
            \gate[2, style={fill=yellow!20}]{\shortstack{$R_{XX}$ \\ $ \alpha x_1$}}& 
            {} &
            {} & 
            {}&
            {}
            &
            \gate[2, style={fill=cyan!20}]{\shortstack{$R_{ZZ}$ \\ $2\alpha x_{1r}$}} & 
            \gate[2, style={fill=brown!20}]{\shortstack{$R_{YY}$ \\ $ 2\alpha x_{1r}$}} & 
            \gate[2, style={fill=yellow!20}]{\shortstack{$R_{XX}$ \\ $ 2\alpha x_{1r}$}}
            & 
            {} & 
            {} & 
            {} & {} & 
            {} &
            \\
            \lstick{$q_3$}& 
            \gate[style={fill=gray!20}]{U3}
            & {}
            & {}
            & {}
            & {}
            & {}
            & {}
            & {}
            & {}
            & {}
            & {} & 
            {} & 
            {} & {} & 
            {} & 
            {} & 
        \end{quantikz}
        }
    \caption{Example of feature encoding on Heisenberg Hamiltonian feature map with data reloading technique. We define $\alpha$ as the Pauli rotation factor and $x_i$ is the i-th data feature. We assume to have input data with 3 features $\{x_0, x_1, x_2\}$ to encode into four qubits $q_0, q_1, q_2,$ and $q_3$. We tile the input $x_{ir}$ (reloaded) features next to the original ones, as if we repeat the Heisenberg Hamiltonian feature map twice. Indeed, it is as if we had the original circuit with the reloaded one $U_r$ next to it. We also scale the reloaded features differently to balance the amount of information encoded in
the circuit.}
    \label{fig: data_reuploading circuit}
\end{figure}

\clearpage
\section{Classical ML classifier performance}
\label{app: classical_ML results}

In Table~\ref{tab: Classical XGB results}, we present the performance of the classical XGBoost and SVC models on the four different datasets used in the quantum experiments. As with the quantum results, we report the average results over 10 different splits for each dataset. Additionally, in Table~\ref{tab: Classical XGB results plastic reduced}, we include the performance of classical XGBoost on the reduced Plastic Astronomy dataset used in the real hardware experiments.

Note that comparing to classical ML modeling was not the focus of this work -- as the goal was to improve a given QML model, that is improve a QML model for a given specified quantum feature map.  Thus we would generally not expect to have a high chance to improve over classical without further tuning or optimizing the particular feature map circuit used as well.

In some cases, such as when using feature selection on the Virtual Screening dataset encoded with a 15-qubit Separate Entangled feature map (with pairwise entanglement) and XGBoost as the classifier, models employing QFEO show a slight performance improvement over classical models. However, overall, we did not observe a significant improvement, even with experiments on the QPU employing a 100-qubits feature map. As noted earlier, performance depends on factors such as the feature map, hyperparameters, noise and error mitigation techniques, and, of course, the data. Therefore, further analysis -- such as testing different feature maps and feature map configurations, different datasets, increasing qubit count, or utilizing more advanced error mitigation techniques -- could potentially help find a benefit from using the quantum models over classical models in some cases.

\begin{table*}[h]
\centering
\begin{tabular}{|c|c|c|}
\hline
Classifier & Dataset & Score \\ 
\Xhline{3\arrayrulewidth}
\multirow{4}{*}{XGBoost} & Churn & $0.7311 \pm 0.011$ \\ 
\cline{2-3} 
& Virtual Screening & $0.7225 \pm 0.004$ \\
\cline{2-3}
& German Numeric & $0.7859 \pm 0.018$\\
\cline{2-3}
& Plastic Astronomy & $0.9243 \pm 0.011$\\
\Xhline{3\arrayrulewidth}
\multirow{4}{*}{SVC} & Churn & $0.7110 \pm 0.007$ \\ 
\cline{2-3} 
& Virtual Screening &  $0.7271 \pm 0.006$ \\ 
\cline{2-3}
& German Numeric & $0.7868 \pm 0.023$\\
\cline{2-3}
& Plastic Astronomy & $0.8952 \pm 0.014$\\
\hline
\end{tabular}
\caption{Mean ± Std Dev AUC Test Scores for Classical XGBoost and SVC Classifier.}
\label{tab: Classical XGB results}
\end{table*}

\begin{table*}[h]
\centering
\begin{tabular}{|c|c|c|}
\hline
Classifier & Dataset & Score \\ 
\Xhline{3\arrayrulewidth}
XGBoost & Plastic Astronomy red. & $0.8968$ \\
\hline
\end{tabular}
\caption{Mean ± Std Dev AUC Test Scores for Classical XGBoost Classifier on reduced Plastic Astronomy dataset.}
\label{tab: Classical XGB results plastic reduced}
\end{table*}